\documentclass[
eqnum, %
nolinenum
]{jfp}

\usepackage[scaled=0.865]{helvet} %
\usepackage{newtxmath}
\usepackage{bold-extra}

\usepackage{listings}
\usepackage{url}
\usepackage{circledsteps}
\usepackage{amsmath}

\usepackage{natbib}
\usepackage[section]{placeins}
\usepackage{textcomp}
\usepackage{tabularx}
\usepackage{mathpartir}
\usepackage{ifthen}
\usepackage{balance}
\usepackage{listings-golang} %
\usepackage{color}
\usepackage[all,cmtip]{xy}
\usepackage{cleveref}
\usepackage{csquotes}
\usepackage[shortlabels]{enumitem}
\usepackage{marginnote}
\usepackage{xstring}
\usepackage{fancyvrb}
\usepackage{dashrule}
\usepackage{tikz}
\usetikzlibrary{positioning}
\usepackage{grabbox}
\usepackage{csquotes}

\hypersetup{hypertexnames=false}

\definecolor{GrayBgColor}{rgb}{0.9, 0.9, 0.9}
\definecolor{GrayFgColor}{rgb}{0.2, 0.2, 0.2}
\definecolor{StringColor}{rgb}{0.0, 0.38039, 0.141176}
\definecolor{DarkBlueColor}{rgb}{0.0, 0.019, 0.41}
\definecolor{DarkRedColor}{rgb}{0.35, 0.0, 0.0}
\usepackage{graphicx}

\makeatother
\makeatletter

\lstdefinelanguage{myhaskell}%
  {morekeywords={if,then,else,case,class,data,default,deriving,%
      hiding,if,in,infix,infixl,infixr,import,instance,let,module,%
      newtype,of,qualified,type,where,do,forall},%
   sensitive,%
   keywordstyle=\color{DarkRedColor}\bf\ttfamily,
   stringstyle=\color{StringColor},
   morecomment=[l]--,%
   morecomment=[n]{\{-}{-\}},%
   morestring=[b]",%
   literate={REC}{{$\texttt{RE}_{\texttt{C}}$}}3 {RC}{{$\texttt{R}_{\texttt{C}}$}}2 {cast_RE}{{$\texttt{cast}_{\texttt{RE}}$}}5 {ListC}{{$\texttt{List}_{\texttt{C}}$}}5 {FunC}{{$\texttt{Fun}_{\texttt{C}}$}}4%
  }[keywords,comments,strings]%

\renewcommand\emph[1]{{\it #1}}

\newcommand\qqquad{\quad\qquad}

\newcommand\go[1]{\lstinline[language=Golang]{#1}}
\newcommand\hs[1]{\lstinline[language=myhaskell,mathescape=true]{#1}}
\lstnewenvironment{golisting}{\lstset{language=Golang,mathescape=true}}{}

\newcommand\FGGm{\ensuremath{\textrm{FGG}^{-}}}

\newcommand{\mathem}{\sf}
\newcommand{\kw}[1]{{\normalfont\sffamily\fontseries{b}\selectfont #1}}
\newcommand{\IN}{\mbox{\kw{in}}}
\newcommand{\LET}{\mbox{\kw{let}}}

\newcommand{\CASE}{\mbox{\kw{case}}}

\newcommand{\OF}{\mbox{\kw{of}}}

\newcommand\Nat{\mathbb{N}}

\newcommand{\Foreach}[2]{\overline{#1}^{#2}} %
\newcommand{\ForeachN}[1]{\overline{#1}}    %
\newcommand\Multi[2][!*NEVER USED ARGUMENT*!]{%
  \ifthenelse{\equal{#1}{!*NEVER USED ARGUMENT*!}}{\ForeachN{#2}}{\Foreach{#2}{#1}}%
}
\newcommand\MultiSubst[3][!*NEVER USED ARGUMENT*!]{\Angle{\Multi[#1]{\subst{#2}{#3}}}}
\newcommand\MultiN[1]{\Multi[n]{#1}}
\newcommand\MultiQ[1]{\Multi[q]{#1}}
\newcommand\MultiP[1]{\Multi[p]{#1}}

\newcommand{\TYPE}{\mbox{\kw{type}}}
\newcommand{\STRUCT}{\mbox{\kw{struct}}}
\newcommand{\INTERFACE}{\mbox{\kw{interface}}}
\newcommand{\FUNC}{\mbox{\kw{func}}}
\newcommand{\RETURN}{\mbox{\kw{return}}}

\newcommand{\MAIN}{\mbox{\mathem main}}

\newcommand\GoSynCatName[1]{\mbox{\small #1}}

\newcommand{\methodSpecificationsRel}{{\mathsf{methods}}}
\newcommand{\methodSpecifications}[1]{\methodSpecificationsRel(#1)}

\newcommand{\methodName}[1]{{\mathem methodName}(#1)}

\newcommand\turnsOk{\turnsG{ok}}
\newcommand\fgTyOk[2]{#1 \turnsOk #2}

\newcommand\fgFormalsOk[2]{#1 \turnsOk #2}
\newcommand\fgDeclOk[1]{\turnsOk #1}

\newcommand{\fgTyEnv}{\Delta}
\newcommand{\fgEnv}{\Gamma}
\newcommand{\EmptyFgTyEnv}{\emptyset}
\newcommand{\EmptyFgEnv}{\emptyset}
\newcommand{\fgEmptyEnv}{\EmptyFgEnv}
\newcommand{\fgEmptyTyEnv}{\EmptyFgTyEnv}
\newcommand{\turnsFGG}{\, \vdash_{\mathsf{G}} \,}
\newcommand{\fgOk}[1]{#1~\mathsf{ok}}

\newcommand{\subtypeOf}[2]{#1 \subtypeOfSym #2}
\newcommand{\subtypeOfTT}[2]{\ensuremath{\subtypeOf{\texttt{#1}}{\texttt{#2}}}}
\newcommand{\subtypeOfSym}{\mathrel{<:}}

\newcommand{\FGGUniqueStructs}{\Restriction{fgg-unique-structs}}
\newcommand{\FGGUniqueFields}{\Restriction{fgg-distinct-fields}} %

\newcommand{\FGGUniqueReceiver}{\Restriction{fgg-unique-method-defs}} %

\newcommand{\EvCtx}{{\mathcal E}}
\newcommand{\reduceSym}{\longrightarrow}
\newcommand\reduceSymFG{\reduceSym}
\newcommand\reduceSymTL{\reduceSym}
\newcommand{\reduce}[2]{#1 \reduceSym  #2}

\newcommand{\reduceStarSym}{\reduceSym^*}
\newcommand{\reduceStar}[2]{#1 \reduceStarSym #2}

\newcommand{\reducek}[3]{#2 \reduceSym^{#1} #3} %
\newcommand{\subst}[2]{#1 \mapsto #2}

\newcommand{\reduceFGk}[3]{#1 \reduceSymFG^{#2} #3} %

\newcommand{\reduceExpTL}[3]{#2 \reduceSymTL_{#1} #3}
\newcommand{\reduceTL}[2]{#1 \reduceSymTL #2}
\newcommand{\reduceTLN}[3]{#2 \reduceSymTL^*_{#1} #3} %

\newcommand{\vbFG}{\theta} %
\newcommand{\tbFG}{\eta} %
\newcommand\fgSubst{\vbFG}
\newcommand\fgTySubst{\tbFG}

\makeatletter
\newcommand{\xRightarrow}[2][]{\ext@arrow 0359\Rightarrowfill@{#1}{#2}}
\makeatother

\newcommand\fgTyVar{\alpha}
\newcommand\fgTyVarAux{\beta}
\newcommand\fgType{\tau}
\newcommand\fgTypeAux{\sigma}
\newcommand\fgGetTyVars[1]{\hat{#1}}

\newcommand\fgDecls{\ForeachN{D}}

\newcommand{\TL}{\mbox{TL}} %

\newcommand\kA{k'}

\newcommand{\kT}{K}         %
\newcommand{\domT}{{\mathem{dom}}} %
\newcommand\Dom[1]{\domT(#1)}

\newcommand\LrAngle[1]{\left\langle#1\right\rangle}
\newcommand\Angle[1]{\langle#1\rangle}
\newcommand{\pair}[2]{\Angle{#1,#2}}

\newcommand{\quadr}[4]{\Angle{#1,#2, #3, #4}}

\newcommand{\program}{\mathit{Prog}}

\newcommand{\clsT}{\mathit{Cls}}
\newcommand{\patT}{\mathit{Pat}}
\newcommand{\expT}{E}

\newcommand{\xT}{X}

\newcommand{\xTy}[1]{\xT_{#1}} %

\newcommand\ttmath[1]{\mbox{\normalfont\tt #1}}

\newcommand\PL{\ttmath{(}}
\newcommand\PR{\ttmath{)}}
\newcommand\Tuple[1]{\PL #1 \PR}
\newcommand\Pair[2]{\Tuple{#1\Comma #2}}
\newcommand\Triple[3]{\Tuple{#1\Comma #2\Comma #3}}
\newcommand\Quadr[4]{\Tuple{#1\Comma #2\Comma #3\Comma #4}}
\newcommand\Comma{\hspace{-1pt}\ttmath{,}} %

\newcommand{\uT}{V}                 %

\newcommand{\ignore}[1]{}

\newcommand{\shorteq}{%
  \settowidth{\@tempdima}{-}%
  \resizebox{\@tempdima}{\height}{=}%
}

\newcommand{\tcaseof}[2]{\CASE\, #1 \,\OF\, #2}

\newcommand{\tletrecin}[2]{\LET\ #1 \ \IN\ #2}  %
\newcommand\tllambda[1]{\lambda #1 \texttt{.}}

\newcommand\turnsSomething[1]{\mathrel{\vdash_{\hspace{-0.1em}\scalebox{0.5}{\ensuremath{\mathsf{#1}}}}}}

\newcommand{\vbTL}{\rho}
\newcommand\tlSubst{\vbTL}
\newcommand{\vbMethTL}{\mu}
\newcommand\tlMethTable{\vbMethTL}

\newcommand{\EvCtxT}{{\mathcal R}}

\newcommand{\REvCtxT}{{\mathcal R}} %
\newcommand\Hole{\scalebox{0.8}{\ensuremath{\square}}}

\newcommand{\turnsG}[1]{\turnsSomething{#1}}

\newcommand{\trans}[4]{#2 \turnsG{#1} #3 \leadsto #4}

\newcommand{\tdExpTrans}[3]{\trans{\mathsf{exp}}{#1}{#2}{#3}}
\newcommand{\tdExpTransEmpty}[2]{\tdExpTrans{\pair{\fgEmptyTyEnv}{\fgEmptyEnv}}{#1}{#2}}

\newcommand{\tdMethTrans}[2]{\trans{\mathsf{meth}}{}{#1}{#2}}

\newcommand{\TurnsTdProg}{\turnsG{\mathsf{prog}}}
\newcommand{\tdProgTrans}[2]{\TurnsTdProg #1 \leadsto #2}

\newcommand\tdCheckSubst[5]{\trans{\mathsf{subst}}{#1}{\subst{#2}{#3} : #4}{#5}}

\newcommand\tdMethods[4]{\pair{#1}{#2} \in \methodSpecifications{#3, #4}}

\newcommand\tdICons[4]{\tdUpcast{#1}{\subtypeOf{#2}{#3}}{#4}}

\newcommand{\tdUpcast}[3]{\trans{\mathsf{coerce}}{#1}{#2}{#3}}

\newcommand{\distinct}[1]{{\mathem distinct}(#1)}

\newcommand{\methodLookup}[2]{{\mathem methodLookup}(#1, #2)}

\newcommand{\mapPerm}{\pi}               %

\newcommand\MultiThree[1]{\Multi[\raisebox{-1pt}{\scalebox{0.6}{\,3}}]{#1}}
\newcommand\TripleY{\Tuple{\MultiThree{Y}}}
\newcommand\QuadrY[1]{\Tuple{#1\Comma\MultiThree{Y}}}
\newcommand\PartialM[2]{\lambda \Triple{Y_1}{Y_2}{Y_3} . #1\ \Quadr{#2}{Y_1}{Y_2}{Y_3}}

\newcommand\Sem[1]{\llbracket#1\rrbracket}
\newcommand{\redLRConfk}[4]{ #3 \approx #4 \in \Sem{#2}_{#1}}

\newcommand\LREquivNoTy[3]{#1 \approx_{#2} #3}
\newcommand\LREquiv[4]{\redLRConfk{#4}{#3}{#1}{#2}}
\newcommand\LREquivVal[4]{#1 \equiv #2 \in \Sem{#3}_{#4}}

\newcommand{\mT}[2]{\xT_{{#1},{#2}}}

\newcommand\sSynCatName[1]{\mbox{\small #1}}

\newcommand\sTransSection[2]{%
  \begin{tabularx}{\textwidth}{Xr}%
    #1\hfill & {\small \it #2}%
  \end{tabularx}%
}
\newcolumntype{Y}{>{\centering\arraybackslash}X}
\newcommand\TransSectionCenter[1]{%
  \begin{tabularx}{\textwidth}{Y}%
    {{\small \it #1}}
  \end{tabularx}%
}

\newcommand\sOver[2][!*NEVER USED ARGUMENT*!]{%
  \ifthenelse{\equal{#1}{!*NEVER USED ARGUMENT*!}}{\overline{#2}}{\overline{#2}^{#1}}%
}

\newcommand\sOptEmpty\bullet

\newcommand\RefTirName[1]{\TirName{#1}}
\newcommand\Rule[1]{\RefTirName{#1}}

\newcommand{\ei}{\end{array}}
\newcommand{\ba}{\begin{array}}
\newcommand{\ea}{\end{array}}
\newcommand{\bda}{\[\ba}
\newcommand{\eda}{\ea\]}
\newcommand{\bp}{\begin{quote}\tt\begin{tabbing}}
\newcommand{\ep}{\end{tabbing}\end{quote}}

\def\ruleform#1{{\setlength{\fboxrule}{0.5pt}\fbox{\normalsize \ensuremath{#1}}}}

\newcommand{\turns}{\, \vdash \,}

\newcommand{\fv}[1]{\mathsf{fv}(#1)}
\newcommand{\ftv}[1]{\mathsf{ftv}(#1)}
\newcommand{\dom}[1]{\mathsf{dom}(#1)}

\newcommand{\Figurebox}[1]
        {\fbox{\begin{minipage}{\textwidth} #1 \end{minipage}}}
\newcommand\boxfigWithExplanation[4]
           {\begin{figure*}\Figurebox{#4}\caption{\label{#1}#2}#3\end{figure*}}
\newcommand{\boxfig}[3]{\boxfigWithExplanation{#1}{#2}{}{#3}}

\newcommand\noteForall[1]{(\forall\,#1)}

\newcommand\DivergeSym{\mathsf{diverge}}
\newcommand\Diverge[1]{\DivergeSym(#1)}
\newcommand\MLookupRes[4]{\quadr{#1}{#2}{#3}{#4}}
\newcommand\DOT{\mathbin{.}}
\newenvironment{mathparNarrow}{%
  
  \begin{mathpar}
}{%
  \end{mathpar}
}

\makeatletter
\newlength{\CaseIndent}
\newcommand{\listmargincase}{
  \ifnum\@listdepth=1
    \setlength{\leftmargin}{\CaseIndent}%
  \else
    \ifnum\@listdepth=2
      \setlength{\leftmargin}{2\CaseIndent}%
    \else
      \ifnum\@listdepth=3
        \setlength{\leftmargin}{3\CaseIndent}%
      \else
        \ifnum\@listdepth=4
        \setlength{\leftmargin}{4\CaseIndent}%
        \fi
      \fi
    \fi
  \fi}

\newsavebox{\CDName}
\newcommand\CaseFormat[1]{\textit{#1}}
\newenvironment{CaseDistinction}[2][no]{%
  \begin{CaseDistinctionExplicit}{#2}{\ifthenelse{\equal{#1}{repeat}}{ #2}{}}%
}{%
  \end{CaseDistinctionExplicit}%
}
\newenvironment{CaseDistinctionExplicit}[2]
  {\par\noindent\CaseFormat{Case distinction} #1.%
   \sbox{\CDName}{\CaseFormat{End case distinction}#2.}%
   \begin{itemize}[topsep=0pt]{}{\listmargincase\setlength\listparindent\parindent}%
  }
  {\end{itemize}\usebox{\CDName}%
  }
\newcommand{\Case}[1]{\item\CaseFormat{Case} #1:}

\newenvironment{igather}{\collect@body\@igather}{\global\@ignoretrue}

\newcommand\@igather[1]{{%
\let\old@label@in@display\label@in@display%
\renewcommand\label@in@display[1]{\stepcounter{equation}\Labelnote{##1}\tag{\theequation}\old@label@in@display{##1}}%
\begin{gather*}
#1
\end{gather*}
}}
\makeatother

\newcommand\Labelnote[1]{} %

\newenvironment{EnumAlph}{%
  
  \renewcommand\theenumi{\alph{enumi}}
  \begin{enumerate}[topsep=0pt]
}{\end{enumerate}}

\newcommand\LabelQualifier{}
\newcommand\QualifyLabel[1]{#1::\LabelQualifier}
\newcommand\QLabel[1]{\label{\QualifyLabel{#1}}}
\newcommand\QRef[1]{\eqref{\QualifyLabel{#1}}}

\makeatletter
\newcommand{\TextLabel}[1]{\refstepcounter{equation}\label{\QualifyLabel{#1}}\Labelnote{\QualifyLabel{#1}}\textup{\tagform@{\theequation}}}
\makeatother

\renewenvironment{proof}[1][**empty**]{%
        \setcounter{equation}{0}\par\addvspace{6pt}\noindent%
        \ifthenelse{\equal{#1}{**empty**}}{\textbf{Proof.}}{\textbf{#1.}}%
        \hskip5.5pt%
        }{\par\addvspace{6pt}}

\newcommand\ReasonAbove[2]{\overset{#2}{#1}}

\makeatletter
\def\smallunderbrace#1{\mathop{\vtop{\m@th\ialign{##\crcr
   $\hfil\displaystyle{#1}\hfil$\crcr
   \noalign{\kern3\p@\nointerlineskip}%
   \tiny\upbracefill\crcr\noalign{\kern3\p@}}}}\limits}
\makeatother
\newcommand\BraceBelow[2]{\smallunderbrace{#1}_{#2}}

\newcommand{\verteq}{\rotatebox{90}{\footnotesize$\,=$}}
\newcommand{\EqualBelow}[2]{\underset{\scriptstyle\overset{\mkern4mu\verteq}{#2}}{#1}}

\newcommand\LabeledRule[1]{{\begin{array}[t]{@{}c@{\qquad}r}#1\end{array}}}

\renewcommand\smallsum{\mathrm{\Sigma}}
\newcommand\MyQED{\hfill$\blacksquare$}

\newcommand\subColored{no}
\newcommand\sub[1]{\ifthenelse{\equal{\subColored}{yes}}{\color{StringColor}{\texttt{#1}}}{\texttt{#1}}}
\newcommand\tyarg[1]{{\color{DarkBlueColor}{\texttt{#1}}}}
\newcommand\transFun[2]{\ensuremath{\texttt{#1}_{\sub{#2}}}}

\newcommand\coerceFun[1]{\transFun{coerce}{#1}}

\newcommand\NameSetFGG{\mathcal{N}}
\newcommand\structNameSet{\NameSetFGG_{\mathit{struct}}}
\newcommand\ifaceNameSet{\NameSetFGG_{\mathit{iface}}}
\newcommand\tyvarSet{\NameSetFGG_{\mathit{tyvar}}}
\newcommand\fieldNameSet{\NameSetFGG_{\mathit{field}}}
\newcommand\methodNameSet{\NameSetFGG_{\mathit{method}}}
\newcommand\expvarSet{\NameSetFGG_{\mathit{var}}}

\newcommand\NameSetTL{\mathcal{V}}
\newcommand\TLVarSet{\NameSetTL_{\mathit{Var}}}
\newcommand\TLConsSet{\NameSetTL_{\mathit{Con}}}

\newcommand\Restriction[1]{{\normalfont\Rule{#1}}}

\newcommand\ImplementationURL{\url{https://github.com/skogsbaer/fgg-translate} and \url{http://doi.org/10.5281/zenodo.8147425}}
\newcommand\Size[1]{|#1|}
\newcommand\Measure{\mathcal M}

\newcommand\this{\mathit{this}}
\newcommand\that{\mathit{that}}
\newcommand\This{\mathit{This}}
\newcommand\That{\mathit{That}}
\newcommand\content{\mathit{content}}
\newcommand\val{\mathit{val}}
\newcommand\IntT{\mathit{int}}
\newcommand\eq{\mathit{eq}}
\newcommand\BoxT{\mathit{Box}}
\newcommand\Eq{\mathit{Eq}}
\newcommand\Any{\mathit{Any}}
\newcommand\Num{\mathit{Num}}
\newcommand\bool{\mathit{bool}}
\newcommand\main{\mathit{main}}
\newcommand\Void{\Tuple{}}
\newcommand\One{\Tuple{1}}
\newcommand\Two{\Tuple{2}}
\newcommand\BoxVal[1]{\Tuple{\Tuple{#1}}}

\newcommand\Fbox{\grabbox0\hbox{\fbox{\usebox0}}}
\newcommand\PFGG{P_{\textrm{FGG}}}
\newcommand\PFGGrestricted{P_{\textrm{FGG$^{\star}$}}}
\newcommand\PFG{P_{\textrm{FG}}}
\newcommand\PTL{P_{\textrm{TL}}}
\newcommand\DiagLabel[1]{{\footnotesize{#1}}}

\newcommand\EraseSym{{\mathem erase}}
\newcommand\EraseTy[2]{\EraseSym(#1, #2)}
\newcommand\Erase[1]{\EraseSym(#1)}
\newcommand\EraseLam{\mathbf{K}_{\lambda}}

\newcommand\AppRef[1]{Appendix~\ref{#1} (page~\pageref{#1})}

\theoremstyle{plain}
\newtheorem{theorem}{Theorem}[subsection]
\newtheorem{lemma}[theorem]{Lemma}
\newtheorem{corollary}[theorem]{Corollary}
\theoremstyle{definition}
\newtheorem{definition}[theorem]{Definition}
\newtheorem{convention}[theorem]{Convention}
\newtheorem{assumption}[theorem]{Assumption}

\begin{document}

\journaltitle{JFP}
\cpr{Cambridge University Press}
\doival{10.1017/xxxxx}

\totalpg{\pageref{lastpage01}}
\jnlDoiYr{2023}

\title{A Type-Directed, Dictionary-Passing Translation of Method Overloading and Structural Subtyping in Featherweight Generic Go}

\begin{authgrp}
\author{Martin Sulzmann}
\affiliation{
  Karlsruhe University of Applied Sciences, Germany\\
  \email{martin.sulzmann@h-ka.de}
}
\author{Stefan Wehr}
\affiliation{
  Offenburg University of Applied Sciences, Germany\\
  \email{stefan.wehr@hs-offenburg.de}
}
\end{authgrp}

\begin{abstract}
Featherweight Generic Go (FGG) is a minimal core calculus modeling the
essential features of the programming language Go. It includes support
for overloaded methods, interface types, structural subtyping and
generics.  The most straightforward semantic description of the dynamic
behavior of FGG programs is to resolve method calls based on runtime
type information of the receiver.

This article shows a different approach by defining a type-directed
translation from \FGGm\ to an untyped lambda-calculus.
\FGGm\ includes all features of FGG but type assertions.
The translation of an \FGGm\ program provides evidence for the availability of methods as
additional dictionary parameters, similar to the dictionary-passing
approach known from Haskell type classes.  Then, method calls can be
resolved by a simple lookup of the method definition in the dictionary.

Every program in the image of the translation has the same dynamic
semantics as its source \FGGm\ program. The proof of this result is based
on a syntactic, step-indexed logical relation. The step-index ensures a
well-founded definition of the relation in the presence of recursive
interface types and recursive methods.
Although being non-deterministic, the translation is coherent.
\end{abstract}

\maketitle

\section{Introduction}
\label{sec:intro}

Go~\citeyearpar{golang} is a statically typed programming language introduced by
Google in 2009.
It supports method overloading by allowing multiple declarations of the same method signature
for different receivers. Receivers are structs, similar to structs in C.
The language also supports interfaces; as in many object-oriented languages, an interface
consists of a set of method signatures. But unlike in many object-oriented
languages, subtyping in Go is structural not nominal.

Earlier work by \citet{FeatherweightGo}
introduces Featherweight Go (FG), a minimal core calculus that
covers method overloading, structs, interfaces and structural subtyping.
Their work specifies static typing rules and a dynamic semantics for FG based on runtime method lookup.
However, the actual Go implementation appears to employ a different dynamic semantics.
Quoting Griesemer and co-workers:
\enquote{\emph{Go is designed to enable efficient implementation. Structures are laid out in memory as a sequence of fields,
while an interface is a pair of a pointer to an underlying structure and a pointer to a dictionary of methods.}}

In our own prior work~\citep{SulzmannWehr-aplas2021,SulzmannWehr-mpc2022},
we formalize a type-directed dictionary-passing translation for FG
and establish its semantic equivalence with FG's dynamic semantics.
Griesemer and coworkers also introduce Featherweight Generic Go (FGG), an extension of FG with generics.
In this work, we show how our translation approach can be extended to deal with generics.
Our focus is on the integration of generics with method overloading and structural subtyping.
Hence, we consider \FGGm{}, which is equivalent to FGG but does not support type assertions.
Our contributions are as follows:

\begin{itemize}
\item We specify the translation of source \FGGm\ programs to
  an untyped $\lambda$-calculus with recursive let-bindings, constructors and pattern matching.
  We employ a dictionary-passing translation scheme \`a la type classes~\citep{Hall:1996:TCH:227699.227700}
  to statically resolve overloaded method calls. The translation
  is guided by the typing of the \FGGm\ program. As the typing rules include a subsumption
  rule, the translation is inherently non-deterministic.
\item We establish the semantic correctness of the dictionary-passing translation.
  The result relies on a syntactic, step-indexed logical relation to ensure well-foundedness of
  definitions in the presence of recursive interface types and recursive methods.
\item We show that values produced by different translations of the same program are identical
  up to dictionaries embedded inside these values.
\item We report on an implementation of the translation.
\end{itemize}

The upcoming \Cref{sec:examples} presents an overview of our translation by example.
\Cref{sec:featherweight-generic-go} gives a recap of the source language \FGGm, whereas
\Cref{sec:type-direct-transl} defines the target language and the translation itself.
Next, \Cref{sec:formal-properties} establishes the formal properties of the translation,
rigorous proofs of our results can be found in the Appendix.
\Cref{sec:discussion} presents the implementation,
\Cref{sec:related-work} covers related work.
Finally, \Cref{sec:future-work-concl} summarizes this work and points out directions for future work.

\section{Overview}
\label{sec:examples}

This section introduces Featherweight Generic Go \cite[FGG][]{FeatherweightGo} and
our type-directed dictionary-passing translation through a
series of examples.
FGG is a tiny model of Go that includes essential typing features
such as method overloading, structs, interfaces, structural subtyping, and the extension with generics.
Its original formulation also includes type assertions (dynamic type casts). As we omitted
this feature from our translation, we use the name \FGGm{} to refer to the
source language of our translation. Except for the omission of type assertions, \FGGm{} and FGG
are equivalent.

An \FGGm\ program consists of declarations for structs, interfaces, methods, and a main function. Function and method bodies
only contain a single return statement, all expression are free from side effects.
For the examples in this section, we extend \FGGm\ with primitive types for integers and strings, with
an operator \go{+} for string concatenation and a builtin function \go{intToString},
with definitions of local variables, and with function definitions.

We will first consider \FGGm\ without generics
to highlight the idea behind our type-directed dictionary-passing translation scheme.
Then, we show how the translation scheme can be adapted to deal with the addition of generics.
All examples have been checked against our implementation\footnote{\ImplementationURL}
of the translation.

\subsection{Starting Without Generics}

\boxfig{f:fgg-format}{String-formatting and its translation}{
  \renewcommand\subColored{yes}
  \vspace{-1ex}
\begin{lstlisting}[name=goexample,numbers=left,escapechar=@,language=Golang,mathescape=true]
type Num    struct { val int }
type Format interface { format() string }
type Pretty interface { format() string; pretty() string }

func (this Num) format() string { return intToString(this.val) }
func (this Num) pretty() string { return this.format() }

func formatSome(x Format) string { return x.format() }

func main() {
  var s1 string = formatSome(Num{1})
  var pr Pretty = Num{2}
  var s2 string = formatSome(pr)
}
\end{lstlisting}
  \hrule{}
\begin{lstlisting}[name=hsexample,firstnumber=15,numbers=left,escapechar=@,language=myhaskell,mathescape=true,lastline=23]
-- Field access for struct Num
val x = x

-- Method calls on interfaces
format$_{\sub{Format}}$ (x, f) = f x$\hspace{10.1mm}$-- call format on receiver of type Format
format$_{\sub{Pretty}}$ (x, (f,p)) = f x$\hspace{3mm}$-- call format on receiver of type Pretty
pretty$_{\sub{Pretty}}$ (x, (f,p)) = p x$\hspace{3mm}$-- call pretty on receiver of type Pretty

-- Method definitions (lines 5 and 6 in the FGG code)
format$_{\sub{Num}}$ this = intToString (val this)
pretty$_{\sub{Num}}$ this = format$_{\sub{Num}}$ this

-- Coercions
toFormat$_{\sub{Num}}$ x  = (x,format$_{\sub{Num}}$)               -- Num <: Format
toPretty$_{\sub{Num}}$ x  = (x,(format$_{\sub{Num}}$,pretty$_{\sub{Num}}$))   -- Num <: Pretty
toFormat$_{\sub{Pretty}}$ (x,(f,p))  = (x,f)            -- Pretty <: Format

-- Function definitions (lines 8 and 10 in the FGG code)
formatSome x = format$_{\sub{Format}}$ x

main = let s1 = formatSome (toFormat$_{\sub{Num}}$ 1)
           pr = toPretty$_{\sub{Num}}$ 2
           s2 = formatSome (toFormat$_{\sub{Pretty}}$ pr)
\end{lstlisting}
  \vspace{-1ex}
  \renewcommand\subColored{no}
}

The upper part of \Cref{f:fgg-format} shows an (extended) \FGGm{} program
for formatting values as strings. The code does not use
generics yet.

Structs in Go are similar to structs in C,
a syntactic difference is the Go convention that field or variable names precede their types.
Here, struct \go{Num} has a single field \go{val} of type \go{int}, so it simply
acts as a wrapper for integers.

Interfaces in Go declare sets of method signatures sharing the same receiver
where method names must be distinct and the receiver is left implicit.
Interfaces are types and describe all receivers that implement the methods declared by the interface.
In our example, interface \go{Format} declares a method \go{format} for rendering its receiver as a string.
The second interface \go{Pretty} also declares \go{format}, but adds
a second method \go{pretty} with the intention to produce a visually more attractive output.

Methods and functions are introduced via the keyword \go{func}.
A method can be distinguished from a
function as the receiver
argument in parenthesis precedes the method name.
Methods can be overloaded on the receiver type.
In lines 5 and 6, we find methods \go{format} and \go{pretty}, respectively, for receiver type
\go{Num}.
In the body of \go{format},
we assume a builtin function \go{intToString} for converting integers to strings.
Lines 8 and 10 define two functions.

An interface only names a set of method signatures, its definition is not required
for a method to be valid. For example,
the methods in lines~5 and~6 could be defined without the interfaces in lines~2 and~3,
or the methods could be placed before the interfaces.

However, interfaces and method definitions imply structural subtype relations.
Interface \go{Format} contains a subset of the methods declared by interface \go{Pretty}.
Hence, \go{Pretty} is a structural subtype of \go{Format}, written (1) $\subtypeOfTT{Pretty}{Format}$.
Line~5 defines method \go{format}
for receiver type \go{Num},
we say that \go{Num} implements method \go{format}.
Hence, \go{Num} is a structural subtype of \go{Format}, written (2) $\subtypeOfTT{Num}{Format}$.
Receiver \go{Num} also implements the \go{pretty} method, see line~6.
Hence, we also find that (3) $\subtypeOfTT{Num}{Pretty}$.
Structural subtype relations play a crucial role when type checking programs.

For example, consider the function call \go{formatSome(Num\{1\})} in line~11.
Here, \go{Num\{1\}} is a value of the \go{Num} struct with \go{val} set to \go{1}.
From above we find that (2) $\subtypeOfTT{Num}{Format}$.
That is, \go{Num} implements the \go{Format} interface and therefore the function call type checks.
Consider the variable declaration and assignment in line~12.
Value \go{Num\{2\}} is assigned to a variable of interface type \go{Pretty}.
Based on the subtype relation  (3) $\subtypeOfTT{Num}{Pretty}$ the assignment type checks.
Consider the function call \go{formatSome(pr)} in line~13, where \go{pr} has type \go{Pretty}.
Based on the subtype relation (1) $\subtypeOfTT{Pretty}{Format}$ the function call type checks.

In~\cite{FeatherweightGo}, the dynamic behavior of programs is explained via runtime lookup of methods, where
based on the receiver's runtime type the appropriate method definition is selected.
The Go (and FGG/\FGGm{})
conditions demand that for each method name and receiver type there can be at most one definition.
This guarantees that method calls can be resolved unambiguously.

\subsection{Type-Directed Translation}

We explain the meaning of extended \FGGm{} programs by translation into
an untyped $\lambda$-calculus with recursive top-level definitions,
let-bindings, pattern matching,
integers, strings,
an operator \texttt{++} for string concatenation, and
a builtin function \texttt{intToString}.
We will use a Haskell-style notation.

Method definitions belonging to an interface are grouped together in a \emph{dictionary} of methods.
Thus,  method calls can be turned into primitive function calls by simply looking up
the appropriate method in the dictionary.
Structural subtype relations are turned into \emph{coercion} functions that
transform, for example, a struct value into an interface value to
make sure that the appropriate dictionaries are available.
Where to insert dictionaries and coercions in the program is guided
by the type checking rules.
Hence, the translation is type-directed.

Our translation strategy can be summarized as follows:
\begin{description}
\item[Struct.] An \FGGm\ value at the type of a struct with $n$ fields is represented by an $n$-tuple
  holding the values of the fields. We call such an $n$-tuple a \emph{struct value}.
\item[Interface.] An \FGGm\ value at the type of an interface is represented as a pair
  $(V, \mathcal{D})$, where $V$ is a struct value
  and $\mathcal D$ is a \emph{method dictionary}.
  Such a method dictionary is a tuple holding implementations of all interface methods for $V$,
  in order of declaration in the interface.
  We call the pair $(V, \mathcal{D})$ an \emph{interface value}.
\item[Coercion.] A structural subtype relation $\subtypeOf{\fgType}{\fgTypeAux}$ implies a \emph{coercion function}
  to transform the target representation of an \FGGm\ value at type $\fgType$
   into a representation at type $\fgTypeAux$.
\end{description}

The lower part of \Cref{f:fgg-format} gives the translation of our running example.
In this overview section, we identify a $1$-tuple with the single value it holds.

For each field name, we assume a helper function to access the field component,
see line~16.
Method calls on interface values lookup the respective method definition in the dictionary
and apply it to the struct value embedded inside the interface value.
See lines~19-21.
Method definitions translate to plain functions,
see lines~24-25. Recall that
for each method name and receiver type there can be at most one definition.
Hence, the generated function names are all distinct.

Structural subtype relations translate to coercions,
see lines~28-30.
For example, (2) $\subtypeOfTT{Num}{Format}$
translates to the $\mathtt{toFormat}_{\sub{Num}}$ coercion.
Input parameter \go{x} represents a target representation of a \go{Num} value.
The output $\Pair{\mathtt{x}}{\mathtt{format}_{\sub{Num}}}$ is an interface value
holding the receiver and the corresponding method definition.
Coercion $\mathtt{toPretty}_{\sub{Num}}$ corresponds to (3) $\subtypeOfTT{Num}{Pretty}$
and coercion $\mathtt{toFormat}_{\sub{Pretty}}$
to (1) $\subtypeOfTT{Pretty}{Format}$.

The translation of the main function, starting at line~35,
is guided by the type checking of the source program.
Each application of a structural subtype relation leads to the insertion
of the corresponding coercion function in the target program.
For example, the function call \go{formatSome(Num\{1\})}
translates to $\texttt{formatSome}~\texttt{(toFormat}_{\sub{Num}}~\texttt{1)}$
because typing of the source requires (2) $\subtypeOfTT{Num}{Format}$.
The other coercions arise for similar reasons.

\subsection{Adding Generics}

We extend our running example by including pairs, see \Cref{f:fgg-format2}.
The struct type \texttt{Pair[T Any, U Any]} is \emph{generic}
in the type of the pair components, \go{T} and \go{U} are \emph{type variables}.
When introducing type variables we must also specify an upper \emph{type bound} to constrain the set of concrete types
that will replace type variables.
The \emph{bounded type parameter} \texttt{T Any} can therefore be interpreted as
$\forall \texttt{T}. \subtypeOf{\texttt{T}}{\texttt{Any}}$.
Upper bounds are always interface types.
The upper bound \texttt{Any} is satisfied by any type because the set of methods that need to be implemented is empty.

\boxfig{f:fgg-format2}{String-formatting with generics (extending code from \Cref{f:fgg-format})}{
  \renewcommand\subColored{yes}
  \vspace{-1ex}
\begin{lstlisting}[name=goexample,numbers=left,escapechar=@,language=Golang,mathescape=true]
type Any interface {}
type Pair$\tyarg{[T Any, U Any]}$ struct { left T; right U }

func (this Pair$\tyarg{[T Format, U Format]}$) format() string {
  return "(" + this.left.format() + "," + this.right.format() + ")"
}

func main2() {
  var p  Pair$\tyarg{[Num, Num]}$ = Pair$\tyarg{[Num, Num]}${ Num{1}, Num{2} }
  var s1 string = p.format()
  var s2 string = formatSome(p)
}
\end{lstlisting}
  \hrule{}
\begin{lstlisting}[name=hsexample,firstnumber=13,numbers=left,escapechar=@,language=myhaskell,mathescape=true,lastline=16]
-- Field access for struct Pair
left (x, _) = x
right (_, x) = x

-- Method definition (line 4 in the FGG code)
format$_{\sub{Pair}}$ (toFormat$_{\sub{T}}$, toFormat$_{\sub{U}}$) this =
  "(" ++ format$_{\sub{Format}}$ (toFormat$_{\sub{T}}$ (left this)) ++
  "," ++ format$_{\sub{Format}}$ (toFormat$_{\sub{U}}$ (right this)) ++ ")"

-- Coercion Pair[T, U] <: Format (given T <: Format and U <: Format)
toFormat$_{\sub{Pair}}$ (toFormat$_{\sub{T}}$,toFormat$_{\sub{U}}$) p = (p,format$_{\sub{Pair}}$ (toFormat$_{\sub{T}}$,toFormat$_{\sub{U}}$))

-- Main function
main2 = let p = (1, 2)
            s1 = format$_{\sub{Pair}}$ (toFormat$_{\sub{Num}}$, toFormat$_{\sub{Num}}$) p
            s2 = formatSome (toFormat$_{\sub{Pair}}$ (toFormat$_{\sub{Num}}$, toFormat$_{\sub{Num}}$) p)
\end{lstlisting}
  \vspace{-1ex}
  \renewcommand\subColored{no}
}

To format pairs, we need to format the left and right component that are of generic types~\go{T} and~\go{U}.
Hence, the method definition for \go{format}
in line~4 states the type bound \go{Format} for type variables~\go{T} and~\go{U}.
In general, bounds of type parameters for the receiver struct of a method declaration must be in a covariant subtype relation
relative to the bounds in the struct declaration.
This is guaranteed in our case as we find $\subtypeOfTT{Format}{Any}$.
Importantly, the type bounds in line~4 imply the subtype relations (4) $\subtypeOfTT{T}{Format}$
and (5) $\subtypeOfTT{U}{Format}$.
Thus, we can show that the method body type checks.
For example, expression \go{this.left} is of type \go{T}.
Based on (4), this expression is also of type \go{Format} and therefore the method call in line~5
\go{this.left.format()} type checks.

We consider type checking the main function.
Instances for generic type variables must always be explicitly supplied.
Hence, when constructing a pair that holds number values, see line~9, we find \texttt{Pair[Num, Num]}.

Consider the method call \go{p.format()} in line~10.
The receiver struct \go{Pair[T Format, U Format]} of the method definition in line~4 matches
\go{p}'s type \go{Pair[Num, Num]}
by replacing \go{T} and \go{U} by \go{Num}.
The type bounds in the receiver type are satisfied as we know from above that (2) $\subtypeOfTT{Num}{Format}$.
Hence, the method call type checks.

By generalizing the above argument we find that
\begin{center}
  (6) \quad $ \{\subtypeOfTT{T}{Format}, \subtypeOfTT{U}{Format} \} \turns \subtypeOfTT{Pair[T, U]}{Format}$.
\end{center}
That is, under the assumptions $\subtypeOfTT{T}{Format}$ and $\subtypeOfTT{U}{Format}$
we can derive that $\subtypeOfTT{Pair[T, U]}{Format}$.
In particular, we find that $\subtypeOfTT{Pair[Num, Num]}{Format}$.
Hence, the function call \go{formatSome(p)} in line~11 type checks.

Extending our type-directed translation scheme to deal with generics turns out to be fairly straightforward.

\begin{description}
\item[Bounded type parameter.] A bounded type parameter \go{T Ifce} where \go{T} is a type variable
  and \go{Ifce} is an interface type becomes a coercion parameter \go{toIfce}$_{\sub{T}}$.
  At instantiation sites, coercions need to be inserted.
\end{description}

The lower part of \Cref{f:fgg-format2}  shows the translated program.
Starting at line~18 we find the translation of the definition of method \go{format} for pairs.
Each bounded type parameter \go{T Format} and \go{U Format} is turned into a coercion parameter
\texttt{toFormat$_{\sub{T}}$} and \texttt{toFormat$_{\sub{U}}$}.
In the target, we use a curried function definition where coercion parameters are collected in a tuple.

A method call of \go{format} needs to supply concrete instances for these coercion parameters.
See line~27 which is the translation of calling \go{format} on receiver type \go{Pair[Num,Num]}.
Hence, we must pass as the first argument
the tuple of coercions \texttt{(toFormat$_{\sub{Num}}$, toFormat$_{\sub{Num}}$)}
to \texttt{format$_{\sub{Pair}}$}.

Subtype relation (6) implies the (parameterized) coercion \texttt{toFormat$_{\sub{Pair}}$} in line~23.
Given coercions \texttt{toFormat$_{\sub{T}}$} and \texttt{toFormat$_{\sub{U}}$}
we can transform a pair \go{p} into an interface value for \go{Format},
where the method dictionary consists of
the partially applied translated method definition
\texttt{format$_{\sub{Pair}}$}.

We make use of \texttt{toFormat$_{\sub{Pair}}$}
in the translation of the function call \go{formatSome(p)}, see line~28.
Based on the specific coercion \texttt{toFormat$_{\sub{Num}}$},
the call \texttt{toFormat$_{\sub{Pair}}$} transforms the pair value \go{p} into
the interface value \texttt{(p, format$_{\sub{Pair}}$ (toFormat$_{\sub{Num}}$,toFormat$_{\sub{Num}}$))}.
Then, we call \go{formatSome} on this interface value.

\subsection{Bounded type parameters of methods}

\boxfig{f:fgg-format3}{Bounded type parameters of methods (extending code from \Cref{f:fgg-format2})}{
  \renewcommand\subColored{yes}
  \vspace{-1ex}
\begin{lstlisting}[name=goexample,numbers=left,escapechar=@,language=Golang,mathescape=true]
func (this Pair$\tyarg{[T Format, U Format]}$) formatSep$\tyarg{[S Format]}$(s S) string {
  return this.left.format() + s.format() + this.right.format()
}

type FormatSep interface { formatSep$\tyarg{[S Format]}$(s S) string }

func formatSepSome(x FormatSep, s Format) string {
	return x.formatSep$\tyarg{[Format]}$(s)
}

func main3 () {
  var p Pair$\tyarg{[Num, Num]}$ = Pair$\tyarg{[Num, Num]}${ Num{1}, Num{2} }
  var s1 string = p.formatSep$\tyarg{[Num]}$(Num{3}) $$ // result: 132
  var s2 string = formatSepSome(p,Num{4})  // result: 142
}
\end{lstlisting}
  \hrule{}
\begin{lstlisting}[name=hsexample,firstnumber=16,numbers=left,escapechar=@,language=myhaskell,mathescape=true,lastline=25]
-- Method call on interface
-- call formatSep on receiver of type FormatSep
formatSep$_{\sub{FormatSep}}$ (x, f) = f x

-- Method definition (line 1 in the FGG code)
formatSep$_{\sub{Pair}}$ (toFormat$_{\sub{T}}$, toFormat$_{\sub{U}}$) this toFormat$_{\sub{S}}$ s =
  format$_{\sub{Format}}$ (toFormat$_{\sub{T}}$ (left this)) ++
  format$_{\sub{Format}}$ (toFormat$_{\sub{S}}$ s) ++
  format$_{\sub{Format}}$ (toFormat$_{\sub{U}}$ (right this))

-- Coercions
toFormat$_{\sub{Format}}$ x = x$\hspace{5mm}$-- Format <: Format
toFormatSep$_{\sub{Pair}}$ (toFormat$_{\sub{T}}$, toFormat$_{\sub{U}}$) p =
  -- Pair[T, U] <: FormatSep (given T <: Format and U <: Format)
  (p, formatSep$_{\sub{Pair}}$ (toFormat$_{\sub{T}}$, toFormat$_{\sub{U}}$))

-- Function definitions (lines 7 and 11 in the FGG code)
formatSepSome (x, s) = (formatSep$_{\sub{FormatSep}}$ x) toFormat$_{\sub{Format}}$ s

main3 =
  let p = (1,2)
      s1 = formatSep$_{\sub{Pair}}$ (toFormat$_{\sub{Num}}$, toFormat$_{\sub{Num}}$) p toFormat$_{\sub{Num}}$ 3
      s2 = formatSepSome
             (toFormatSep$_{\sub{Pair}}$ (toFormat$_{\sub{Num}}$, toFormat$_{\sub{Num}}$) p,
              toFormat$_{\sub{Num}}$ 4)
\end{lstlisting}
  \vspace{-1ex}
  \renewcommand\subColored{no}
}

There may be bounded type parameters local to methods.
Consider Figure~\ref{f:fgg-format3} where we further extend our running example.
Starting at line~1 we find a definition of method \go{formatSep} for pairs.
This method takes an argument \go{s} that acts as a separator when formatting pairs.
Argument \go{s} is of the generic type \go{S} constrained by the type bound \go{Format}.
Type parameter \go{S} is local to the method and not connected to the receiver struct.
Type arguments for \go{S} must also be explicitly specified in the program text,
see method calls in lines 8 and 13.

In the translation, bounded type parameters of methods simply become additional coercion parameters.
Consider the translation of \go{formatSep} defined on pairs starting at line~21.
The translated method definition first expects the coercion parameters
\texttt{(toFormat$_{\sub{T}}$, toFormat$_{\sub{U}}$)} that result from the bounded type
parameters \texttt{T~Format} and \texttt{U~Format} of the receiver.
Then, we find the receiver argument \go{this} followed
by the coercion parameter \texttt{toFormat$_{\sub{S}}$} resulting from \texttt{S~Format},
and finally the method argument \go{s}.
The translation of the method body follows the scheme we have seen so far, see lines~22-24.
When calling method \go{formatSep} on a pair we need to provide
the appropriate coercions, see line~37.

From the method definition of \go{formatSep} for pairs and from the definition of interface
\go{FormatSep}, we find that the following subtype relation holds:
\begin{center}
  (7)\quad $ \{\subtypeOfTT{T}{Format}, \subtypeOfTT{U}{Format} \} \turns \subtypeOfTT{Pair[T, U]}{FormatSep}$.
\end{center}
Subtype relation (7) implies the coercion \texttt{toFormatSep$_{\sub{Pair}}$} in line~28.
Thus, the function call of \go{formatSepSome} from line 14
translates to the target code starting in line~38.

The point to note is that a coercion parameter corresponding to a bounded type parameter of a method is
not part of the dictionary; it is only supplied at the call site of the method.
Consider the call \go{x.formatSep[Format](s)} in line~8.
In the translation (line 33),  we first partially apply the respective dictionary entry on the receiver.
This is done via the target expression \texttt{(formatSep$_{\sub{FormatSep}}$ x}\texttt{)}.
Type \go{Format} is a valid instantiation for
type parameter \go{S} of \go{formatSep} because
$\subtypeOfTT{Format}{Format}$ in \FGGm. In the translation, this
corresponds to the (identity) coercion \texttt{toFormat$_{\sub{Format}}$}.
Hence, we supply the remaining arguments \texttt{toFormat$_{\sub{Format}}$} and \go{s}.

\subsection{Bounded type parameters of structs and interfaces}
\label{sec:bound-type-param}

\boxfig{f:fgg-format-fpair}{Bounded type parameters of structs and interfaces (extending code from \Cref{f:fgg-format})}{
  \renewcommand\subColored{yes}
  \vspace{-1ex}
\begin{lstlisting}[name=goexample,numbers=left,escapechar=@,language=Golang,mathescape=true]
type FPair$\tyarg{[T Format, U Format]}$ struct { left T; right U }
type Factory$\tyarg{[T Format]}$ interface { create() FPair$\tyarg{[T, T]}$ }

type MyFactory struct {}
func (this MyFactory) create() FPair$\tyarg{[Num, Num]}$ {
  return FPair$\tyarg{[Num, Num]}${Num{1}, Num{2}}
}

func doWork$\tyarg{[T Format]}$(factory Factory$\tyarg{[T]}$) string {
  var p FPair[T, T] = factory.create()
  var t T = p.left
  return t.format()
}

func main4() {
  var s = doWork[Num](MyFactory{})
}
\end{lstlisting}
  \hrule{}
\begin{lstlisting}[name=hsexample,firstnumber=16,numbers=left,escapechar=@,language=myhaskell,mathescape=true,lastline=25]
-- Field access for struct FPair
left$_{\sub{FPair}}$ (x, _) = x
right$_{\sub{FPair}}$ (_, x) = x

-- Method call on interface
create$_{\sub{Factory}}$ (x, f) = f x -- call create on receiver of type Factory

-- Method definition (line 5 in the source program)
create$_{\sub{MyFactory}}$ this = (1, 2)

-- Coercion
toFactory$_{\sub{MyFactory}}$ x = (x, create$_{\sub{MyFactory}}$) -- MyFactory <: Factory[Num]

-- Function definition (line 9 in the source program)
doWork toFormat$_{\sub{T}}$ factory =
  let p = create$_{\sub{Factory}}$ factory
      t = left$_{\sub{FPair}}$ p
  in format$_{\sub{Format}}$ (toFormat$_{\sub{T}}$ t)

main4 = doWork toFormat$_{\sub{Num}}$ (toFactory$_{\sub{MyFactory}}$ ())
\end{lstlisting}
  \vspace{-1ex}
  \renewcommand\subColored{no}
}

Structs and interfaces may also carry bounded type parameters.
In \FGGm{} and FGG, these type parameters do not have a meaning at runtime as their purpose is only
to rule out more programs statically.
Hence, in our translation approach they do not translate into additional dictionary parameters or coercions.

Let us explain with the example in \Cref{f:fgg-format-fpair}. Struct \go{FPair} (short for \enquote{formatted pairs})
requires the type bound \go{Format} on its type parameters.
The generic interface \go{Factory} defines a factory method returning formatted
pairs. It requires a type bound \go{T Format} for the type \go{FPair[T, T]} in its method
signature to be well-formed. The need for this type bound arises because
FGG's type system does not allow to conclude from just an occurrence of \go{FPair[T, T]}
that \go{T} is already a subtype of
\go{Format}.

Struct \go{MyFactory} defines a concrete factory implementation for \go{FPair[Num, Num]},
function \go{doWork} accepts a generic \go{Factory[T]} for arbitrary \go{T}. Again,
\go{doWork} requires a type bound \go{T Format} for type \go{Factory[T]} to be well-formed.
The main function may then call \go{doWork} with a \go{MyFactory} value because
\go{MyFactory} is a subtype of \go{Factory[Num]}.

The translated code (lower part of \Cref{f:fgg-format-fpair}) demonstrates that bounded type parameters of
structs and interfaces have no representation at runtime, so the translation
effectively ignores them. A struct value is still just a tuple with the
fields of the struct (lines 17, 18), and an interface value just combines a struct value with a method
dictionary (e.g.\ line~27). Only bounded type parameters of receiver structs (\Cref{f:fgg-format2}),
methods (\Cref{f:fgg-format3}) and functions
(\Cref{f:fgg-format-fpair})
lead to additional coercion parameters.

An important point to note is that there is a difference between generic interfaces and interfaces with
generic methods. Interface \go{Factory[T]} in \Cref{f:fgg-format-fpair} is generic in \go{T}, a
subtype of \go{Factory[U]} must provide an implementation of the \go{create} method for some
fixed type \go{U}. In contrast, interface \go{FormatSep} from \Cref{f:fgg-format3} is not generic
but contains a method \go{formatSep} that is generic in \go{S}.
A subtype of \go{FormatSep} must provide an implementation of \go{formatSep}
that is also generic in \go{S}.

\subsection{Outlook}

Next, \Cref{sec:featherweight-generic-go} formalizes \FGGm\ following the description by~\cite{FeatherweightGo}.
Then, we give the details of our type-directed translation scheme in \Cref{sec:type-direct-transl}
and establish that the meaning of \FGGm\ programs is preserved in \Cref{sec:formal-properties}.

\section{Featherweight Generic Go$^{-}$}
\label{sec:featherweight-generic-go}

Featherweight Go \cite[FG,][]{FeatherweightGo} is a small subset of the full Go language~\citeyearpar{golang}
supporting only essential features such as structs, interfaces, method
overloading and structural subtyping.
In the same article, the authors add generics to FG with the goal to scale
the design to full Go. The resulting
calculus is called Featherweight Generic Go (FGG). Since version 1.18, full Go includes
generics as well, but with limited expressivity compared to the FGG proposal
(see \Cref{sec:generics-go}).
For the translation presented in this article, we stick to the original FGG language
with minor differences in presentation but excluding dynamic type assertions.
We refer to this language as \FGGm.

The next two subsections introduce the syntax and the dynamic semantics of \FGGm. We defer
the definition of its static semantics until \Cref{sec:type-directed}, where we specify
it as part of the type-directed dictionary-passing translation.

\subsection{Syntax}
\label{sec:syntax}

\boxfig{f:fgg}{Syntax of \FGGm}{
  \vspace{-1ex}
\bda{ll}
\ba{lr@{~}l}
   \GoSynCatName{Struct name}       & t_S, u_S & \in \structNameSet
\\ \GoSynCatName{Interface name}    & t_I, u_I & \in \ifaceNameSet
\\ \GoSynCatName{Type variable name}          & \fgTyVar, \fgTyVarAux & \in \tyvarSet
\\[\medskipamount]
\GoSynCatName{Type name}              & t, u & ::= t_S \mid t_I
\\ \GoSynCatName{Type} & \fgType, \fgTypeAux & ::= \fgTyVar \mid t[\ForeachN{\fgType}]
\ea
&
\ba{lr@{~}l}
   \GoSynCatName{Field name}     & f & \in \fieldNameSet
\\ \GoSynCatName{Method name}   & m & \in \methodNameSet
\\ \GoSynCatName{Variable name} & x,y & \in \expvarSet
\\[\medskipamount]
\GoSynCatName{Struct type} & \fgType_S, \fgTypeAux_S & ::= t_S[\ForeachN{\fgType}]
\\
\GoSynCatName{Interface type} & \fgType_I, \fgTypeAux_I & ::= t_I[\ForeachN{\fgType}]
\ea
\eda
\bda{lr@{~}r@{~}l}
\GoSynCatName{Expression}    & e, g & ::=  &
  x \mid e.m[\Multi{\fgType}](\overline{e})
  \mid \fgType_S \{ \Multi{e} \} \mid e.f
\\
\GoSynCatName{Method signature}       & R & ::= &
  m[\Multi{\fgTyVar\,\fgType_I}](\overline{x \ \fgType}) \ \fgType
\\
\GoSynCatName{Declaration} & D & ::= &
\TYPE\ t_S[\Multi{\fgTyVar\,\fgType_I}] \ \STRUCT\ \{ \overline{f \ \fgType} \} \\
&& \mid & \TYPE\ t_I[\Multi{\fgTyVar\,\fgType_I}] \ \INTERFACE\ \{ \overline{R} \} \\
&& \mid & \FUNC\ (x \ t_S[\Multi{\fgTyVar\,\fgType_I}])\,R\, \{ \RETURN\ e \}
\\
\GoSynCatName{Program} & P & ::= & \overline{D} \ \FUNC\ \MAIN () \{ \_ = e \}
\eda
}

\Cref{f:fgg} introduces the syntax of \FGGm.
We assume several countably infinite, pairwise disjoint sets
for names, ranged over by $\NameSetFGG$ with some subscript (upper part of the figure).
Meta variables $t_S$ and $u_S$ denote struct names,
$t_I$ and $u_I$ interface names,
$\alpha$ and $\beta$
type variables,
$f$ field names, $m$ method names,
and $x, y$ denote names for variables in expressions.
Overbar notation $\sOver[n]{\mathfrak s}$ is a shorthand for the
sequence $\mathfrak s_1 \ldots \mathfrak s_n$ where $\mathfrak s$ is some
syntactic construct. In some places, commas separate the sequence
items.
If irrelevant, we omit the $n$ and simply write
$\sOver{\mathfrak s}$. Using the index variable $i$ under an overbar
marks the parts that vary from sequence item to sequence
item; for example, $\sOver[n]{\mathfrak s'\,\mathfrak s_i}$ abbreviates
$\mathfrak s'\,\mathfrak s_1\ldots\mathfrak s'\,\mathfrak s_n$
and $\sOver[q]{\mathfrak s_j}$ abbreviates
$\mathfrak s_{j1}\,\ldots\,\mathfrak s_{jq}$.

The middle part of \Cref{f:fgg} shows the syntax of types in \FGGm.
A type name $t,u$ is either a struct or interface name. Types
$\fgType, \fgTypeAux$ include types variables $\fgTyVar$ and instantiated types
$t[\Multi{\fgType}]$.
For non-generic structs or interfaces, we often
write just $t$ instead of $t[]$.
Struct types $\fgType_S$, $\fgTypeAux_S$ and interface types $\fgType_I$, $\fgTypeAux_I$
denote syntactic subsets of the full type syntax.

The lower part of \Cref{f:fgg} defines the syntax of \FGGm\ expressions, declarations, and programs.
Expressions, ranged over by $e$ and $g$, include variables $x$, method calls, struct literals,
and field selections.
A method call $e.m[\Multi{\fgType}](\Multi{e})$ invokes method $m$ on receiver $e$ with type
arguments $\Multi{\fgType}$ and arguments $\Multi{e}$. If $m$ does not take type arguments,
we often write just $e.m(\Multi{e})$. A struct literals $\fgType_S\{\Multi[n]{e}\}$
creates an instance of a struct with $n$ fields,
the arguments $\Multi[n]{e}$ become the values of the fields in order of appearance
in the struct definition.
A field selection $e.f$ projects the value of some struct field $f$ from
expression $e$.

A method signature $R ::= m[\Multi{\fgTyVar\,\fgType_I}](\Multi{x\ \fgType})\,\fgType$
consists of a name $m$, bounded type parameters $\fgTyVar_i$ with interface type $\fgType_{Ii}$
as upper bounds, parameters $x_i$ of type $\fgType_i$,
and return type $\fgType$. It binds $\Multi{\fgTyVar}$ and $\Multi{x}$.
The scope of a type variable $\fgTyVar_i$ is $\Multi{\fgType}$, $\fgType$, and all upper bounds
$\Multi{\fgType_I}$, so \FGGm\ supports F-bounded quantification \citep{Canning1989}.
For non-generic methods, we often write just $m(\Multi{x_i\ \fgType_i})\,\fgType$.

A declaration $D$ comes in three forms: a struct
$\TYPE\ t_S[\Multi{\fgTyVar\,\fgType_I}] \ \STRUCT\ \{ \overline{f \ \fgType} \}$
with fields $f_i$ of type $\fgType_i$; an interface
$\TYPE\ t_I[\Multi{\fgTyVar\,\fgType_I}] \ \INTERFACE\ \{ \overline{R} \}$
with method signatures $\Multi{R}$; or a method
$\FUNC\ (x \ t_S[\Multi{\fgTyVar\,\fgType_I}])\,R\, \{ \RETURN\ e \}$
providing an
implementation of method $R$ for struct $t_S$.
All three forms bind the type variables $\Multi{\fgTyVar}$, a method implementation additionally binds
the receiver parameter $x$.
The scope of a type variable $\fgTyVar_i$ includes all upper bounds
$\Multi{\fgType_I}$, the body of the declaration enclosed in $\{\ldots\}$, and for method declarations also the
signature $R$. We omit the $[\Multi{\fgTyVar\,\fgType_I}]$ part completely
if $\Multi{\fgTyVar\,\fgType_I}$ is empty.
Finally, a program $P$ consists of a sequence of declarations together
with a main function. Method and function bodies only contain a single expression.
We follow the usual convention and identify syntactic constructs up to renaming of bound variables or
type variables.

The syntax of \FGGm\ as presented here differs slightly from its original form~\citep{FeatherweightGo}.
The original article encloses
type parameters in parenthesis, an additional $\TYPE$ keyword
starts a list of type parameters. Here,
we follow the syntax of full Go and use square brackets without any keyword.
Further, the original article prepends \kw{package main} to each program, something
we omit for succinctness. Finally, we reduce the number of syntactic meta-variables to improve readability.

\subsection{Dynamic Semantics}
\label{sec:dynamic-semantics}

\boxfig{f:fgg-dynamic}{Dynamic semantics of \FGGm}{
  \vspace{-1ex}
  \bda{lr@{~}r@{~}l}
  \sSynCatName{Value} &
  v, u, w & ::= & \fgType_S \{ \ForeachN{v} \}
  \\
  \sSynCatName{Evaluation context} &
  \EvCtx & ::= & \Hole
   \mid \fgType_S \{ \ForeachN{v}, \EvCtx, \ForeachN{e} \}
   \mid \EvCtx.f
   \mid \EvCtx.m[\Multi{\fgType}](\ForeachN{e})
   \mid v.m[\Multi{\fgType}](\ForeachN{v}, \EvCtx, \ForeachN{e})
   \\
   \sSynCatName{Value substitution} &
   \vbFG & ::= & \Angle{\ForeachN{\subst{x}{v}}}
   \\
   \sSynCatName{Type substitution} &
   \tbFG & ::= & \Angle{\ForeachN{\subst{\fgTyVar}{\fgType}}}
   \eda
  \sTransSection{
    \ruleform{\reduce{e}{e}}
  }{
    Reductions
  }
  \begin{mathpar}
    \inferrule[fg-context]
    {\reduce{e}{e'}
    }
    {\reduce{\EvCtx [e]}{\EvCtx [e']}
    }

    \inferrule[fg-field]
    { \TYPE\ t_S[\Multi{\fgTyVar\,\fgType_I}] \ \STRUCT\ \{ \Foreach{f \ \fgTypeAux}{n} \} \in  \ForeachN{D}\\
    }
    { \reduce{t_S[\Multi{\fgType}] \{ \Foreach{v}{n} \}.f_i}{v_i}
    }

    \inferrule[fg-call]
    { v = t_S[\Multi{\fgType}] \{ \ForeachN{u} \}
      \\ \FUNC\ (x \ t_S[\Multi{\fgTyVar\,\fgType_I}]) \ m[\Multi{\fgTyVar'\,\fgType_I'}]
             (\ForeachN{x \ \fgTypeAux}) \ \fgTypeAux \ \{ \RETURN\ e \} \in \ForeachN{D}
    }
    {\reduce{v.m[\Multi{\fgType'}](\ForeachN{v})}
      {\Angle{\subst{x}{v}, \ForeachN{\subst{x}{v}}}
       \Angle{\Multi{\subst{\fgTyVar}{\fgType}}, \Multi{\subst{\fgTyVar'}{\fgType'}}}
       e}
    }
  \end{mathpar}
}

\Cref{f:fgg-dynamic} defines a call-by-value dynamic semantics for \FGGm\ using a small-step reduction semantics
with evaluation contexts. The definition is largely taken from \citet{FeatherweightGo}.

We use $v, u, w$ to denote values, where
a value is a struct literal with all fields being values.
A call-by-value evaluation context $\EvCtx$ is an expression with a hole $\Hole$ such that
the hole marks the point where the next evaluation step should happen.
We write $\EvCtx[e]$ to denote the replacement of the hole in $\EvCtx$ with expression $e$.
A value substitution $\vbFG$ is a finite mapping $\Angle{\Multi{\subst{x}{v}}}$ from variables
to values, whereas a type substitution $\tbFG$ is a finite mapping
$\Angle{\Multi{\subst{\fgTyVar}{\fgType}}}$ from type variables to types.
The (type) variables in the domain of a substitution must be distinct.
Substitution application, written in prefix notation as
$\vbFG e$ or $\tbFG e$ or $\tbFG \fgType$, is defined in the usual, capture-avoiding way.
When combining two sequences, we implicitly assume that both sequences have the same length.
For example, combining variables $\Multi{x}$ and values $\Multi{v}$ to a substitution
$\Angle{\Multi{\subst{x}{v}}}$ implicitly assumes that there are as many variables as values.

The reduction relation $\reduce{e}{e'}$ denotes that expression $e$ reduces to expression $e'$.
To avoid clutter, the sequence of declarations $\fgDecls$ of the underlying program is implicitly available
in the rules defining this reduction relation. Rule \Rule{fg-context} applies a reduction step
inside an expression. Rule \Rule{fg-field} reduces a field selection $t_S[\Multi{\fgType}]\{\Multi{v}\}.f_i$
by extracting value $v_i$ corresponding to field $f_i$ from the struct literal.
Rule \Rule{fg-call} reduces a method call $t_S[\Multi{\fgType}]\{\Multi{u}\}.m[\Multi{\fgType'}](\Multi{v})$.
It retrieves a method definition for $m$ and $t_S$ and substitutes
type arguments, receiver, and value arguments in the method body.

Reduction in \FGGm\ is deterministic (see \Cref{lem:determ-eval-fgg} in \Cref{sec:determ-eval-fgg}
for a formal proof), assuming the following three
restrictions:
\begin{description}
\item[\FGGUniqueStructs] Each struct $t_S$ is defined at most once in the program.
\item[\FGGUniqueFields] Each struct definition
  $\TYPE\ t_S[\Multi{\fgTyVar\,\fgType_I}]\ \STRUCT\ \{ \overline{f \ \fgType} \}$
  has distinct
  field names $\Multi f$.
\item[\FGGUniqueReceiver] Each method definition
  $\FUNC\ (x \ t_S[\Multi{\fgTyVar\,\fgType_I}]) \ m[\Multi{\fgTyVar'\,\fgType_I'}]
             (\ForeachN{x \ \fgTypeAux}) \ \fgTypeAux \ \{ \RETURN\ e \}$
  is uniquely identified by struct name $t_S$ and method name $m$.
\end{description}
The first two restrictions ensures that the value for a field in rule \Rule{fg-field} is
unambiguous. The third restriction avoids multiple matching method definitions in
rule \Rule{fg-call}.

\section{Type-directed translation}
\label{sec:type-direct-transl}

This section defines a type-directed, dictionary-passing translation from \FGGm\ to an untyped $\lambda$-calculus
extended with recursive let-bindings, constructors and pattern matching. We first introduce
the target language, then specify the translation itself, and last but not least
give some examples. Formal properties of the translation
are deferred until~\Cref{sec:formal-properties}.

\subsection{Target Language}
\label{sec:target-language}

\boxfig{f:target-lang}{Target language (TL)}{
  \vspace{-1ex}
  \bda{ll}
  {
    \ba[t]{lr@{~}l}
    \sSynCatName{Variable} & X,Y & \in \TLVarSet
    \\
    \sSynCatName{Constructor} & K & \in \TLConsSet
    \\
    \sSynCatName{Expression} & \expT, G & ::=
    X \mid K \mid \expT \ \expT \mid
    \lambda \xT. \expT
    \\ & & \phantom{::=}
    \mid \tcaseof{\expT}{\ForeachN{\clsT}}
    \ea
  }
  &
  {
    \ba[t]{lr@{~}l}
    \sSynCatName{Pattern clause} &
    \clsT & ::=  \patT \rightarrow \expT
    \\
    \sSynCatName{Pattern} &
    \patT & ::= \kT \ \ForeachN{\xT}
    \\
    \sSynCatName{Program}  &
    \program & ::=
    \LET\ \ForeachN{X = V} \ \IN\ \expT
    \ea
   }
   \eda
  \bda{l@{\quad}r@{~}r@{~}l}
  \sSynCatName{Value} &
  \uT, U, W & ::= & \kT\ \ForeachN{\uT} \mid \lambda \xT . \expT
  \\
  \sSynCatName{Evaluation context} &
  \REvCtxT & ::= & \Hole
   \mid \tcaseof{\REvCtxT}{\Multi{\clsT}}
   \mid \REvCtxT\ \expT
   \mid \uT\ \REvCtxT
  \\
     \sSynCatName{Substitution} &
   \vbTL, \vbMethTL & ::= & \Angle{\ForeachN{\subst{\xT}{\uT}}}
   \eda

\vspace{1ex}
\sTransSection{
  \ruleform{
    \reduceExpTL{\vbMethTL}{\expT}{\expT'}
  }
}{
  TL expression reductions
}
\vspace{-0.5ex}
\begin{mathpar}
 \inferrule[tl-context]
 {\reduceExpTL{\vbMethTL}{\expT}{\expT'}
 }
 {\reduceExpTL{\vbMethTL}{\REvCtxT [\expT]}{\REvCtxT [\expT']}
 }

 \inferrule[tl-lambda]
 {}
 { \reduceExpTL{\vbMethTL}{(\lambda \xT. \expT) \ \uT}{\Angle{\subst{\xT}{\uT}} \expT}
 }

 \inferrule[tl-case]
 { \kT \ \Multi[n]{\xT} \rightarrow \expT \in \Multi{\clsT}
 }
 {\reduceExpTL{\vbMethTL}{\tcaseof{\kT \ \Multi[n]{\uT}}{\Multi{\clsT}}}
         {\Angle{\Multi[n]{\subst{\xT}{\uT}}} \expT}
 }

 \inferrule[tl-method]{}{
   \reduceExpTL{\vbMethTL}{\xT}{\vbMethTL(\xT)}
 }
\end{mathpar}

\vspace{1ex}
\sTransSection{
  \ruleform{\reduceTL{\program}{\program'}}
}{
  TL reductions
}
\vspace{-0.5ex}
\begin{mathpar}
    \inferrule[tl-prog]{
      \vbMethTL = \Angle{\ForeachN{\subst{X}{V}}}\\
      \reduceExpTL{\vbMethTL}{\expT}{\expT'}
    }{
      \reduceTL{\tletrecin{\ForeachN{X = V}}{\expT}}{\tletrecin{\ForeachN{X = V}}{\expT'}}
    }
\end{mathpar}
} %

\Cref{f:target-lang} defines the syntax and the call-by-value dynamic semantics of the target language (TL).
We use uppercase letters for constructs of the target language.
Variables $X, Y$ and constructors $K$ are drawn from countably infinite, pairwise disjoint sets
$\TLVarSet$ and $\TLConsSet$, respectively.
Expressions, ranged over by $E$ and $G$, include variables $X$, constructors $K$, function applications $E\,E'$,
$\lambda$-abstractions $\lambda X.E$, and pattern matching via case-expressions
$\tcaseof{\expT}{\ForeachN{\patT \to \expT}}$. Patterns $\patT$ have the form $K\,\Multi{X}$, they do not nest.
We assume that all constructors
in $\Multi{\patT}$ are distinct. To avoid some parentheses, we use the conventions that application
binds to the left and that the body of a $\lambda$ extends to the right as far as possible.

A program $\LET\ \ForeachN{X = V} \ \IN\ \expT$
consists of a sequence of (mutually recursive) definitions and a (main) expression,
where we assume that the variables $\Multi X$ are distinct.
In the translation from \FGGm, the values $\Multi V$ are always
functions resulting as translations of \FGGm\ methods.
We identify expressions, pattern clauses and programs up to renaming of bound variables.
Variables are bound by $\lambda$ expressions, patterns, and let-bindings of programs.

Some syntactic sugar simplifies the construction of patterns, expressions and programs.
(a) We use nested patterns to abbreviate nested case-expressions.
(b) We assume data constructors for tuples up to some fixed but arbitrary size. The syntax
$\Tuple{\Foreach{E}{n}}$ constructs an $n$-tuple when used as an expression, and
$\Tuple{\Multi[n]{\patT}}$ deconstructs it when used in a pattern context.
(c) We use patterns in $\lambda$-expressions; that is, the notation $\lambda \patT. \expT$ stands for
$\lambda X . \tcaseof{X}{\patT \rightarrow \expT}$ where $X$ is fresh.

Target values $V,U,W$ are either $\lambda$-expressions or constructors applied to values.
A constructor value $K\,\Foreach{V}{n}$ is short for $(\ldots (K\,V_1) \ldots)\,V_n$.
A call-by-value evaluation context $\REvCtxT$ is an expression with a hole $\Hole$ such that
the hole marks the point where the next evaluation step should happen.
We write $\REvCtxT[E]$ to denote the replacement of the hole in $\REvCtxT$ with expression $E$.

A substitution $\vbTL, \vbMethTL$ is a finite mapping $\Angle{\Multi{\subst{X}{V}}}$ from variables
to values. The variables $\Multi{X}$ in the domain must be distinct.
Substitution application, written in prefix notation
$\vbTL E$, is defined in the usual, capture-avoiding way.
We use two different meta variables $\vbMethTL$ and $\vbTL$ for substitutions in the target language
with the convention that
the domain of $\vbMethTL$ contains only top-level variables bound by $\LET$.
As top-level variables result from translating \FGGm\ methods,
we sometimes call $\vbMethTL$ a \emph{method substitution}.

The reduction semantics for the target language is defined by two relations:
$\reduceExpTL{\vbMethTL}{\expT}{\expT'}$ reduces expression $E$ to $E'$ under method substitution $\vbMethTL$,
and $\reduceTL{\program}{\program'}$ reduces $\program$ to $\program'$.
The definition of the latter simply forms a method substitution $\vbMethTL$ from the top-level bindings
of $\program$ and then reduces the main expression of $\program$ under $\vbMethTL$ (rule \Rule{tl-prog}).
We defer the substitution of top-level--bound variables because they might be recursive.

The definition of the reduction relation for expressions extends over four rules.
Rule \Rule{tl-context} uses evaluation context $\REvCtxT$ to reduce inside an expression,
rule \Rule{tl-lambda} reduces function application in the usual way. Pattern matching
in rule \Rule{tl-case} assumes that the scrutinee is a constructor value $K\,\Multi[n]{V}$; the lookup
of a pattern clause matching $K$ yields at most one result as we assume that clauses
have distinct constructors. During a sequence of reduction steps, a variable bound by
$\LET$ at the top-level might become a redex, as only $\lambda$-bound variables are substituted right away.
Thus, rule \Rule{tl-method} finds the value for the variable in the method substitution $\vbMethTL$.

\subsection{Translation}
\label{sec:type-directed}

Before we dive into the technical details, we summarize our translation strategy.

\begin{description}
\item[Struct.] An \FGGm\ value of some struct type is represented in the TL as a
  \emph{struct value}; that is, a tuple $\Tuple{\Multi[n]{V}}$ where $n$ is the number of
  fields and $V_i$ represents the $i$-th field of the struct.
\item[Interface.] An \FGGm\ value of some interface type is represented in the TL as an
  \emph{interface value}; that is a pair
  $\Pair{V}{\mathcal{D}}$, where $V$ is a struct value realizing the interface
  and $\mathcal D$ is a \emph{dictionary.}
\item[Dictionary.] A dictionary $\mathcal D$ for an interface with methods $\Multi[n]{R}$
  is a tuple $\Tuple{\Multi[n]{V}}$ such that $V_i$ is a \emph{dictionary entry} for method $R_i$.
\item[Dictionary entry.] A dictionary entry for a method with signature
  $R = m[\Multi{\fgTyVar\,\fgType_I}](\Multi{x\,\fgTypeAux})\fgTypeAux$
  is a function accepting a triple:
  (1) receiver, (2) tuple with coercions corresponding to the bounded type
  parameters $\Multi{\fgTyVar\,\fgType_I}$ of the method,
  (3) tuple for parameters $\Multi{x}$.
\item[Coercion.] A structural subtype relation $\subtypeOf{\fgType}{\fgTypeAux}$ implies a \emph{coercion function}
  to transform the target representation of an \FGGm\ value at type $\fgType$
  into a representation at type $\fgTypeAux$.
\item[Bounded type parameter.] A bounded type parameter $\fgTyVar\,\fgType_I$
  becomes a coercion parameter $X_{\fgTyVar}$ transforming
  the type supplied for $\fgTyVar$ to its bound $\fgType_I$.
  At instantiation sites, coercions need to be inserted.
\item[Method declaration.]
  A method declaration
  $\FUNC\ (x\,t_S[\Multi{\fgTyVar\,\fgType_I}])\,m[\Multi{\fgTyVar'\,\fgType_I'}](\Multi{x\,\fgTypeAux})\,\fgTypeAux\,\{\RETURN\,e\}$
  is represented as a top-level function $\mT{m}{t_S}$ accepting a quadruple:
  (1) tuple with coercions corresponding to the bounded type parameters $\Multi{\fgTyVar\,\fgType_I}$
  of the receiver,
  (2) receiver $x$,
  (3) tuple with coercions corresponding to bounded type parameters $\Multi{\fgTyVar'\,\fgType_I'}$
  of the method,
  (4) tuple for parameters~$\Multi{x}$.
\end{description}

In essence, the above is a more detailed description of the translation scheme motivated in \Cref{sec:examples}.
The only difference is that dictionary entries and translations of methods are now represented as uncurried functions.
For example, instead of the curried representation in Figure~\ref{f:fgg-format3}

\renewcommand\subColored{yes}
\begin{lstlisting}[name=hsexample,escapechar=@,language=myhaskell,mathescape=true]
formatSep$_{\sub{Pair}}$ (toFormat$_{\sub{T}}$, toFormat$_{\sub{U}}$) this toFormat$_{\sub{S}}$ x = ...

toFormatSep$_{\sub{Pair}}$ (toFormat$_{\sub{T}}$, toFormat$_{\sub{U}}$) p =
   (p, formatSep$_{\sub{Pair}}$ (toFormat$_{\sub{T}}$, toFormat$_{\sub{U}}$))
\end{lstlisting}
our actual translation scheme uses uncurried functions, as in the following code:
\begin{lstlisting}[name=hsexample,escapechar=@,language=myhaskell,mathescape=true]
-- translation of method
formatSep$_{\sub{Pair}}$ ((toFormat$_{\sub{T}}$, toFormat$_{\sub{U}}$), this, toFormat$_{\sub{S}}$, x) = ...

toFormatSep$_{\sub{Pair}}$ (toFormat$_{\sub{T}}$, toFormat$_{\sub{U}}$) p =
   (p, \(this,locals,arg) ->    -- dictionary entry
           formatSep$_{\sub{Pair}}$ ((toFormat$_{\sub{T}}$, toFormat$_{\sub{U}}$),locals,arg))
\end{lstlisting}
\renewcommand\subColored{no}

Using an uncurried representation instead of a curried representation is just a matter taste.
As we have carried out the semantic equivalence proof initially based on the uncurried representation,
we stick to it from now on.

\subsubsection{Conventions and Notations}

The translation relies on three total, injective functions with pairwise disjoint ranges
for mapping
\FGGm\ names to TL variables. The first function
$\expvarSet \to \TLVarSet$ translates a \FGGm\ variable $x$
to a TL variable $X$. To avoid clutter, we do not spell out
the translation function explicitly but use the abbreviation that a lowercase
$x$ always translates into its uppercase counterpart $X$.
The second function $\tyvarSet \to \TLVarSet$ translates
an \FGGm\ type variable $\fgTyVar$ into a TL variable, abbreviated
$\xTy{\fgTyVar}$. The third function $\methodNameSet \times \structNameSet \to \TLVarSet$
gives us the TL variable $\mT{m}{t_S}$ representing the translation
of a method $m$ for struct $t_S$.
Here is a summary of the shorthand notations for name translation functions, where
$\methodName{R}$ denotes the name part of method signature $R$.
\begin{mathpar}
  x \leadsto \xT
  \quad\qquad
  \fgTyVar \leadsto \xTy{\fgTyVar}
  \quad\qquad
  \inferrule[]{
    m = \methodName{R}
  }{
    \FUNC\ (x \ {t_S}[\Multi{\fgTyVar\,\fgType_I}]) \, R \, \{ \RETURN\ e \} \leadsto
    \mT{m}{t_S}
  }
\end{mathpar}

The notation for translating names slightly differs from the approach used in
the examples of \Cref{sec:examples}.
For instance, the coercion \texttt{toFormat$_{\sub{T}}$} from \Cref{f:fgg-format3}
is now named $X_{\sub{T}}$ and
method \texttt{formatSep$_{\sub{Pair}}$} becomes
$X_{\texttt{formatSep},\texttt{Pair}}$.
The notation of the formal translation stresses that
$X_{\sub{T}}$ and $X_{\texttt{formatSep},\texttt{Pair}}$ are variables of the target
language.

An \FGGm\ type environment $\fgTyEnv$ is a mapping $\{\Multi{\fgTyVar : \fgType_{I}}\}$
from type variables $\fgTyVar_i$ to their upper bounds $\fgType_{Ii}$.
An \FGGm\ value environment $\fgEnv$ is a mapping
$\{\Multi{x : \fgType}\}$ from \FGGm\ variables $x_i$ to their types $\fgType_i$.
An environment may contain at most one binding for a type variable or variable.
We write $\emptyset$ for the empty environment, $\Dom{\cdot}$ for the domain
of an environment, and $\cup$ for the disjoint union of two environments.
The notation $\distinct{\Multi{\mathfrak s}}$ asserts that $\Multi{\mathfrak{s}}$ is a sequence of disjoint items.
We let $[n]$ denote the set $\{1,\ldots,n\}$.

In the following, we assume that the declarations $\fgDecls$ of the \FGGm\ program being translated
are implicitly available in all rules. This avoids the need for
threading the declarations through all translation rules.

\subsubsection{Auxiliary Judgments}
\label{sec:auxiliary-judgments}

\boxfig{f:trans-auxiliaries}{Auxiliary judgments for the translation}{
  \sTransSection{
    \ruleform{
      \tdCheckSubst{\fgTyEnv}{\Multi{\fgTyVar\,\fgType_I}}{\Multi{\fgTypeAux}}{\fgTySubst}{V}
    }
  }{Instantiation of bounded type parameters}
  \begin{mathpar}
    \inferrule[type-inst-checked]{
      \fgTySubst = \Angle{\Multi[n]{\subst{\fgTyVar}{\fgTypeAux}}} \\
      \tdUpcast{\fgTyEnv}{\subtypeOf{\fgTypeAux_i}{\fgTySubst \fgType_{Ii}}}{V_i} \\
      \noteForall{i \in [n]}
    }{
      \tdCheckSubst{\fgTyEnv}{\Multi[n]{\fgTyVar\,\fgType_I}}
        {\Multi[n]{\fgTypeAux}}{\fgTySubst}{\Tuple{\Multi[n]{V}}}
    }
  \end{mathpar}

  \sTransSection{
    \ruleform{
      \tdMethods{R}{V}{\fgTyEnv}{\fgType_S} \qquad
      \methodSpecifications{\fgType_I} = \{ \ForeachN{R} \}
    }
  }{
    Method access
  }

  \begin{mathpar}
    \inferrule[methods-struct]{
      \FUNC\ (x \ {t_S}[\Multi{\fgTyVar\,\fgType_I}]) \, R \, \{ \RETURN\ e \} \in \fgDecls\\
      \tdCheckSubst{\fgTyEnv}{\Multi{\fgTyVar\,\fgType_I}}{\Multi{\fgTypeAux}}{\fgTySubst}{V}
    }{
      \tdMethods{\fgTySubst R}{V}{\fgTyEnv}{t_S[\Multi{\fgTypeAux}]}
    }

    \inferrule[methods-iface]{
      \TYPE\ t_I[\Multi{\fgTyVar\,\fgType_I}] \ \INTERFACE\ \{ \ForeachN{R} \} \in \ForeachN{D}\\
      \fgTySubst = \Angle{\Multi{\subst{\fgTyVar}{\fgTypeAux}}}
    }
    {
      \methodSpecifications{t_I[\Multi{\fgTypeAux}]} = \{ \fgTySubst\ForeachN{R} \}
    }
  \end{mathpar}
}

\Cref{f:trans-auxiliaries} defines some auxiliary judgments.
The judgment $\tdCheckSubst{\fgTyEnv}{\Multi{\fgTyVar\,\fgType_I}}{\Multi{\fgTypeAux}}{\fgTySubst}{V}$,
defined by rule \Rule{type-inst-checked},
constructs a type substitution $\fgTySubst = \MultiSubst{\fgTyVar}{\fgTypeAux}$ and checks
that the $\Multi{\fgTypeAux}$ conform to their upper bounds $\Multi{\fgType_I}$
under type environment $\fgTyEnv$.
In the tuple $\Tuple{\Multi[n]{V}}$ of $\lambda$-abstractions each $V_i$ coerces the actual type
argument to its upper bound. The relevant
premise for checking upper bounds is
$\tdUpcast{\fgTyEnv}{\subtypeOf{\fgTypeAux_i}{\fgTySubst \fgType_{Ii}}}{V_i}$, which
asserts that $\fgTypeAux_i$ is a structural subtype of $\fgTySubst \fgType_{Ii}$ giving raise
to a coercion function $V_i$.
The judgment will be defined and explained in the next subsection.

The lower part of \Cref{f:trans-auxiliaries} defines two judgments for looking up
methods defined for a struct or interface type.
Judgment $\tdMethods{R}{V}{\fgTyEnv}{\fgType_S}$
states that method signature $R$ is available for struct type $\fgType_S$ under
type environment $\fgTyEnv$, see rule \Rule{methods-struct}.
The value $V$ is a tuple of coercion functions resulting
from checking the bounds of the receiver's type parameters.
Judgment $\methodSpecifications{\fgType_I} = \{ \ForeachN{R} \}$
states that the set of method signatures
available for interface type $\fgType_I$ is $\{ \Multi{R} \}$, see rule \Rule{methods-iface}.
As stated before, this rule forms the substitution
$\Angle{\Multi{\subst{\fgTyVar}{\fgTypeAux}}}$ by implicitly assuming that
$\Multi{\fgTyVar}$ and $\Multi{\fgTypeAux}$ have the same length.

\subsubsection{Translation of Structural Subtyping}
\label{sec:transl-struct-subtyp}

\boxfig{f:upcast}{Translation of structural subtyping}{
  \sTransSection{
  \ruleform{\tdUpcast{\fgTyEnv}{\subtypeOf{\fgType}{\fgTypeAux}}{V}}
}{
  Translation of structural subtyping
}
\begin{mathpar}
  \inferrule[coerce-tyvar]{
   Y~\textrm{fresh}\\
    (\fgTyVar : \fgTypeAux_I) \in \fgTyEnv\\
    \tdUpcast{\fgTyEnv}{\subtypeOf{\fgTypeAux_I}{\fgType}}{V}
  }{
    \tdUpcast{\fgTyEnv}{\subtypeOf{\fgTyVar}{\fgType}}
      {\tllambda{Y} V\, (\xTy{\fgTyVar}\, Y)}
  }

  \inferrule[coerce-struct-iface]{
    X,\MultiThree{Y}~\textrm{fresh}\\
    \TYPE\ t_I[\Multi{\fgTyVar\,\fgType_I}] \ \INTERFACE\ \{ \Foreach{R}{n} \} \in \ForeachN{D}\\
    \fgTySubst = \Angle{\Multi{\subst{\fgTyVar}{\fgType}}}\\
    \tdMethods{\fgTySubst R_i}{V_i}{\fgTyEnv}{\fgType_S}\\
    m_i = \methodName{R_i}\\
    U_i = \tllambda{\TripleY} X_{m_i,t_S}\,\QuadrY{V_i}
    \quad\noteForall{i \in [n]}
  }{
    \tdUpcast{\fgTyEnv}{\subtypeOf{\fgType_S}{t_I[\Multi{\fgType}]}}
    {\tllambda{\xT} \Tuple{\xT\Comma \Tuple{\Multi[n]{U}}} }
  }

  \inferrule[coerce-iface-iface]{
    Y,\Foreach{X}{n}~\textrm{fresh}\\
    \mapPerm : [q] \to [n]~\textrm{total}\\
    \TYPE\ t_I[\Multi{\fgTyVar\,\fgType_I}] \ \INTERFACE\ \{ \Foreach{R}{n} \} \in \ForeachN{D}  \\
    \TYPE\ u_I[\Multi{\fgTyVarAux\,\fgTypeAux_I}] \ \INTERFACE\ \{ \Foreach{R'}{q} \} \in \ForeachN{D}  \\
    \Angle{\Multi{\subst{\fgTyVarAux}{\fgTypeAux}}} R'_i =
    \Angle{\Multi{\subst{\fgTyVar}{\fgType}}} R_{\mapPerm(i)} \quad\noteForall{i \in [q]}
  }{
    \tdUpcast{\fgTyEnv}{\subtypeOf{t_I[\Multi{\fgType}]}{u_I[\Multi{\fgTypeAux}]}}
    { \tllambda{\Tuple{Y\Comma \Tuple{\Foreach{X}{n}}}}
      \Tuple{Y\Comma \Tuple{\xT_{\mapPerm(1)}\Comma \ldots\Comma \xT_{\mapPerm(q)}}}
    }
  }
\end{mathpar}
}%

\Cref{f:upcast} defines the relation
$\tdUpcast{\fgTyEnv}{\subtypeOf{\fgType}{\fgTypeAux}}{V}$
for asserting that $\fgType$ is a structural subtype of $\fgTypeAux$, yielding
a coercion function $V$ to convert the target representations of $\fgType$ to $\fgTypeAux$.

Rule \Rule{coerce-tyvar} covers the case of a type variable $\fgTyVar$.
The premise states that type bound $(\fgTyVar : \fgTypeAux_I)$ exists in the environment.
By convention, $X_{\fgTyVar}$ is the name of the corresponding coercion function.
We further find that $\tdUpcast{\fgTyEnv}{\subtypeOf{\fgTypeAux_I}{\fgTypeAux}}{V}$.
Hence, we obtain the coercion function for $\subtypeOf{\fgTyVar}{\fgTypeAux}$
by composition of coercion functions $V$ and $X_{\fgTyVar}$.

Rule \Rule{coerce-struct-iface} covers structs.
The premise $\tdMethods{\fgTySubst R_i}{V_i}{\fgTyEnv}{\fgType_S}$
asserts that each method with name $\methodName{R_i}$ of interface $t_I$ is defined for $\fgType_S$. Value
$V_i$ is a tuple with coercion parameters corresponding to the bounds of the receiver's type parameters.
Thus, $U_i = \tllambda{\TripleY} X_{m_i,t_S}\,\QuadrY{V_i}$
is the dictionary entry for the $i$-th method:
a function accepting receiver $Y_1$, coercion parameters $Y_2$ corresponding to
bounded type parameters of the method,
and the argument tuple $Y_3$. As written earlier, dictionary entries and top-level functions
$X_{m_i,t_S}$ are uncurried. Thus, we need to deconstruct the argument triple $\TripleY$
and construct a new quadruple $\QuadrY{V}$ for calling $X_{m_i,t_S}$.

Rule \Rule{coerce-iface-iface} covers structural subtyping between interface types
$t_I[\Multi{\fgType}]$ and $u_I[\Multi{\fgTypeAux}]$. In this case, $t_I$ must declare all methods of $u_I$,
so we can build a dictionary for $u_I$ from
the methods in the dictionary for $t_I$.
Thus, the premise of the rule
requires the total function $\pi$ to be chosen
in such a way that the $i$-th method of $u_I$ has the same signature as the
$\pi(i)$-th method of $t_I$. The translation uses pattern matching to deconstruct
the dictionary of $t_I$ as
$\Tuple{\Multi[n]{X}}$. Then the $i$-th method in the dictionary of $u_I$ is
$X_{\pi(i)}$, so we construct the wanted dictionary as
$\Triple{X_{\pi(1)}}{\ldots}{X_{\pi(q)}}$.

\subsubsection{Translation of Expressions}
\label{sec:transl-expr}

\boxfig{f:trans-exp1}{Translation of expressions}{
  \sTransSection{
    \ruleform{\tdExpTrans{\pair{\fgTyEnv}{\fgEnv}}{e : \fgType}{\expT}}
  }{
    Translating expressions
  }
  \begin{mathpar}
    \inferrule[var]{
      (x : \fgType) \in \fgEnv
    }{
      \tdExpTrans{\pair{\fgTyEnv}{\fgEnv}}{x : \fgType}{\xT}
    }

    \inferrule[struct]{
      \fgTyOk{\fgTyEnv}{t_S[\Multi{\fgType}]}\\
      \TYPE\ t_S[\Multi{\fgTyVar\,\fgType_I}] \ \STRUCT\ \{ \Foreach{f \ \fgTypeAux}{n} \} \in  \ForeachN{D}\\\\
      \tdExpTrans{\pair{\fgTyEnv}{\fgEnv}}{e_i : \Angle{\Multi{\subst{\fgTyVar}{\fgType}}}\fgTypeAux_i}{\expT_i}
      \quad\noteForall{i \in [n]}
    }{
      \tdExpTrans{\pair{\fgTyEnv}{\fgEnv}}
      {t_S[\Multi{\fgType}] \{ \Foreach{e}{n} \} : t_S[\Multi{\fgType}] }{\Tuple{\Foreach{\expT}{n}}}
    }

  \inferrule[access]{
    \tdExpTrans{\pair{\fgTyEnv}{\fgEnv}}{e : t_S[\Multi{\fgType}]}{\expT} \\
    \TYPE\ t_S[\Multi{\fgTyVar\,\fgType_I}] \ \STRUCT\ \{ \Foreach{f \ \fgTypeAux}{n} \} \in  \ForeachN{D}
  }{
    \tdExpTrans{\pair{\fgTyEnv}{\fgEnv}}
    {e.f_i : \Angle{\Multi{\subst{\fgTyVar}{\fgType}}} \fgTypeAux_i }
    { \CASE\ \expT\ \OF\ \Tuple{\Foreach{\xT}{n}} \rightarrow \xT_i}
  }

  \inferrule[call-struct]{
    \tdExpTrans{\pair{\fgTyEnv}{\fgEnv}}{e : t_S[\Multi{\fgType}]}{\expT} \\
    \tdMethods{m[\Multi{\fgTyVar'\,\fgType_I'}](\Foreach{x\,\fgTypeAux}{n})\fgTypeAux}{V}{\fgTyEnv}{t_S[\Multi{\fgType}]}\\
    \tdCheckSubst{\fgTyEnv}{\Multi{\fgTyVar'\,\fgType_I'}}{\Multi{\fgType'}}{\fgTySubst}{V'}\\
    \tdExpTrans{\pair{\fgTyEnv}{\fgEnv}}{e_i: \fgTySubst \fgTypeAux_i}{\expT_i}
    \quad\noteForall{i \in [n]}
  }{
    \tdExpTrans{\pair{\fgTyEnv}{\fgEnv}}{e.m[\Multi{\fgType'}](\Foreach{e}{n}) : \fgTySubst \fgTypeAux}
    {\mT{m}{t_S}~\Quadr{V}{\expT}{V'}{\Tuple{\Foreach{\expT}{n}}} }
  }

  \inferrule[call-iface]{
    \tdExpTrans{\pair{\fgTyEnv}{\fgEnv}}{e : \fgType_I}{\expT}\\
    \methodSpecifications{\fgType_I} = \Foreach{R}{q}\\
    R_j = m[\Multi{\fgTyVar'\,\fgType_I'}](\Foreach{x\,\fgTypeAux}{n})\fgTypeAux \quad(\textrm{for some}~j \in [q])\\
    \tdCheckSubst{\fgTyEnv}{\Multi{\fgTyVar'\,\fgType_I'}}{\Multi{\fgType'}}{\fgTySubst}{V'}\\
    \tdExpTrans{\pair{\fgTyEnv}{\fgEnv}}{e_i : \fgTySubst \fgTypeAux_i}{\expT_i}
    \quad\noteForall{i \in [n]}\\
    Y, \Foreach{X}{q}\textrm{~fresh}
  }{
    \tdExpTrans{\pair{\fgTyEnv}{\fgEnv}}
    {e.m[\Multi{\fgType'}](\Foreach{e}{n}) : \fgTySubst\fgTypeAux}
    {\CASE\ \expT \ \OF\ \Tuple{Y\Comma \Tuple{\Foreach{X}{q}}} \to X_j\Triple{Y}{V'}{\Tuple{\Foreach{E}{n}}}}
  }

  \inferrule[sub]{
    \tdExpTrans{\pair{\fgTyEnv}{\fgEnv}}{e : \fgType}{\expT}\\
    \tdUpcast{\fgTyEnv}{\subtypeOf{\fgType}{\fgTypeAux}}{V}
  }{
    \tdExpTrans{\pair{\fgTyEnv}{\fgEnv}}{e : \fgTypeAux}{V \ \expT}
  }
\end{mathpar}
}

\Cref{f:trans-exp1} defines the typing and translation relation for expressions.
The judgment $\tdExpTrans{\pair{\fgTyEnv}{\fgEnv}}{e : \fgType}{E}$ states that
under type environment $\fgTyEnv$ and value environment $\fgEnv$ the \FGGm\ expression $e$
has type $\fgType$ and translates to TL expression $E$.

Rule \Rule{var} retrieves the type
of \FGGm\ variable $x$ from the environment and translates $x$ to its TL counterpart $X$.
The context makes variable $X$ available, see the translation of method definitions in \Cref{sec:transl-meth-progr}.
Rule \Rule{struct} type checks and translates a struct literal $t_S[\Multi{\fgType}](\Multi{e})$.
Premise $\fgTyOk{\fgTyEnv}{t_S[\Multi{\fgType}]}$ checks that type $t_S[\Multi{\fgType}]$ is well-formed;
the definition of the judgment $\fgTyOk{\fgTyEnv}{\fgType}$ is given in \Cref{f:well-formedness} and will be explained
in the next subsection. Each argument $e_i$ translates to $E_i$, so the result is
$\Tuple{\Multi[n]{E}}$.
Rule \Rule{access} deals with field access $e.f_i$, where expression $e$ must have
struct type $t_S[\Multi{\fgType}]$ such that $t_S$ defines field $f_i$. Thus, $e$ translates to a tuple $E$,
from which we extract the $i$-th component via pattern matching.

Rule \Rule{call-struct} handles a method call $e.m[\Multi{\fgType'}](\Multi{e})$, where
receiver $e$ has struct type $t_S[\Multi{\fgType}]$ and translates to $E$.
The $V$ in the premise corresponds to a tuple of coercion functions that
result from checking the bounds of the receiver's type parameters, whereas
$V'$ is a tuple of coercion functions for the bounds of the type parameters of the method.
Argument $e_i$ translates to $E_i$.
According to our translation strategy, a method declaration for $m$ and $t_S$ is
represented as a top-level function $X_{m,t_S}$ accepting a quadruple:
coercions for the receiver's type parameters,
receiver,
coercions for the bounded type parameters local to the method, and method arguments. Thus,
the result of the translation is
$\mT{m}{t_S}\ \Quadr{V}{\expT}{V'}{\Tuple{\Multi{\expT}}}$.

Rule \Rule{call-iface} handles a method call $e.m[\Multi{\fgType'}](\Multi{e})$, where
receiver $e$ has interface type $\fgType_I$ and translates to $E$.
Similar to \Rule{call-struct}, $V'$ is a tuple of coercion functions that result from checking the bounds of the
type parameters local to the method.
Expressions $E_i$ are the translation of the arguments $e_i$.
Following our translation strategy, receiver $E$ is a pair where the first component
is a struct value and the second component is a
dictionary for the interface. Thus, we use pattern matching to extract the struct as $Y$
and the wanted method as $X_j$. This $X_j$ is a function accepting a triple:
receiver, coercions for bounded type parameters of the method, and method arguments.
Hence, the translation result is $X_j\,\Triple{Y}{V'}{\Tuple{\Multi{E}}}$.
The difference to rule \Rule{call-struct} is that there is no need to supply coercions for the bounded
type parameters of the receiver.
These coercions have already been supplied when building the dictionary,
see rule \Rule{coerce-struct-iface} of \Cref{f:upcast}.

The last rule \Rule{sub} is a subtyping rule allowing an expression $e$
with translation $E$ at type $\fgType$ to be assigned some (structural) supertype $\fgTypeAux$.
Premise
$\tdUpcast{\fgTyEnv}{\subtypeOf{\fgType}{\fgTypeAux}}{V}$
serves two purposes: it ensures that $\fgTypeAux$ is a supertype of $\fgType$
and it yields a coercion function $V$ from
$\fgType$ to $\fgTypeAux$. The translation
of $e$ at type $\fgTypeAux$ is then $V\,E$.
In~\citet{FeatherweightGo}, the subtype check is included for each form of expression.
For clarity, we choose to have a separate subtyping rule
as in our translation scheme each subtyping relation implies a coercion function.

\subsubsection{Well-formedness}
\label{sec:well-form-judgm}

\boxfig{f:well-formedness}{Well-formedness}{
  \sTransSection{
    \ruleform{
      \fgTyOk{\fgTyEnv}{\fgType}\qquad
      \fgTyOk{\fgTyEnv}{\Multi{\fgType}}
    }
  }{
    Well-formedness of types
  }
  \begin{mathpar}
    \inferrule[ok-tyvar]{
      (\fgTyVar : \fgType_I) \in \fgTyEnv
    }{
      \fgTyOk{\fgTyEnv}{\fgTyVar}
    }

    \inferrule[ok-tynamed]{
      \fgTyOk{\fgTyEnv}{\Multi{\fgType}}\\
      \TYPE\,t[\Multi{\fgTyVar\,\fgType_I}]\ldots \in \fgDecls\\\\
      \tdCheckSubst{\fgTyEnv}{\Multi{\fgTyVar\,\fgType_I}}{\Multi{\fgType}}{\fgTySubst}{V}
    }{
      \fgTyOk{\fgTyEnv}{t[\Multi{\fgType}]}
    }

    \inferrule[ok-many-ty]{
      \fgTyOk{\fgTyEnv}{\fgType_i} \quad (\forall i \in [n])
    }{
      \fgTyOk{\fgTyEnv}{\Multi[n]{\fgType}}
    }
  \end{mathpar}

  \sTransSection{
    \ruleform{
      \fgFormalsOk{\fgTyEnv}{\Multi{\fgTyVar\,\fgType_I}}\qquad
      \fgTyOk{\fgTyEnv}{R}
    }
  }{
    Well-formedness of type parameters and method signatures
  }
  \begin{mathpar}
    \inferrule[ok-bounded-typarams]{
      \Dom{\fgTyEnv} \cap \{ \Multi{\fgTyVar} \} = \emptyset\\
      \distinct{\Multi{\fgTyVar}}\\\\
      \fgTyOk{\fgTyEnv \cup \{ \Multi{\fgTyVar : \fgType_I} \}}{\Multi{\fgType_I}}
    }{
      \fgFormalsOk{\fgTyEnv}{\Multi{\fgTyVar\,\fgType_I}}
    }

    \inferrule[ok-msig]{
      \fgFormalsOk{\fgTyEnv}{\Multi{\fgTyVar\,\fgType_I}}\\
      \distinct{\Multi{x}}\\\\
      \fgTyOk{\fgTyEnv \cup \{ \Multi{\fgTyVar : \fgType_I} \}}{\Multi{\fgTypeAux}\fgTypeAux}
    }{
      \fgTyOk{\fgTyEnv}{m[\Multi{\fgTyVar\,\fgType_I}](\Multi{x\,\fgTypeAux})\fgTypeAux}
    }
  \end{mathpar}

  \sTransSection{
    \ruleform{
      \fgDeclOk{D}
    }
  }{
    Well-formedness of declarations
  }
  \begin{mathpar}
    \inferrule[ok-struct]{
      t_S \textrm{~defined once in~}\fgDecls\\
      \fgFormalsOk{\fgEmptyEnv}{\Multi{\fgTyVar\,\fgType_I}}\\
      \fgTyOk{\{ \Multi{\fgTyVar : \fgType_I} \}}{\Multi{\fgTypeAux}}\\
      \distinct{\Multi{f}}
    }{
      \fgDeclOk{
        \TYPE\ t_S[\Multi{\fgTyVar\,\fgType_I}]\
        \STRUCT\ \{ \overline{f \ \fgTypeAux} \}
      }
    }

    \inferrule[ok-iface]{
      t_I \textrm{~defined once in~}\fgDecls\\
      \fgFormalsOk{\fgEmptyEnv}{\Multi{\fgTyVar\,\fgType_I}}\\
      (\forall i \in [n])~\fgTyOk{ \{\Multi{\fgTyVar : \fgType_I} \} }{R_i}\\\\
      \distinct{\Multi{\methodName{R_i}}}
    }{
      \fgDeclOk{
        \TYPE\ t_I[\Multi{\fgTyVar\,\fgType_I}]\
        \INTERFACE\ \{ \Multi[n]{R} \}
      }
    }

  \inferrule[ok-method]{
    \ForeachN{D}~\textrm{contains one}~\FUNC\textrm{-declaration for}~t_S\textrm{~and~}\methodName{R}\\
    \fgFormalsOk{\fgEmptyEnv}{\Multi{\fgTyVar\,\fgType_I}}\\
    \fgTyOk{ \{ \Multi{\fgTyVar : \fgType_I} \} }{R}\\
    (\TYPE\ t_S[\Multi[n]{\fgTyVar\,\fgType_I'}]\,\STRUCT\ldots) \in \fgDecls\\
    \methodSpecifications{{\fgType_I'}_i} \subseteq \methodSpecifications{{\fgType_I}_i}\quad(\forall i \in [n])
  }{
    \fgDeclOk{
      \FUNC\ (x \ t_S[\Multi[n]{\fgTyVar\,\fgType_I}]) R \ \{ \RETURN\ e \}
    }
  }
  \end{mathpar}
}

\Cref{f:well-formedness} defines several well-formedness judgments.
The judgments $\fgTyOk{\fgTyEnv}{\fgType}$ and $\fgTyOk{\fgTyEnv}{\Multi{\fgType}}$
assert that a single type and multiple types, respectively, are well-formed
under type environment $\fgTyEnv$. A type variable is well-formed if it is contained
in $\fgTyEnv$ (rule \Rule{ok-tyvar}). A named type $t[\Multi{\fgType}]$ is well-formed
if its type arguments $\Multi{\fgType}$ are well-formed and if they are subtypes of the upper bounds
in the definition of $t$. The latter is checked by the premise
$\tdCheckSubst{\fgTyEnv}{\Multi{\fgTyVar\,\fgType_I}}{\Multi{\fgType}}{\fgTySubst}{V}$
of rule \Rule{ok-tynamed}, thereby ignoring the type substitution $\fgTySubst$ and the
coercion functions $V$. We have already seen in \Cref{sec:bound-type-param} that these
coercions $V$ are not represented in the translated program because
type bounds of structs and interfaces have no operational meaning.

Judgment $\fgTyOk{\fgTyEnv}{\Multi{\fgTyVar\,\fgType_I}}$ asserts that bounded type parameters
$\Multi{\fgTyVar\,\fgType_I}$ are well-formed
under type environment $\fgTyEnv$ (rule \Rule{ok-bounded-typarams}).
Judgment $\fgTyOk{\fgTyEnv}{R}$
ensures that a method signature
is well-formed (rule \Rule{ok-msig}). To form the combined environment
$\fgTyEnv \cup \{\Multi{\fgTyVar : \fgType_I}\}$ in the premise requires disjointness of the type variables in
$\Dom{\fgTyEnv}$ and $\Multi{\fgTyVar}$.
This can always be achieved by $\alpha$-renaming the
type variables bound by $R$.

Judgment $\fgDeclOk{D}$ validates declaration $D$.
A struct declaration is well-formed if it is defined only once
(restriction \FGGUniqueStructs{} in \Cref{sec:dynamic-semantics}), if
all field names are distinct (restriction \FGGUniqueFields{}),
and if the field types are well-formed.
An interface declaration is well-formed if it is defined only once, if all its method signatures are well-formed, and if
all methods have distinct names.

A method declaration for $t_S$ and $m$ is well-formed if there is no other declaration
for $t_S$ and $m$ (restriction \FGGUniqueReceiver{}), if the method signature is well-formed,
and if each bound $\fgType_{Ii}$ of the method declaration is a structural subtype of the
corresponding bound $\fgType_{Ii}'$ in the declaration of $t_S$.
In \FGGm, this boils down to checking that the methods
of $\fgType_{Ii}'$ are a subset of the methods of $\fgType_{Ii}$.
The well-formedness conditions for method declarations
do not type check the method body. We will deal with this in the upcoming translation
rule for methods.

\subsubsection{Translation of Methods and Programs}
\label{sec:transl-meth-progr}

\boxfig{f:trans-meth-prog}{Translation of methods and programs}{
\sTransSection{
  \ruleform{\tdMethTrans{\FUNC\ (x \ t_S[\Multi{\fgTyVar\,\fgType_I}])\,R\,\{\RETURN\,e\}}{X = V} }
}{
  Translating method declarations
}
\begin{mathpar}
  \inferrule[method]{
    \fgTyEnv = \{\Multi{\fgTyVar\,\fgType_I}, \Multi{\fgTyVarAux\,\fgTypeAux_I}\}\\
    \fgEnv = \{ x : t_S[\Multi{\fgTyVar}], \Multi{x : \fgTypeAux} \}\\
    x \notin \{\Multi{x}\}\\
    \tdExpTrans
      {\pair{\fgTyEnv}{\fgEnv}}
      {e : \fgTypeAux}{\expT}
      \\
      V = \lambda \Quadr{\Tuple{\Multi{\xTy{\fgTyVar_i}}}}
                  {X}
                  {\Tuple{\Multi{\xTy{\fgTyVarAux_i}}}}
                  {\Tuple{\Multi{X}}}
           . E
  }{
    \tdMethTrans
    {\FUNC\ (x \ t_S[\Multi{\fgTyVar\,\fgType_I}]) \,
      m[\Multi{\fgTyVarAux\,\fgTypeAux_I}](\Multi{x \ \fgTypeAux})\, \fgTypeAux \ \{ \RETURN\ e \}}
    {\mT{m}{t_S} = V}
  }
\end{mathpar}

\sTransSection{
  \ruleform{\tdProgTrans{P}{\program}}
}{
  Translating programs
}
\begin{mathpar}
  \inferrule[prog]{
    \Multi{D} \textrm{~implicitly available in all subderivations}\\
    \tdExpTrans{\pair{\EmptyFgEnv}{\EmptyFgEnv}}{e : \fgType}{\expT} \\\\
    \fgDeclOk{D_i} \\
    (\textrm{for all}~D_i \in \Multi{D})\\\\
    \tdMethTrans{D_i}{X_i = V_i}\\
    (\textrm{for all}~D_i = \FUNC\,\ldots \in \Multi{D})\\
 }
 { \TurnsTdProg
   {\ForeachN{D} \ \FUNC\ \MAIN () \{ \_ = e \}} \leadsto
     \LET\ \ForeachN{X_i = V_i}\ \IN\ \expT
 }
\end{mathpar}
} %

\Cref{f:trans-meth-prog} defines the translation for method declarations and programs.
Rule \Rule{method} deals with method declarations
$\FUNC\ (x \ t_S[\Multi{\fgTyVar\,\fgType_I}]) \, m[\Multi{\fgTyVarAux\,\fgTypeAux_I}](\Multi{x \ \fgTypeAux}) \, \fgTypeAux \ \{ \RETURN\ e \}$.
The translation of such a declaration is the binding $\mT{m}{t_S} = V$.
According to our translation
strategy, $V$ must be a function accepting a quadruple:
coercions $\Tuple{\Multi{X_{\fgTyVar_i}}}$ for the bounded type parameters of the receiver,
receiver $X$ corresponding to $x$,
coercions $\Tuple{\Multi{X_{\fgTyVarAux_i}}}$ for the bounded type parameters local to the method,
and finally method arguments $\Multi{X}$ corresponding to $\Multi{x}$. Binding
all these variables with a $\lambda$ makes them available in the translated body $E$.

Judgment $\tdProgTrans{P}{\program}$ denotes the translation of an \FGGm\ program $P$ to a TL program $\program$.
Rule \Rule{prog}
type checks the main expression $e$ under empty environments against some type
$\fgType$ to get its translation $E$. Next, the rule requires all struct or interface declarations
to be well-formed. Finally, it translates each method declaration to a binding $X_i = V_i$.
The resulting TL program is then
$\LET\ \ForeachN{X_i = V_i}\ \IN\ \expT$.

\subsection{Example}
\label{sec:example-translation}

\boxfig{f:trans-example-src}{Example: \FGGm\ code (top) and its translation (middle)
  with abbreviations (bottom)
}{
  \vspace{-1ex}
\begin{flalign*}
  \begin{array}{l@{}}
    \TYPE\ \Any\ \INTERFACE\ \{ \}\\
    \TYPE\ \Num\ \STRUCT\ \{\, \val\ \IntT\, \}\\
    \TYPE\ \BoxT[\fgTyVar\, \Any]\ \STRUCT\ \{\, \content\ \fgTyVar\, \}\\
    \TYPE\ \Eq[\fgTyVar\, \Any]\ \INTERFACE\ \{\, \eq(\that\,\fgTyVar)\,\bool\, \}\\
    \FUNC\ (\this \ \Num) \, \eq(\that\,\Num) \ \bool\ \{\,
       \RETURN\ \this.\val\,\mathtt{==}\,\that.\val \,\}
    \\
    \FUNC\ (\this \ \BoxT[\fgTyVar\,\Eq[\fgTyVar]]) \,
    \eq(\that\,\BoxT[\fgTyVar]) \ \bool\ \{\,
       \RETURN\ \this.\content.\eq(\that.\content) \,\}
    \\
    \FUNC\ \main() \{\, \_ =
    \BoxT[Num]\{\Num\{1\}\}.\eq(\BoxT[Num]\{\Num\{2\}\}) \,\}
  \end{array}
  &&%
\end{flalign*}
\hrule{}
\vspace{-1ex}
\begin{flalign*}
  \begin{array}[t]{l@{~}r@{}l@{}}
    \LET &
           X_{\eq,\Num} & {}=
                          \tllambda{\Quadr{\Void}{\This}{\Void}{\Tuple{\That}}} E_2 \,\mathtt{==}\, E_3
    \\
         & X_{\eq,\BoxT} & {}=
      \tllambda{\Quadr{\Tuple{X_{\fgTyVar}}}{\This}{\Void}{\Tuple{\That}}} E_1
    \\
    \multicolumn{3}{l}{
    \IN~
    X_{\eq,\BoxT}~\Quadr{\Tuple{V_3}}{\BoxVal{1}}{\Void}{\BoxVal{2}}
    }
  \end{array}
  &&%
\end{flalign*}
\newcommand\Comment[1]{\multicolumn{2}{l@{}}{\mbox{\small\color{GrayFgColor} \texttt{--}~\ensuremath{#1}}}}
\par\noindent\hdashrule{\textwidth}{0.4pt}{1pt}
\begin{minipage}[t]{0.5\linewidth}
\vspace{-4ex}
  \begin{flalign*}
  \begin{array}[t]{r@{}l@{}}
    \Comment{\textrm{translated body of~}\eq\textrm{~for~}\BoxT}
    \\
    E_1 & {}= \CASE\,V_1\,E_2\, \OF\, \Pair{Y}{\Tuple{X_1}} \to
    \\ & \qquad X_1~\Triple{Y}{\Void}{\Tuple{E_3}}
    \\[\smallskipamount]
    \Comment{\textrm{selectors for field~}\mathit{content}\textrm{~of~}\BoxT}
    \\
    E_2 & {}= \CASE\ \This\ \OF\ \Tuple{X_1} \to X_1
    \\
    E_3 & {}= \CASE\ \That\ \OF\ \Tuple{X_1} \to X_1
  \end{array}
  &&%
\end{flalign*}
\end{minipage}
\begin{minipage}[t]{0.5\linewidth}
\vspace{-4ex}
\begin{flalign*}
  \begin{array}[t]{r@{}l@{}}
    \Comment{\textrm{coercion}~\ensuremath{\subtypeOf{\fgTyVar}{\Eq[\fgTyVar]}}}
    \\
    V_1 & {}= \tllambda{Y} V_2\,(X_{\fgTyVar}\,Y)
    \\[\smallskipamount]
    \Comment{\textrm{identity coercion}~\ensuremath{\subtypeOf{\Eq[\fgTyVar]}{\Eq[\fgTyVar]}}}
    \\
    V_2 & {}= \tllambda{\Pair{Y}{\Tuple{X}}} \Pair{Y}{\Tuple{X}}
    \\[\smallskipamount]
    \Comment{\textrm{coercion}~\ensuremath{\subtypeOf{\Num}{\Eq[\Num]}}}
    \\
    V_3 & {}= \tllambda{X}\Pair{X}{\Tuple{\tllambda{\TripleY}{X_{\eq,\Num}\,\QuadrY{\Void}}}}
  \end{array}
  &&%
\end{flalign*}
\end{minipage}
}

We now give an example of the translation. The \FGGm\ code in the top part of \Cref{f:trans-example-src}
defines equality for numbers $\Num$ and for generic boxes $\BoxT[\fgTyVar\,\Any]$.
Interface $\Any$ defines no methods, it serves as an upper bound for otherwise unrestricted
type variables.
We take the liberty to assume a basic type $\IntT$ and an operator
$\mathtt{==}$ for equality.
Interface $\Eq[\fgTyVar]$ requires a method $\eq$
for comparing the receiver with a value of type $\fgTyVar$.
We provide implementations of $\eq$ for $\Num$ and $\BoxT[\fgTyVar]$. Comparing the content of
a box requires the F-bound $\Eq[\fgTyVar]$ \citep{Canning1989}.
The main function compares two boxes for equality.

\boxfig{f:trans-example}{Example: translation of the method declaration for $\BoxT$ and $\eq$}{
  \begin{mathpar}
  \inferrule*[rightstyle=\tiny\sc,leftstyle=\tiny\sc,right=method]{
    \inferrule*[Right=call-iface]{
      {\begin{array}[b]{@{}c@{}}
         \inferrule*[Left=sub]{
         \CircledText{1}\qquad
           \inferrule*[Right=access]{
             \ldots
           }{
             \tdExpTrans{\pair{\fgTyEnv}{\fgEnv}}{\this.\content : \fgTyVar}{E_2}
           }
         }{
           \tdExpTrans{\pair{\fgTyEnv}{\fgEnv}}{\this.\content : \Eq[\fgTyVar]}{V_1\,E_2}
         }
       \end{array}
     }
     \quad
     \inferrule*[Right=access]{
       \ldots
     }{
       \tdExpTrans{\pair{\fgTyEnv}{\fgEnv}}{\that.\content : \fgTyVar}{E_3}
     }
     \\
     \methodSpecifications{\Eq[\fgTyVar]} = \eq(\that\,\fgTyVar)\bool
   }{
     \tdExpTrans{\pair{\fgTyEnv}{\fgEnv}}{\this.\content.\eq(\that.\content) : \bool}{E_1}
   }
   \\
   \fgTyEnv = \{ \fgTyVar : \Eq[\fgTyVar] \}\\
   \fgEnv = \{ \this : \BoxT[\fgTyVar], \that : \BoxT[\fgTyVar] \}
 }{
   \tdMethTrans{
     \begin{array}[t]{@{}l@{}}
       \FUNC\ (\this \ \BoxT[\fgTyVar\,\Eq[\fgTyVar]]) \,
       \eq(\that\,\BoxT[\fgTyVar]) \ \bool\ \{\\
       \quad \RETURN\ \this.\content.\eq(\that.\content) \\ \}
     }
       {X_{\eq,\BoxT} =
       \tllambda{\Quadr{\Tuple{X_{\fgTyVar}}}{\This}{\Void}{\Tuple{\That}}} E_1
       }
   \end{array}
}
\end{mathpar}
\vspace{0.1cm}

\TransSectionCenter{
  Subderivation~\CircledText{1}
}
\begin{mathpar}
  \inferrule*[rightstyle=\sc\tiny,right=coerce-tyvar]{
    \inferrule*[right=coerce-iface-iface]{
    }{
      \tdUpcast{\fgTyEnv}{\subtypeOf{\Eq[\fgTyVar]}{\Eq[\fgTyVar]}}{V_2}
    }
    \\
    {
      \begin{array}[b]{c}
        (\fgTyVar : \Eq[\fgTyVar]) \in \fgTyEnv
      \end{array}
    }
  }{
    \tdUpcast{\fgTyEnv}{\subtypeOf{\fgTyVar}{\Eq[\fgTyVar]}}{V_1}
  }
\end{mathpar}
} %

\boxfig{f:trans-example2}{Example: translation of the main function}{
\begin{mathpar}
  \inferrule*[rightstyle=\sc\tiny,right=call-struct]{
    \inferrule*[Right=methods-struct]{
      \inferrule*[Right=type-inst-checked]{
        \inferrule*[Right=coerce-struct-iface]{
          \inferrule*[Right=methods-struct]{
            \ldots
          }{
            \tdMethods{\eq(\That\,\Num)\,\bool}{\Void}{\fgEmptyTyEnv}{\Num}
          }
        }{
          \tdUpcast{\fgEmptyTyEnv}{\subtypeOf{\Num}{\Eq[\Num]}}{V_3}
        }
      }{
        \tdCheckSubst{\fgEmptyTyEnv}{\fgTyVar\,\Eq[\fgTyVar]}{\Num}{\LrAngle{\subst{\fgTyVar}{\Num}}}{\Tuple{V_3}}
      }\\
      \FUNC\ (\this \ \BoxT[\fgTyVar\,\Eq[\fgTyVar]]) \, \eq(\that\,\BoxT[\fgTyVar]) \ \bool\,\ldots \in \fgDecls
    }{
      \tdMethods{\eq(\That\,\BoxT[\Num])\,\bool}{\Tuple{V_3}}{\fgEmptyTyEnv}{\BoxT[\Num]}
    }
    \\
    \tdExpTransEmpty{\BoxT[\Num]\{\Num\{1\}\} : \BoxT[\Num]}{\Tuple{\Tuple{1}}}\\
    \tdExpTransEmpty{\BoxT[\Num]\{\Num\{2\}\} : \BoxT[\Num]}{\Tuple{\Tuple{2}}}
  }{
    \begin{array}[t]{@{}l@{}}
    \tdExpTransEmpty{\BoxT[\Num]\{\Num\{1\}\}.\eq(\BoxT[\Num]\{\Num\{2\}\}) : \bool\\ \qquad\qquad}{
      X_{\eq,\BoxT}~\Quadr{\Tuple{V_3}}{\BoxVal{1}}{\Void}{\BoxVal{2}}
    }
    \end{array}
  }
\end{mathpar}
} %

The middle part of the figure shows the translation of the \FGGm\ code, using abbreviations in the bottom part.
Variable $X_{\eq,\Num}$ holds the translation of the declaration of $\eq$ for $\Num$; it simply
compares $E_2$ (translation of $\this.\val$) with $E_3$ (translation of $\that.\val$).
Remember that the translation of a method declaration takes a quadruple with coercions for the
bounded type parameters of the receiver, the receiver itself, coercions for the bounded type parameters of the
method, and the method arguments.
Here, $\Void$ is a tuple of size zero, corresponding to the non-existing type
parameters, $\Tuple{\That}$ denotes a tuple of size one, corresponding to the single argument
$\that$.

The translation of $\eq$ for $\BoxT$ is more involved. \Cref{f:trans-example} shows
its derivation.
We omit \enquote{obvious} premises
and some trivial details from the derivation trees.
Rule \Rule{call-iface} translates the body of the method. It coerces the
receiver to the interface type $\Eq[\fgTyVar]$ and then extracts the
method to be called via pattern matching, see $E_1$.
The construction of the coercion is done via
$\tdUpcast{\fgTyEnv}{\subtypeOf{\fgTyVar}{\Eq[\fgTyVar]}}{V_1}$,
see subderivation \CircledText{1}.
Coercion $V_1$ is slightly more complicated then necessary because the translation does not optimize
the identity coercion $V_2$.
Inside of $V_1$, we use $X_{\fgTyVar}$.
This variables denotes a coercion from $\fgTyVar$ to the representation of $\Eq[\fgTyVar]$;
it is bound by the $\lambda$-expression in the definition of $X_{\eq,\BoxT}$.

The translation of the main expression calls $X_{\eq,\BoxT}$ with appropriate
arguments, see \Cref{f:trans-example2} for the derivation.
The values $\Tuple{\Tuple{1}}$ and $\Tuple{\Tuple{2}}$ are nested tuples of size one,
representing numbers wrapped in $\Num$ and $\BoxT$ structs.
The method call of $\eq$ is translated by rule \Rule{call-struct},
relying on rule \Rule{methods-struct} to instantiate the type variable $\fgTyVar$
to $\Num$, as witnessed by the coercion $V_3$.

\section{Formal Properties}
\label{sec:formal-properties}

In this section, we establish that the type-directed translation from
\Cref{sec:type-directed} preserves the static and dynamic semantics of \FGGm\ programs.
The translation as formalized is non-deterministic:
for the same source program we may derive syntactically different target programs.
Thus, we further show that all target programs resulting from the same source program behave
equivalently.
Detailed proofs for all lemmas and theorems are given in the appendix.

\subsection{Preservation of Static Semantics}
\label{sec:pres-stat-semant-1}

It is straightforward to verify that the type system originally defined for FGG
is equivalent to the type system induced by the type-directed translation presented
in \Cref{sec:type-directed}, provided the FGG program does not contain type assertions.
In the following, we write $\fgTyEnv \turnsFGG \subtypeOf{\fgType}{\fgTypeAux}$ for FGG's subtyping relation,
$\fgTyEnv; \fgEnv \turnsFGG e : \fgType$ for its typing relation on expressions, and
$\turnsFGG \fgOk{P}$ for the FGG typing relation on programs. These three relations were specified
by \citet{FeatherweightGo}. The original article on FGG also includes support for dynamic type assertions,
something we do not consider for our translation. Hence, we assume that FGG expressions do not
contain type assertions.

\begin{lemma}[FGG typing equivalence]
  \label{thm:fgg-typing-equiv}
  Typing in FGG is equivalent to the type system
  induced by the translation, provided there are no type assertions.
  \begin{EnumAlph}
  \item If $\fgTyEnv \turnsFGG \subtypeOf{\fgType}{\fgTypeAux}$ then
    either $\tdICons{\fgTyEnv}{\fgType}{\fgTypeAux}{V}$ for some $V$ or
    $\fgTypeAux = \fgType$ and $\fgType$ is not an interface type.
  \item If $\tdICons{\fgTyEnv}{\fgType}{\fgTypeAux}{V}$ then
    $\fgTyEnv \turnsFGG \subtypeOf{\fgType}{\fgTypeAux}$.
  \item If $\fgTyEnv; \fgEnv \turnsFGG e : \fgType$ then
    $\tdExpTrans{\pair{\fgTyEnv}{\fgEnv}}{e : \fgType}{E}$ for some $E$.
  \item If $\tdExpTrans{\pair{\fgTyEnv}{\fgEnv}}{e : \fgType}{E}$ then
    $\fgTyEnv; \fgEnv \turnsFGG e : \fgType'$ for some $\fgType'$ and $\fgTyEnv \turnsFGG \subtypeOf{\fgType'}{\fgType}$.
  \item $\turnsFGG \fgOk{P}$ iff $\tdProgTrans{P}{\program}$.
  \end{EnumAlph}
\end{lemma}

Claims (a) and (b) state that structural subtyping in FGG is equivalent to the relation
from \Cref{f:upcast}, except that the latter is not reflexive for type variables and struct types.
Claims (c) and (d) establish that expression typing in FGG and our expression typing
from \Cref{f:trans-exp1} are equivalent modulo subtyping.
The exposition in~\citet{FeatherweightGo} includes a subtyping check for each form of expression
whereas we choose to have a separate subtyping rule.
Hence, the type computed by the original rules for FGG might be a subtype of the type deduced by our system.

FGG enjoys type soundness (see Theorem~4.3 and~4.4 of \citealt{FeatherweightGo}). The reduction rules
for FGG and \FGGm{} are obviously equivalent. Thus,
\Cref{thm:fgg-typing-equiv} gives the following type soundness result for our type system:
\begin{corollary}
  \label{cor:fgg-soundness}
  Assume $\tdExpTrans{\pair{\fgEmptyEnv}{\fgEmptyEnv}}{e : \fgType}{E}$ for some $e$, $\fgType$, and $E$.
  Then either $e$ reduces to some value of type $\fgType$ or $e$ diverges.
\end{corollary}

\subsection{Preservation of Dynamic Semantics}
\label{sec:pres-dynam-semant-1}

This section proves that evaluating a well-typed \FGGm\ program yields the same behavior as evaluating one of its
translations. Thereby, we consider all possible outcomes of evaluation: reduction to a value or divergence.
Further, we show that different translations of the same program have equivalent behavior.

The proof of semantic equivalence is enabled by a syntactic, step-indexed logical relation that
relates an \FGGm\ expression and a TL expression at some \FGGm\ type.
We write $\reducek{k}{e}{e'}$ if $e$ reduces to $e'$ in \emph{exactly} $k \in \Nat$ steps,
where $\Nat$ denotes the natural numbers including zero.
By convention, we write $\reducek{0}{e}{e'}$ to denote $e = e'$.
The notation $\reduceStar{e}{e'}$ states that $\reducek{k}{e}{e'}$ for some $k \in \Nat$.
We write $\Diverge{e}$ to denote that $e$ does not terminate; that is, for all
$k \in \Nat$ there exists some $e'$ with $\reducek{k}{e}{e'}$.
The same definitions apply analogously to reductions in the target language.

\subsubsection{The Logical Relation}

\boxfig{f:reduce-rel-fg-tl-exp}{Relating \FGGm\ to \TL\ expressions}{
  \sTransSection{
    \ruleform{\LREquiv{e}{\expT}{\fgType}{k}}
  }{
    Expressions
  }
  \begin{mathpar}
    \inferrule[equiv-exp]{
      {
        \ba{c}
        (
          \forall k' < k,
          v ~.~ \reduceFGk{e}{\kA}{v} \implies \exists \uT.
          \reduceTLN{\vbMethTL}{\expT}{\uT} \wedge \LREquivVal{v}{\uT}{\fgType}{k - \kA}
        )
        \\
        (
          \forall k' < k,
          e' ~.~ \reduceFGk{e}{k'}{e'} \wedge \Diverge{e'}
          \implies
          \Diverge{E}
        )
        \ea
      }
    }{
      \LREquiv{e}{\expT}{\fgType}{k}
    }
  \end{mathpar}

  \sTransSection{
    \ruleform{\LREquivVal{v}{\uT}{\fgType}{k}}
  }{Values}
  \begin{mathpar}
    \inferrule[equiv-struct]{
      \TYPE\ t_S[\Multi{\fgTyVar\,\fgType_I}] \ \STRUCT\ \{ \Foreach{f \ \fgTypeAux}{n} \} \in  \ForeachN{D} \\
      \forall i \in [n] . \LREquivVal{v_i}{V_i}{\MultiSubst{\fgTyVar}{\fgType} \fgTypeAux_i}{k}
    }{
      \LREquivVal{t_S[\Multi{\fgType}] \{ \Foreach{v}{n} \}}{\Tuple{\Foreach{V}{n}}}{t_S[\Multi{\fgType}]}{k}
    }

    \inferrule[equiv-iface]{
      \exists \fgTypeAux_S . \forall k_1 < k . \LREquivVal{v}{U}{\fgTypeAux_S}{k_1} \\
      \methodSpecifications{\fgType_I} = \{ \Foreach{R}{n} \} \\
      \forall i \in [n], k_2 < k \,.\,
      \LREquiv{\methodLookup{\methodName{R_i}}{\fgTypeAux_S}}{\uT_i}{R_i}{k_2}
    }
    { \LREquivVal{v}{\Pair{U}{\Tuple{\Foreach{\uT}{n}}}}{\fgType_I}{k}
    }
  \end{mathpar}
  \sTransSection{
    \ruleform{\methodLookup{m}{\fgType_S} = \MLookupRes{x}{\fgType_S}{R}{e}}
  }{
    Method lookup
  }
  \begin{mathpar}
    \inferrule[method-lookup]{
      \FUNC\ (x \ t_S[\Multi{\fgTyVar\,\fgType_I}]) \ R \ \{ \RETURN\ e \} \in  \ForeachN{D}\\
      m = \methodName{R}\\
      \fgTySubst = \MultiSubst{\fgTyVar}{\fgType}
    }{
      \methodLookup{m}{t_S[\Multi{\fgType}]} = \MLookupRes{x}{t_S[\Multi{\fgType}]}{\fgTySubst R}{\fgTySubst e}
    }
  \end{mathpar}

  \sTransSection{
    \ruleform{
      \LREquiv{\MLookupRes{x}{\fgType_S}{R}{e}}{(\tllambda{X} E)}{R}{k}
    }
  }{
    Method dictionary entries
  }
  \begin{mathpar}
    \inferrule[equiv-method-dict-entry]{
      \forall \kA \leq k, \Multi[p]{\fgType}, W, v, V, \Multi[n]{v}, \Multi[n]{V}.\\\\
      (\fgTySubst = \MultiSubst[p]{\fgTyVar}{\fgType} \wedge
      \LREquiv{\Multi[p]{\fgType}}{W}{\Multi[p]{\fgTyVar\,\fgType_I}}{\kA} \wedge
      \LREquiv{v}{V}{\fgType_S}{\kA}
      \wedge (\forall i \in [n]. \LREquiv{v_i}{\uT_i}{\fgTySubst \fgTypeAux_i}{\kA})) \\\\
      \implies \LREquiv{
        \Angle{\subst{x}{v},\Multi[n]{\subst{x}{v}}} \fgTySubst e
      }{
        (\tllambda{X} E)~\Triple{V}{W}{\Tuple{\Multi[n]{V}}}
      }{\fgTySubst \fgTypeAux}{\kA}
    }
    {
      \LREquiv{
        \MLookupRes{x}{\fgType_S}{
          m[\Multi[p]{\fgTyVar\,\fgType_I}](\Foreach{x \ \fgTypeAux}{n}) \, \fgTypeAux
        }{e}
      }{(\tllambda{X} E)}{
        m[\Multi[p]{\fgTyVar\,\fgType_I}](\Foreach{x \ \fgTypeAux}{n}) \, \fgTypeAux
      }{k}
    }
  \end{mathpar}

  \sTransSection{
    \ruleform{
      \LREquiv{\Multi{\fgTypeAux}}{V}{\Multi{\fgTyVar\,\fgType_I}}{k}
    }
  }{
    Bounded type parameters
  }
  \begin{mathpar}
    \inferrule[equiv-bounded-typarams]{
      \fgTySubst = \Angle{\Multi[n]{\subst{\fgTyVar}{\fgTypeAux}}}\\
      \forall k' \leq k, i \in [n], u_i, U_i \,.\,
      \LREquiv{u_i}{U_i}{\fgTypeAux_i}{k'} \implies
      \LREquiv{u_i}{V_i\ U_i}{\fgTySubst \fgType_{Ii}}{k'}
    }{
      \LREquiv{\Multi[n]{\fgTypeAux}}{\Tuple{\Multi[n]{V}}}{\Multi[n]{\fgTyVar\ \fgType_I}}{k}
    }
  \end{mathpar}
}

The definition of the logical relation spreads over two figures~\ref{f:reduce-rel-fg-tl-exp}
and~\ref{f:reduce-rel-fg-tl-decl}.
In these figures, we assume that the declarations $\fgDecls$ of the \FGGm\ program being translated
are implicitly available in all rules. Also, we assume that an arbitrary but fixed method
substitution $\tlMethTable$ is implicitly available to all rules. This $\tlMethTable$ is used
in the reduction rules of the target language to resolve let-bound variables (i.e. translations of methods).
In our main theorem (\Cref{thm:prog-equiv}), we will then require that $\tlMethTable$ results from
translating the methods in $\fgDecls$.

We now
explain the logical relation on expressions, see \Cref{f:reduce-rel-fg-tl-exp}.
The relation $\LREquiv{e}{\expT}{\fgType}{k}$ denotes that \FGGm\ expression $e$
and TL expression $E$ are equivalent at type $\fgType$ for at most $k$ reduction
steps. We call $k$ the \emph{step index}.
Rule \Rule{equiv-exp} has two implications as its premises. The first states that if $e$ reduces to a value $v$
in $k' < k$ steps, then $E$ reduces to some value $V$ in an arbitrary number
of steps and $v$ is equivalent to $V$ at type $\fgType$ for the remaining $k - k'$ steps.
The second premise is for diverging expressions: if $e$ reduces in less than $k$ steps to
some expression $e'$ and $e'$ diverges, then $E$ diverges as well.

The relation $\LREquivVal{v}{\uT}{\fgType}{k}$ defines equivalence of \FGGm\ value $v$ and
TL value $V$ at type $\fgType$ with step index $k$.
Rule \Rule{equiv-struct} handles the case where $\fgType$ is a struct type. Then
$v$ must be a value of this struct type and $V$ must be a struct value
such that all field values of $v$ and $V$ are equivalent.
Rule \Rule{equiv-iface} deals with the case that $\fgType$ is an interface type.
Hence, $V$ must be an interface value $\Pair{U}{\Tuple{\Multi{V}}}$
with two requirements. First, $v$ and $U$ are equivalent for all step indices $k_1 < k$ at some struct type
$\fgTypeAux_S$. Second, $\Tuple{\Multi{V}}$ must be
an appropriate dictionary for the methods of the interface
with receiver type $\fgTypeAux_S$.
To check this requirement, rule \Rule{method-lookup} defines the auxiliary
$\methodLookup{m_i}{\fgTypeAux_S}$ to
retrieve a quadruple $\MLookupRes{x}{\fgTypeAux_S}{R}{e}$ from the declaration
of $m_i$ for $\fgTypeAux_S$. This quadruple has to be equivalent to dictionary entry $V_i$ for all step indices $k_2 < k$
at the signature of the method.

A dictionary entry is always a function value. We write
$\LREquiv{\MLookupRes{x}{\fgType_S}{R}{e}}{(\tllambda{X}{E})}{R}{k}$
to denote equivalence between a quadruple for a method declaration and
some dictionary entry $\tllambda{X} E$.
Rule \Rule{equiv-method-dict-entry}
defines this equivalence such that method body $e$ and $\tllambda{X}E$ take related
arguments to related outputs. Thus, the premise of the rule requires for all step indices $k' \leq k$,
all related type parameters $\Multi{\fgType}$ and $W$, all related receiver values $v$ and $V$,
and all related arguments $\Multi{v}$ and $\Multi{V}$ that $e$ and $\tllambda{X}E$ yield related
results when applied to the respective arguments.

The judgment $\LREquiv{\Multi{\fgTypeAux}}{V}{\Multi{\fgTyVar\,\fgType_I}}{k}$ denotes equivalence between
concrete type arguments
$\Multi{\fgTypeAux}$ and their TL realization $V$ when checking the bounds of type parameters $\Multi{\fgTyVar\,\fgType_I}$.
The definition in rule \Rule{equiv-bounded-typarams} relies on our translation strategy that
bounded type parameters are represented by coercions.

Having explained all judgments from \Cref{f:reduce-rel-fg-tl-exp}, we verify that the recursive definitions
of $\LREquiv{e}{\expT}{\fgType}{k}$ and $\LREquivVal{v}{\uT}{\fgType}{k}$ are well-founded.
Often, logical relations are defined by induction on the structure of
types. In our case, this approach does not work because interface types in \FGGm\ might be recursive,
see our previous work \citep{SulzmannWehr-mpc2022} for an example.
Thus, we use the step index as part of a decreasing measure $\Measure$.
Writing $\Size{V}$ for the size of some target value $V$, we define
$\Measure(\LREquiv{e}{\expT}{\fgType}{k}) = (k, 1, 0)$ and
$\Measure(\LREquivVal{v}{\uT}{\fgType}{k}) = (k, 0, \Size{V})$.
In \Rule{equiv-exp}, either $k$ decreases or stays constant but the second component of $\Measure$ decreases.
In \Rule{equiv-struct}, $k$ and the second component
stay constant but $\Size{V}$ decreases, and
in \Rule{equiv-iface} together with \Rule{equiv-method-dict-entry} and \Rule{equiv-bounded-typarams} step index $k$
decreases. Note that \Rule{equiv-method-dict-entry} and \Rule{equiv-bounded-typarams} only require $k' \leq k$.
This is ok because we already have $k_2 < k$ in \Rule{equiv-iface}.

\boxfig{f:reduce-rel-fg-tl-decl}{Relating \FGGm\ to \TL\ substitutions and declarations}{
  \sTransSection{
    \ruleform{
      \LREquiv{\fgTySubst}{\vbTL}{\fgTyEnv}{k}\qquad
      \LREquiv{\vbFG}{\vbTL}{\fgEnv}{k}
    }
  }{Substitutions}
  \begin{mathpar}
    \inferrule[equiv-ty-subst]{
      \LREquiv{
        \Multi{\fgTySubst \fgTyVar_i}
      }{
        \Tuple{\Multi{\tlSubst \xT_{\fgTyVar_i}}}
      }{\Multi{\fgTyVar\ \fgType}}{k}
    }
    {
      \LREquiv{
        \fgTySubst
      }{
        \tlSubst
      }{\{\Multi{\fgTyVar : \fgType}\}}{k}
    }

    \inferrule[equiv-val-subst]
    { \forall (x : \fgType) \in \fgEnv .\
      \LREquiv{\vbFG(x)}{\vbTL(X)}{\fgType}{k}
    }
    {
      \LREquiv{\vbFG}{\vbTL}{\fgEnv}{k}
    }
  \end{mathpar}

  \sTransSection{
    \ruleform{
      \LREquivNoTy{\FUNC\ (x \ t_S[\Multi{\fgTyVar\,\fgType_I}]) \ R \ \{ \RETURN\ e \}}{k}{X}
    }
  }{
    Method declarations
  }
  \begin{mathpar}
    \inferrule[equiv-method-decl]{
      \forall k' < k, \Multi[p]{\fgType}, \Multi[q]{\fgType'}, \Multi[p]{W},
      \Multi[q]{W'}, v, V, \Multi[n]{v}, \Multi[n]{V} .\\
      \fgTySubst = \Angle{\Multi[p]{\subst{\fgTyVar}{\fgType}}, \Multi[q]{\subst{\fgTyVar'}{\fgType'}}} \wedge
      \LREquiv{
        \Multi[p]{\fgType} \Multi[q]{\fgType'}
      }{
        \Tuple{\Multi[p]{W}, \Multi[q]{W'}}
      }{\Multi[p]{\fgTyVar\,\fgType_I}, \Multi[q]{\fgTyVar'\,\fgType_I'}}{k'}
      \wedge\\
      \LREquiv{v}{V}{t_S[\fgTySubst \Multi[p]{\fgTyVar}]}{k'} \wedge
      (\forall i \in [n] . \LREquiv{v_i}{V_i}{\fgTySubst \fgTypeAux_i}{k'})
      \implies\\
      \LREquiv{
        \Angle{\subst{x}{v}, \Multi[n]{\subst{x}{v}}} \fgTySubst e
      }{
        X~\Quadr{\Tuple{\Multi[p]{W}}}{V}{\Tuple{\Multi[q]{W'}}}{\Tuple{\Multi[n]{V}}}
      }{
        \fgTySubst \fgTypeAux
      }{k'}
    }{
      \LREquivNoTy{\FUNC\ (x \ t_S[\Multi[p]{\fgTyVar\,\fgType_I}]) \
        m[\Multi[q]{\fgTyVar'\,\fgType_I'}](\Multi[n]{x\ \fgTypeAux})\, \fgTypeAux \
        \{ \RETURN\ e \}}{k}{X}
    }
  \end{mathpar}

  \sTransSection{
    \ruleform{\LREquivNoTy{\Multi{D}}{k}{\tlMethTable}}
  }{
    Programs
  }
  \begin{mathpar}
    \inferrule[equiv-decls]{
      \Multi{D}, \tlMethTable \textrm{~are implicitly available in all subderivations}\\
      \forall D_i \in \Multi{D} .
      D_i = \FUNC\ (x \ t_S[\Multi{\fgTyVar\,\fgType}]) \ m M \ \{ \RETURN\ e \}
      \implies \LREquivNoTy{D_i}{k}{\mT{m}{t_S}}
    }{
      \LREquivNoTy{\Multi{D}}{k}{\tlMethTable}
    }

  \end{mathpar}
}

\Cref{f:reduce-rel-fg-tl-decl} extends the logical relation to whole programs.
Judgment $\LREquiv{\fgTySubst}{\vbTL}{\fgTyEnv}{k}$ denotes how a \FGGm\ type
substitution $\fgTySubst$ intended to substitute the type variables from
$\fgTyEnv$ is related to a TL substitution $\vbTL$.
The definition in rule \Rule{equiv-ty-subst} falls back to equivalence of type parameters.
Judgment $\LREquiv{\vbFG}{\vbTL}{\fgEnv}{k}$ similarly relates a \FGGm\ value substitution $\fgSubst$
intended for value environment $\fgEnv$ with a TL substitution $\vbTL$. See rule
\Rule{equiv-val-subst}.

Judgment $\LREquivNoTy{\FUNC\ (x \ t_S[\Multi{\fgTyVar\,\fgType_I}]) \ R \ \{ \RETURN\ e \}}{k}{X}$
states equivalence of a function declaration with a TL variable $X$.
Rule \Rule{equiv-method-decl} takes an approach similar as in rule \Rule{equiv-method-dict-entry}:
method body $e$ and variable $X$ must yield related outputs when applied to
related arguments. Thus, for all related type arguments $\Multi{\fgType}$, $\Multi{\fgType'}$ and
$\Pair{\Multi{W}}{\Multi{W'}}$, all related receiver values $v$ and $V$, and
all related arguments $\Multi{v}$ and $\Multi{V}$, the expression $e$ and variable
$X$ must be related when applied to the appropriate arguments.
However, different than in \Rule{equiv-method-dict-entry}, we only requires
this to hold for all $k' < k$.

Judgment $\LREquivNoTy{\Multi{D}}{k}{\tlMethTable}$ defines equivalence between
\FGGm\ declarations $\fgDecls$ and TL method substitution $\tlMethTable$. The definition in
rule \Rule{equiv-decls} is straightforward: each method declaration for some
method $m$ and struct $t_S$ must be equivalent to variable $\mT{m}{t_S}$.

\subsubsection{Equivalence Between Source and Translation}

To establish the desired result of semantic equivalence between a source program and one of
its translations,
we implicitly make the following assumptions about the globally available declarations
$\fgDecls$ and method substitution $\tlMethTable$.

\begin{assumption}\label{conv:fg-decls}
  We assume that the globally available declarations $\fgDecls$ are well-formed; that is,
  $\fgDeclOk{D_i}$ for all $D_i \in \Multi{D}$
  and $\tdMethTrans{D_i'}{X_i = V_i}$ for some $X_i$ and $V_i$ and all $D_i' = \FUNC\,\ldots \in \Multi{D}$.
  Further, we assume that the globally available method substitution $\vbMethTL$
  has only variables of the form $\mT{m}{t_S}$ in its domain.
\end{assumption}

Several basic properties hold for our logical relation. For example,
monotonicity gives us that with $\LREquiv{e}{E}{\fgType}{k}$ and $k' \leq k$
we also have $\LREquiv{e}{E}{\fgType}{k'}$.
Another property is how target and source reductions preserve equivalence:
\begin{lemma}[Target reductions preserve equivalence]\label{lem:target-reduce}
  If $\LREquiv{e}{E}{\fgType}{k}$ and $\reduceStar{E_2}{E}$ then
  $\LREquiv{e}{E_2}{\fgType}{k}$.
\end{lemma}
\begin{lemma}[Source reductions preserve equivalence]\label{lem:source-reduce}
  If $\LREquiv{e}{E}{\fgType}{k}$ and $\reduce{e_2}{e}$ then
  $\LREquiv{e_2}{E}{\fgType}{k+1}$.
\end{lemma}
The lemmas for monotonicity and several other properties
are stated in \Cref{sec:pres-dynam-semant}, together with all proofs.
We can then establish that an \FGGm\ expression $e$ is semantically equivalent
to its translation $E$.

\begin{lemma}[Expression equivalence]\label{lem:exp-equiv}
  Assume $\LREquivNoTy{\fgDecls}{k}{\tlMethTable}$ and
  $\LREquiv{\fgTySubst}{\tlSubst}{\fgTyEnv}{k}$ and
  $\LREquiv{\fgSubst}{\tlSubst}{\fgTySubst \fgEnv}{k}$.
  If $\tdExpTrans{\pair{\fgTyEnv}{\fgEnv}}{e : \fgType}{E}$
  then $\LREquiv{\fgSubst \fgTySubst e}{\tlSubst E}{\fgTySubst \fgType}{k}$.
\end{lemma}
The proof is by induction on the derivation of
$\tdExpTrans{\pair{\fgTyEnv}{\fgEnv}}{e : \fgType}{E}$, see \AppRef{sec:proof-exp-equiv}
for the full proof.
We next establish semantic equivalence for method
declarations.

\begin{lemma}[Method equivalence]\label{lem:method-equiv}
  Let $\fgDecls$ and $\vbMethTL$ such that for each
  $D = \FUNC\ (x\, t_S[\Multi{\fgTyVar\,\fgType_I}])\,R\,\{\RETURN\, e\} \in \fgDecls$ with $m = \methodName{R}$ we have
  $\tdMethTrans{D}{\mT{m}{t_S} = V}$ and
  $\vbMethTL(\mT{m}{t_S}) = V$ for some $V$.
  Then $\LREquivNoTy{\fgDecls}{k}{\vbMethTL}$ for any $k$.
\end{lemma}

The proof of this lemma is by induction on $k$, see \AppRef{sec:lem:method-equiv} for the full proof.
Finally, the following theorem states our desired result:
semantic equivalence between an \FGGm\ program and its translation.

\begin{theorem}[Program equivalence]\label{thm:prog-equiv}
  Let $\tdProgTrans{\ForeachN{D} \ \FUNC\ \MAIN () \{ \_ = e \}} {\LET\ \Multi{X_i = V_i} \ \IN\ E}$
  with $e$ having type $\fgType$.
  Let $\vbMethTL = \Angle{\Multi{\subst{X_i}{V_i}}}$.
  Then both of the following holds:
  \begin{enumerate}
  \item If $\reduceStar{e}{v}$ for some value $v$ then there exists
    a target language value $V$ such that
    $\reduceTLN{\vbMethTL}{E}{V}$ and
    $\LREquivVal{v}{V}{\fgType}{k}$ for any $k$.
  \item If $e$ diverges then so does $E$.
  \end{enumerate}
\end{theorem}

The \enquote{with $e$ having type $\fgType$} part means that the last rule in the derivation
of the program translation has $\tdExpTrans{\pair{\EmptyFgEnv}{\EmptyFgEnv}}{e : \fgType}{\expT}$
as a premise.
Obviously, $\fgDecls$ and $\tlMethTable$
meet the requirements of Assumption~\ref{conv:fg-decls}.
The theorem then follows from \Cref{lem:exp-equiv} and
\Cref{lem:method-equiv}. See \AppRef{sec:proof-crefthm:pr-equ}
for the full proof.

\subsubsection{Equivalence Between Different Translations}
\label{sec:equiv-betw-diff}

Our translation is non-deterministic because different translations of the same
expression may contain distinct sequences of applications of the subsumption rule
\Rule{sub}.
Recall the example from \Cref{f:fgg-format}. There are (at least) two different ways
to translate expression \go{Num\{1\}} at type \go{Format}.

\begin{enumerate}
\item Use rules \Rule{coerce-struct-iface} and \Rule{sub} to go directly from \go{Num} to
  supertype \go{Format}. The translation is then
  $\mathtt{toFormat}_{\sub{Num}}~\mathtt{1}$.
\item First use \Rule{coerce-struct-iface} and \Rule{sub} to go from \go{Num} to \go{Pretty},
  then use \Rule{coerce-iface-iface} and \Rule{sub} to go from \go{Pretty} to \go{Format}.
  The translation is then
  $\mathtt{toFormat}_{\sub{Pretty}}~(\mathtt{toPretty}_{\sub{Num}} \mathtt{1})$.
\end{enumerate}

Each choice leads to a syntactically distinct target expression. In general,
evaluating the target expressions might lead to syntactically different target values
because target values might contain dictionaries (i.e. tuple of $\lambda$-expressions),
and different translations might produce syntactically different dictionaries.

Another source of non-determinism is that rule \Rule{prog} for typing programs is allowed to
choose the type $\fgType$ of the main expression. For example, instead of typing
\go{Num\{1\}} at type \go{Format}, another translation might pick type \go{Pretty} or \go{Num}
for the main expression. The choice for the type of the main expression might also lead to
syntactically different target expressions.

To summarize, different translations of the same source program might lead
to syntactically different target language programs, and the syntactic differences
might persist in the final values after evaluation.
But a slightly weaker property holds: if we remove all dictionaries in the
final values, then the results are syntactically equal.

\boxfig{f:erase}{Erasure of dictionaries}{
  \sTransSection{
    \ruleform{
      \EraseTy{\fgType}{V} = V \quad
      \Erase{V} = V\quad
    }
  }{
    Erasure of dictionaries
  }
  \begin{minipage}[t]{0.5\textwidth}
    \begin{align*}
      \EraseTy{\fgType_S}{V} &= \Erase{V} \\
      \EraseTy{\fgType_I}{\Pair{V}{U}} &= \Erase{V}
    \end{align*}
  \end{minipage}
  \begin{minipage}[t]{0.5\textwidth}
    \begin{align*}
      \Erase{K\ \Multi{V}} &= K\ \Multi{\Erase{V_i}}                    \\
      \Erase{\tllambda{X}E} &= \EraseLam
    \end{align*}
  \end{minipage}
}

\Cref{f:erase} defines a function $\EraseSym$ that removes all dictionaries from
a target-language value. The function comes in two variations:
\begin{itemize}
\item $\EraseTy{\fgType}{V}$
  removes all dictionaries from value $V$ when viewed at type $\fgType$. Its duty
  is to remove the topmost dictionary if $\fgType$ is an interface type.
  In this case, the function is partial but this is not an issue: a value viewed at an interface type
  is always a pair of values.
\item $\Erase{V}$ removes all dictionaries from $V$ by replacing the $\lambda$-expressions
  with a fixed, otherwise unused, nullary constructor $\EraseLam$. This definition relies
  on then fact that the translation only produces $\lambda$-expressions for dictionary
  entries. We replace $\lambda$-expressions with a dedicated constructor $\EraseLam$ instead
  of the nullary tuple $\Tuple{}$ so as to avoid confusion between an erased $\lambda$ and
  a struct value without fields.
\end{itemize}

The following theorem states that evaluating the outcomes of two translations of the same source
program yields values that are identical up to removal of dictionaries (or both diverge).
This holds even if the two translations assign different types to the main expression of
the source program.
That is, there are no semantic ambiguities and we can establish that our translation is
coherent~\citep{conf/tacs/Reynolds91}.

\begin{theorem}[Coherence]\label{thm:determ-result}
  Let $P = \ForeachN{D} \ \FUNC\ \MAIN () \{ \_ = e \}$.
  Assume $\tdProgTrans{P}{\LET\ \Multi{X_i = V_i} \ \IN\ E}$ with $e$ having type $\fgType$ and
  $\tdProgTrans{P}{\LET\ \Multi{X_i' = V_i'} \ \IN\ E'}$ with $e$ having type $\fgType'$.
  Define $\vbMethTL = \Angle{\Multi{\subst{X_i}{V_i}}}$
  and $\vbMethTL' = \Angle{\Multi{\subst{X_i'}{V_i'}}}$.
  Then both of the following holds:
  \begin{enumerate}
  \item If $\reduceTLN{\vbMethTL}{E}{V}$ for some $V$, then
    $\reduceTLN{\vbMethTL'}{E'}{V'}$ for some $V'$ with $\EraseTy{\fgType}{V} = \EraseTy{\fgType'}{V'}$.
  \item If $E$ diverges then so does $E'$.
  \end{enumerate}
\end{theorem}

See \AppRef{sec:proof-thm-determ-result} for the proof.

\subsection{Getting the step index right}
\label{sec:step-index}

The logical relation in Figures~\ref{f:reduce-rel-fg-tl-exp} and~\ref{f:reduce-rel-fg-tl-decl}
requires at some places the step index in the premise to be strictly
smaller than in the conclusion ($<$), other places require only less-than-or-equal ($\leq$).
In \Rule{equiv-exp}, we have $<$ to keep the definition of the LR well-founded.
The $<$ in rule \Rule{equiv-method-decl} is required for the inductive argument in the proof of
\Cref{lem:method-equiv}.
Rule \Rule{equiv-iface} also has $<$, but rule
\Rule{equiv-method-dict-entry} only requires $\leq$. For well-foundedness, it is crucial
that one of these two rules decreases the step index. However, equally important is that the step index
is not forced to decrease more than once, so we need $<$ in one rule and $\leq$ in the other.
If both rules had $<$, then the proof of \Cref{lem:exp-equiv} would not go through for case \Rule{call-iface}.

\renewcommand\LabelQualifier{lr-design-choices}
Consider the following example in the context of \Cref{f:trans-example-src}:
\[
\begin{array}{r@{\quad}c@{\quad}l}
w_1 =  \Num\{1\} \textrm{~at type~} \Eq[\Num] &
\leadsto
&
W_1 = \Pair{\Tuple{1}}{\Tuple{U}}\\
&& \quad \textrm{where~} U = \tllambda{\TripleY} X_{\eq,\Num}\,\QuadrY{\Void}
\\
w_2 = \Num\{2\} \textrm{~at type~} \Num &
\leadsto
&
W_2 = \Tuple{\Tuple{2}}
\\
w_1.\eq(w_2)
&
\leadsto
&E = \CASE\,W_1\,\OF\,\Pair{Y}{\Tuple{X_1}} \to X_1\,\Triple{Y}{\Void}{\Tuple{W_2}}
\end{array}
\]
For values $w_1$ and $w_2$, we may assume \TextLabel{eq:equiv-val} $\LREquivVal{w_1}{W_1}{\Eq[\Num]}{k}$
and $\LREquiv{w_2}{W_2}{\Num}{k}$ for some $k$.
To verify that the translation yields related expressions,
we must show
\begin{igather}
  \LREquiv{w_1.\eq(w_2)}{E}{\bool}{k} \QLabel{eq:toshow}
\end{igather}
From \QRef{eq:equiv-val}, via inversion of rule \Rule{equiv-iface}, we can derive
\begin{igather}
  \LREquiv{\methodLookup{\eq}{\Num}}{U}{\eq(\that\,\Num)\,\bool}{k-1} \QLabel{eq:methodLookup}
\end{igather}
because the premise of the rule requires this to hold for all $k_2 < k$.
Let $e$ be the body of the method declaration of $\eq$ for $\Num$.
Inverting rule \Rule{equiv-method-dict-entry} for \QRef{eq:methodLookup} yields
\begin{igather}
\LREquiv{
  \Angle{\subst{\this}{w_1}, \subst{\that}{w_2}} e
}{
  U\,\Triple{\One}{\Void}{\Tuple{\Two}}
}{
  \bool
}{
  k'
}\QLabel{eq:equiv-W}
\end{igather}
for $k' = k - 1$ because rule \Rule{equiv-method-dict-entry} has $\leq$
in its premise. Also, we have
$\reducek{1}{w_1.\eq(w_2)}{\Angle{\subst{\this}{w_1}, \subst{\that}{w_2}}} e$ and
$\reduceStar{E}{U\,\Triple{\One}{\Void}{\Tuple{\Two}}}$.
Thus, with \QRef{eq:equiv-W}, \Cref{lem:target-reduce}, and \Cref{lem:source-reduce} we
get $\LREquiv{w_1.\eq(w_2)}{E}{\bool}{k' + 1}$. For $k' = k - 1$, this is exactly
\QRef{eq:toshow}, as required. But if rule \Rule{equiv-method-dict-entry}
required $<$ in its premise, then \QRef{eq:equiv-W} would only hold
for $k' = k - 2$ and we could not derive \QRef{eq:toshow}.

Whether we have $<$ in \Rule{equiv-iface} and $\leq$ in \Rule{equiv-method-dict-entry} or vice versa
is a matter of taste.
In our previous work at MPC \citep{SulzmannWehr-mpc2022}, we established a
dictionary-passing translation for Featherweight Go without generics. The situation
is slightly different there. With generics, we need two rules with respect to methods:
\Rule{equiv-method-decl} for method declarations and
\Rule{equiv-method-dict-entry} for dictionary entries where the coercions for the bounds of the
receiver's type parameters have already been supplied. Without generics, there are no type parameters,
so a single rule suffices (rule \Rule{red-rel-method} in MPC). So in the article at MPC,
we use $<$ for rule \Rule{red-rel-method} and $\leq$ for rule \Rule{red-rel-iface},
the pendant to rule \Rule{equiv-iface} of the current article.

\section{Implementation}
\label{sec:discussion}

We provide an implementation of the translation\footnote{\ImplementationURL}
written in Haskell~\citeyearpar{Haskell}. All examples in this article were
checked against the implementation.
Competitive runtime performance of the translated code was not our goal.
Hence, we took a convenient route and used
Racket~\citeyearpar{Racket} as the target language.
The implementation features all language concepts from \Cref{sec:featherweight-generic-go},
as well as type assertions, generic functions, and several base types (integers, characters, strings, and booleans).

Generic functions and base types are straightforward to support. Implementing the typing
and translation rules from \Cref{f:trans-exp1} requires some care because the presence of
subsumption rule \Rule{sub} renders the translation non-deterministic
(see \Cref{sec:equiv-betw-diff}). We solved this problem by \enquote{inlining}
the subsumption step when checking the arguments of a method call against the parameter types
(rules \Rule{call-struct} and \Rule{call-iface}) and when checking the field values of a struct
against the declared field types (rule \Rule{struct}). On formal grounds, this
is justified as Featherweight Go~\citeyearpar{FeatherweightGo} inlines the subsumption step in similar ways
and typing in \FGGm\ is equivalent the type system induced by our translation
rules (\Cref{thm:fgg-typing-equiv}).
The realization of type assertions (dynamic type casts) uses
type tags~\citep{conf/aplas/OhoriU21}. At runtime,
a type assertion $e.(\fgType)$ checks compatibility between $e$'s type tag and the type tag
corresponding to $\fgType$.

\boxfig{f:tests}{Summary of the test suite for the implementation}{
  \begin{center}
    \includegraphics[width=7cm]{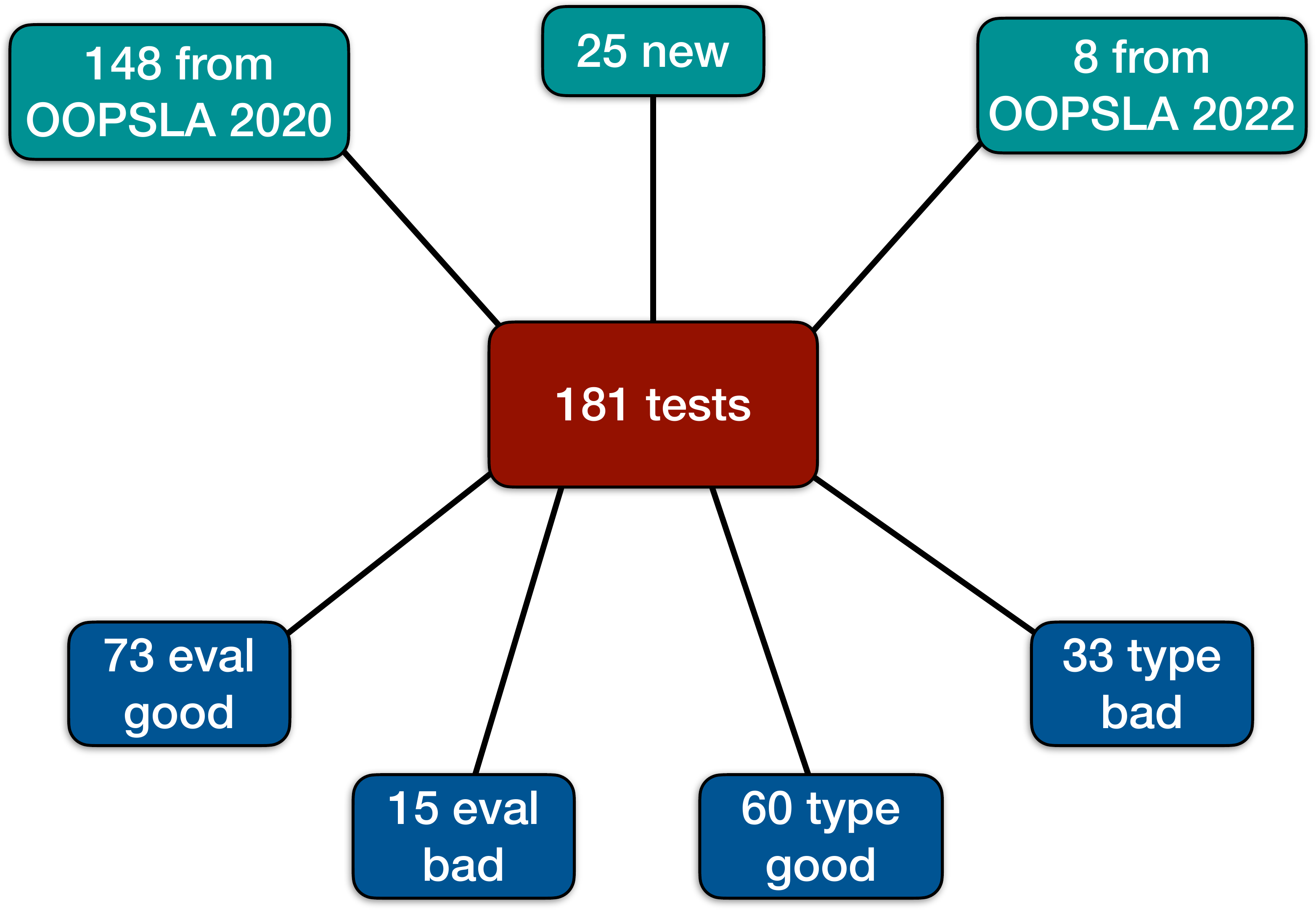}
  \end{center}
}

Our implementation comes with a large test suite and contains in total 181 tests,
covering all main features of the source language. See \Cref{f:tests} for a summary.
We wrote 25 new tests and included all 148 tests and examples from the
OOPSLA 2020 implementation\footnote{\cite{FeatherweightGo}, \url{https://github.com/rhu1/fgg}}.
Moreover, we also included 8 examples from the OOPSLA 2022
implementation\footnote{\cite{Ellis2022}, \url{https://github.com/sfzhu93/fgg2go}}.
(The implementation for OOPSLA 2022 builds on the one for OOPSLA 2020, adding 12 new test cases.
We ignored 4 test cases because they use concurrency features not supported by our
implementation.)
Most tests from OOPSLA 2020/2022 could be integrated in our test suite without changes,
for some we had to perform minor syntactic adjustments.

Each test is a source file in one of the following categories:
\newcommand\Category[1]{{\bfseries\sffamily #1}}
\begin{enumerate}
\item \Category{eval good}, 73 tests: The test type checks and evaluates successfully. For the test
  to succeed, the result of evaluation must match the expected result. We arrived at the expected
  result by inspecting the program and (if applicable) comparing it against the run of the
  OOPSLA 2020/2022 implementation.
\item \Category{eval bad}, 15 tests: The test type checks successfully but fails at runtime.
  For the test to succeed, the error message must match an expected string. We determined the expected
  string in similar way as for the \textsf{eval good} category.
\item \Category{type good}, 60 tests: The test type checks successfully but is not executed
  because it has no interesting operational behavior.
\item \Category{type bad}, 33 tests: The test fails to type check.
  For the test to succeed, the error message must match an expected string.
\end{enumerate}

Some of the tests behave differently under our implementation when compared with the original implementations
for OOPSLA 2020/2022:

\begin{itemize}
\item The OOPSLA 2020 implementation compiles generics by monomorphization;
  that is, generic code is specialized for all type arguments appearing in the program. But monomorphization
  cannot deal with all programs, so their type checker rejects several programs based on some syntactic
  condition (see \Cref{sec:generics-go} for details). Our implementation type checks these programs
  successfully.
\item The OOPSLA 2020 implementation statically rejects type assertions $e.(\fgType)$ where the type of $e$
  is a struct type, even though evaluation might succeed at runtime.  Our implementation is more liberal
  and only rejects type assertions statically that are guaranteed to fail at runtime.
\item The OOPSLA 2020 implementation rejects recursive definitions of structs. For simplicity, we
  omitted this check from our implementation.
\item The OOPSLA 2020 implementation runs several tests only for
  a fixed number of reduction steps because these tests would diverge otherwise. Our implementation only type checks
  such tests.
\end{itemize}

\section{Related Work}
\label{sec:related-work}

The related work section covers generics in Go, type classes in Haskell, logical relations,
and a summary of our own prior work.\footnote{%
  Several points in Sections~\ref{sec:generics-go}, \ref{sec:type-classes-haskell}, and~\ref{sec:logical-relations}
  were already included
  in own prior work \citep{SulzmannWehr-aplas2021,SulzmannWehr-mpc2022}.}
At the end, we give an overview of the existing
translations with source language Featherweight Generic Go.

\subsection{Generics in Go}
\label{sec:generics-go}
The results of this work rest on the definition of Featherweight Generic Go (FGG)
provided by Griesemer and colleagues~\citeyearpar{FeatherweightGo}.
FGG is a minimal core calculus modeling the
essential features of the programming language Go \citeyearpar{golang}.
It includes support for overloaded methods, interface types, structural subtyping,
generics, and type assertions.
Our formalization of FGG ignores dynamic type assertions but otherwise sticks to
the original definition of FGG, apart from some minor cosmetic changes in
presentation. We prove that the type system implied by our
translation is equivalent to the original type system
of FGG, and that translated programs behave the same way as under the original
dynamic semantics.

The original dynamic semantics of FGG uses
runtime method lookup, in the same way as we did
in \Cref{sec:featherweight-generic-go}. The authors define an alternative
semantics via monomorphisation; that is, they specialize generic code for
all type arguments appearing in the program.
This alternative semantics is equivalent
to the one based on runtime method lookup, but there exist type-correct FGG programs
that cannot be monomorphized. For instance, polymorphic recursion leads to infinitely many
type instantiations, so programs with a polymorphic-recursive method cannot be monomorphized.\footnote{See
  \cite{FeatherweightGo}, Figure~10 for an example.}
Further, monomorphization
often leads to a blowup in code size. In contrast, our translation handles all
type-correct FGG programs, and instantiations of generic code with different
type arguments do not increase the code size. However, we expect that monomorphized code
will offer better performance than code generated by our dictionary-passing translation,
because method dictionaries imply several indirections not present in
monomorphized programs.

The current implementation of
generics~\citep{GoGenerics2021} in Go versions 1.18, 1.19, and 1.20~\citeyearpar{golang}
differs significantly from the formalization in FGG.
For example, full Go requires a method declaration for a generic struct to have
exactly the same type bounds as the struct.
In FGG, bounds of the receiver struct in a method declaration
might be stricter than the bounds in
the corresponding struct declaration. In ~\Cref{f:fgg-format2}, we used this feature
to implement formatting for the fully generic \go{Pair} type, provided the
type parameters can be formatted as well. Go cannot express this
scenario without falling back to dynamic type assertions.

\cite{Ellis2022} formalize a dictionary-passing translation from a restricted
subset of FGG to FG. The restriction for FGG is the same as previously explained for
full Go: a method declaration must have the same type bounds as its receiver struct.
The translation utilizes this restriction to translate an FGG struct together with all its methods
into a single FG struct (dictionary). This approach would scale to full Go even with
separate compilation because a struct and all its methods must be
part of the same package. Further, the translation of Ellis and coworkers
replaces all types in method signatures
with the top-type \go{Any}, relying on dynamic type assertions to enable
type checking of the resulting FG program. The authors provide a working implementation
and a benchmark suite to compare their translation against several other approaches,
including the current implementation of generics in full Go.
Our translation targets an extended $\lambda$-calculus and
does not restrict the type bounds of the receiver struct in a method declaration.
We also provide an implementation but no evaluation of its performance.

Method dictionaries bear some resemblance to virtual method tables
(vtables) used to implement virtual method dispatch in
object-oriented languages~\citep{DBLP:conf/oopsla/DriesenH96}.
The main difference between vtables and
dictionaries is that there is a fixed connection between an object and its
vtable (via the class of the object), whereas the connection between a
value and a dictionary may change at runtime, depending on the type
the value is used at. Dictionaries allow access to a method at a fixed
offset, whereas vtables in the presence of multiple inheritance require
a more sophisticated lookup algorithm~\citep{DBLP:conf/oopsla/AlpernCFGL01}.

Generics in class-based languages such as Java~\citep{10.1145/286942.286957,journals/toplas/IgarashiPW01},
and C$\sharp$~\citep{conf/pldi/KennedyS01,conf/ecoop/EmirKRY06}
do not require a dictionary-passing translation
because all methods are part of the virtual method table of an object.
In Go, however, methods are not necessarily attached to the receiver struct, so
additional evidence in form of dictionaries must be passed for such methods. Further,
subtyping in Java and C$\sharp$ is nominal, whereas Go has structural subtyping.

A possible optimization to the dictionary-passing
translation is selective code specialization~\citep{conf/pldi/DeanCG95}. With this
approach, the dictionary-passing translation generates code
that runs for all type arguments. In addition, specialized code is generated
for frequently used combinations of type arguments. This approach allows
to trade code size against runtime performance. The GHC compiler for Haskell
supports a \texttt{SPECIALIZE} pragma~\citep[][\S\,6.20.11.]{GHC941}
that allows developers to specialize a polymorphic function to a particular type.
The specialization also supports type-class dictionaries.

\subsection{Type Classes in Haskell}
\label{sec:type-classes-haskell}

The dictionary-passing translation is well-studied in the context of
Haskell type classes
\citep{Wadler:1989:MAP:75277.75283,Hall:1996:TCH:227699.227700}.
A type-class
constraint translates to an extra function parameter, constraint resolution
provides a dictionary with the methods of the type class for this parameter.
In FGG, structural subtyping relations imply coercions and bounded type parameters translate to coercion parameters.
An interface value pairs a struct value with a dictionary for the methods
of the interface.
Thus, interface values can be viewed as
representations of
existential types~\citep{DBLP:journals/toplas/MitchellP88,laeufer_1996,ThiemannWehr2008}.

Another important property in the type class context is coherence.
Bottu and coworkers~\citeyearpar{10.1145/3341695} make use of logical relations
to state equivalence among distinct target terms resulting
from the same source type class program.
In the setting of FGG-, we first establish semantic equivalence among source and target programs,
see \Cref{thm:prog-equiv}.
From this property, we can derive the coherence property (\Cref{thm:determ-result}) almost for free.
We believe it is worthwhile to establish a property similar to this theorem for type classes.
We could employ a simple denotational semantics
for source type class programs
similar as \cite{DBLP:conf/lfp/Thatte94} or \cite{10.1145/2633357.2633364},
which would then be related to target programs obtained via the dictionary-passing translation.

\Cref{sec:bound-type-param} demonstrated that type bounds on generic structs and interfaces
have no operational meaning.
This situation is similar to contexts of data type definitions in Haskell~2010~\citep{Haskell2010}.
A data type such as
\begin{lstlisting}[language=myhaskell]
  data Eq a => Set a = NilSet | ConsSet a (Set a)
\end{lstlisting}
may require the
context \texttt{Eq a}. However, an occurrence of type \texttt{Set a} does not imply that \texttt{Eq a} holds
but always requires the constraint to be justified elsewhere. The GHC manual states that
\enquote{this is widely considered a misfeature}~\cite[Section~6.4.2]{GHCUserGuide}.

\subsection{Logical Relations}
\label{sec:logical-relations}

Logical relations have a long tradition of proving properties of
typed programming languages.
Such properties include termination~\citep{DBLP:journals/jsyml/Tait67,journals/iandc/Statman85},
type safety~\citep{Skorstengaard}, and program equivalence~\citep[Chapters~6,~7]{ATAPL}.
A logical relation (LR) is often defined inductively, indexed by type.
If its definition is based on an operational semantics, the LR is called
syntactic \citep{conf/icalp/Pitts98,journals/entcs/CraryH07}.
With recursive types, a step-index \citep{journals/toplas/AppelM01,10.1007/11693024_6}
provides a decreasing measure
to keep the definition well-founded.
See \citet[Chapter~8]{books/daglib/0085577} and \citet{Skorstengaard} for introductions to the topic.

LRs are often used to relate two terms of the same language. For our translation, the two terms
are from different languages, related at a type from the source language.
\citet{conf/icfp/BentonH09} prove correctness of compiler transformations.
They use a step-indexed LR
to relate a denotational semantics of the $\lambda$-calculus with recursion to configurations of a SECD-machine.
\citet{conf/popl/HurD11} build on this idea to show equivalence between
an expressive source language (polymorphic $\lambda$-calculus with references, existentials, and recursive types)
and assembly language. Their biorthogonal, step-indexed Kripke LR does not directly relate the two languages
but relies on abstract language specifications.

Our setting is different in that we consider a source language with support for overloading.
Besides structured data and functions, we need to cover recursive interface values.
This leads to some challenges to get the step index right \citep{SulzmannWehr-mpc2022}.

Simulation or bisimulation (see e.g. \citealt{journals/jacm/SumiiP07}) is another common technique for showing
program equivalences. In our setting, using this technique amounts to proving that reduction and translation
commutes:
if source term $e$ reduces to $e'$ and translates to target term $E$, then $e'$ translates
to $E'$ such that $E$ reduces to $E''$ (potentially in several steps) with $E' = E''$.
One challenge is that the two target
terms $E'$ and $E''$ are not necessarily syntactically equal but only semantically.
In our setting, this might be the case if $E'$ and $E''$ contain coercions for structural subtyping.
Even if such coercions behave the same, their syntax might be different.
With LR, we abstract away certain details of single step reductions, as we only compare values, not intermediate
results. A downside of the LR is that getting the step index right is sometimes not trivial.

\citet{journals/pacmpl/Paraskevopoulou21a} combine simulation
and an untyped, step-indexed LR \citep{conf/popl/AcarAB08}
to relate the translation of a reduced expression (the $E'$ from the preceding paragraph)
with the reduction result of the translated expression (the $E''$). They use this technique
to prove correctness of CPS transformations using small-step and big-step operational
semantics. Resource invariants connect the number of steps
a term and its translation might take, allowing them to prove that
divergence and asymptotic runtime is preserved by the transformation. Our LR
does not support resource invariants but includes a case for divergence directly.

\subsection{Prior Work}

Our own work published at APLAS \citep{SulzmannWehr-aplas2021} and MPC
\citep{SulzmannWehr-mpc2022} laid the foundations for the dictionary-passing
translation and its correctness proof of the present article.
For the APLAS paper, we defined a dictionary-passing translation for Featherweight Go
\citep[FG,][]{FeatherweightGo}, the non-generic variant of FGG. That translation
is similar in spirit to the translation presented here, it supports type assertions but not
generics. The APLAS paper includes a proof for
the semantic equivalence between the source FG program and its translation. The result is,
however, somewhat limited as semantic equivalence only holds for terminating programs
whose translation is also known to terminate.

In the MPC paper, we addressed the aforementioned limitation by extending the proof
of semantic equivalence to all possible outcomes of an FG program: termination,
panic (failure of a dynamic type assertion), and divergence. The proof
uses a logical relation similar to the one used here, but without support for generics.
We have already shown more differences in \Cref{sec:step-index}.

\subsection{Summary of Translations}

\begin{figure*}
  \Fbox{
    \begin{minipage}{\textwidth}
      \begin{center}
        \begin{tikzpicture}
          \matrix [column sep=25mm, row sep=12mm] {
            &
            \node (pfg1) {$\PFG'$}; &
            \node (ptl1) {$\PTL'$}; \\
            \node (pfgg) {$\PFGGrestricted$}; &
            \node (pfg) {$\PFG$}; &
            \node (ptl) {$\PTL$}; \\
            &
            &
            \node (ptl2) {$\PTL''$};
            \\
          };
          \draw[->, thick] (pfg1) -- (ptl1) node[above,midway] {\DiagLabel{MPC 2022}};
          \draw[->, thick] (pfgg) -- (pfg) node[above,midway] {\DiagLabel{OOPSLA 2020}};
          \draw[->, thick] (pfg) -- (ptl) node[above,midway] {\DiagLabel{MPC 2022}};
          \draw[->, thick] (pfgg) -- (ptl2) node[midway,sloped,below] {\DiagLabel{this article}};
          \draw[->, thick] (pfgg) -- (pfg1) node[midway,sloped,above] {\DiagLabel{OOPSLA 2022}};
        \end{tikzpicture}
      \end{center}
    \end{minipage}
  }
  \caption{Summary of translations. Arrows represent translations,
    $P_{\ell}$ is a program in language $\ell$.
    Program $\PFGGrestricted$ is subject to certain restrictions, depending on the translation being performed.
  }
  \label{f:summary-diag}
\end{figure*}
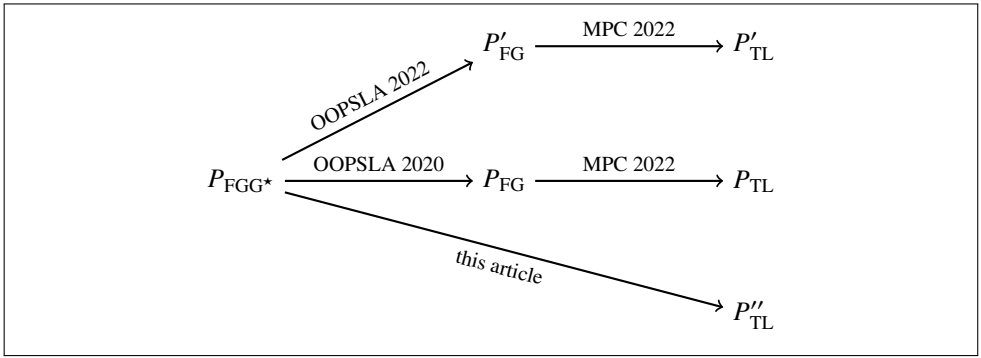

The diagram in \Cref{f:summary-diag} summarizes the existing translations
by \citet[OOPSLA][]{FeatherweightGo},
by \citet[OOPSLA][]{Ellis2022}, from our MPC 2022 paper
\citep{SulzmannWehr-mpc2022}, and from the article at hand.
The three resulting target language programs $\PTL$, $\PTL'$, and $\PTL''$ are
semantically equivalent because all translations preserve the dynamic semantics.
Each translation with $\PFGGrestricted$ as its source has different restrictions.
OOPSLA 2022 requires the receiver struct of some method declaration to have exactly the same
type bounds as the struct declaration itself.
OOPSLA 2020 requires $\PFGG$ to be monomorphizable, checked by a simple syntactic condition.
The translation of this article does not support type assertions.

\section{Conclusion and Future Work}
\label{sec:future-work-concl}

This article defined a type-directed dictionary-passing translation from Featherweight Generic Go (FGG)
without type assertions to an extension of the untyped $\lambda$-calculus. The translation represents a value
at the type of an interface as a combination of a concrete struct value
with a dictionary for all methods of the interface. Bounded type parameters become
extra function arguments in the target. These extra arguments are coercions from the
instantiation of a type variable to its upper bound.

Every program in the image of the translation has the same dynamic
semantics as its source program. Different translations of the same
source program may result in syntactically different but equivalent target programs.
The proof of semantic equivalence is based
on a syntactic, step-indexed logical relation. The step-index ensures a
well-founded definition of the relation in the presence of recursive
interface types and recursive methods. We also reported on an implementation
of the translation.

In this article, we relied on FGG as defined by Griesemer and
coworkers \citeyearpar{FeatherweightGo}, without reconsidering design
decisions. But our translation raises several questions with respect to the
design of generics in FGG and more generally also in Go.
For example, the translation clearly shows that type bounds
in structs and interfaces have no operational meaning. Should we eliminate these type bounds?
Or should we give them a meaning inspired by Haskell's type class mechanism?
Further, a method declaration in full Go must reuse the type bounds of its struct and must
be defined in the same package as the struct. Clearly, this limits
extensibility and flexibility. Can we provide a more flexible design to solve
the expression problem~\citep{ExpressionProblemWadler} in Go, without resorting to
unsafe type assertions?
We would like to use the insights gained through this article to answer these and similar questions in future
work.

A somewhat related point is performance. As explained earlier,
generics in Go are compiled by monomorphization. This gives the best possible performance because
the resulting code is specialized for each type argument. However, not all programs can be monomorphized
and the increase in code size is often considered problematic. This raises another interesting question for
future work. Could selective monomorphization or specialization offer a viable trade-off between
performance, code size, and the ability to compile Go programs which are not monomorphizable?

A statically-typed target language typically offers more room for compiler optimization~\citep{conf/popl/HarperM95}.
Thus, another
interesting direction for future work is a translation to a typed backend, for example
System F~\citep{girard-72,721503}.

The work presented here does not include type assertions (dynamic type casts), although
FGG supports them. We omitted type assertions from our theory for two reasons:
Firstly, type assertions are largely orthogonal to the dictionary-passing translation,
so their inclusion would obscure the working of the translation.
Secondly, type assertions would require some extra design choices to consider. In our implementation
we construct
the check for a type assertion at the same place as in the source program, relying
on dynamic type-tag--passing for gathering all information necessary.
But other approaches are possible. For example, one could construct downcast coercions at call-sites
and pass these coercions around.
The second option could make the treatment of
type assertions more lightweight, but would require significant
research in this direction.

\section*{Acknowledgements}

We would like to thank the anonymous reviewers for their valuable feedback
and constructive suggestions, which greatly improved the quality of this
paper.
We acknowledge support by the Open Access Publication Fund of the
Offenburg University of Applied Sciences, Germany.

\section*{Declaration of interests}
The authors report no conflict of interest.

\section*{Supplementary material}

For supplementary material for this article, please visit \ImplementationURL{}.


\bibliographystyle{jfplike}
\bibliography{main}

\renewcommand\thesection{\Alph{section}}
\appendix

\section{Proofs}
\label{sec:more-form-prop}

\subsection{Deterministic Evaluation in \FGGm\ and TL}
\label{sec:determ-eval-fgg}

\begin{lemma}[Deterministic evaluation in \FGGm]\label{lem:determ-eval-fgg}
  If $\reduce{e}{e'}$ and $\reduce{e}{e''}$ then $e' = e''$.
  If $\reduce{E}{E'}$ and $\reduce{E}{E''}$ then $E' = E''$.
\end{lemma}
\begin{proof}
  We first state and prove three sublemmas:
  \begin{EnumAlph}
  \item If $e = \EvCtx_1[\EvCtx_2[e']]$ then there exists $\EvCtx_3$ with $e = \EvCtx_3[e']$.
    The proof is by induction on $\EvCtx_1$.
  \item If $\reduce{e}{e'}$ then there exists a derivation of $\reduce{e}{e'}$ that ends with
    at most one consecutive application of rule \Rule{fg-context}. The proof is by induction
    on the derivation of $\reduce{e}{e'}$. From the IH, we know that this derivation ends with
    at most \emph{two} consecutive applications of rule \Rule{fg-context}. If there are
    two such consecutive applications, (a) allow us to merge the two
    evaluation contexts involved, so that we need only one consecutive application
    of \Rule{fg-context}.
  \item We call an \FGGm\ expression \emph{directly reducible} if it reduces but not by rule \Rule{fg-context}.
    If $e_1$ and $e_2$ are now directly reducible and $\EvCtx_1[e_1] = \EvCtx_2[e_2]$ then $\EvCtx_1 = \EvCtx_2$
    and $e_1 = e_2$. For the proof, we first note that $\EvCtx_1 = \Hole$ iff $\EvCtx_2 = \Hole$. This
    holds because directly reducible expressions have no inner redexes. The rest of the
    proof is then a straightforward induction on $\EvCtx_1$.
  \end{EnumAlph}
  Now assume $\reduce{e}{e'}$ and $\reduce{e}{e''}$. By (b) we may assume that both derivations
  ends with at most one consecutive application of rule \Rule{fg-context}. It is easy to see
  (as values do not reduce) that both derivations must end with the same rule. If this rule
  is \Rule{fg-field}, then $e' = e''$ by restrictions \FGGUniqueStructs{} and
  \FGGUniqueFields{}. If this rule is \Rule{fg-call}, then $e' = e''$ by
  \FGGUniqueReceiver{}.
  If the rule is \Rule{fg-context},
  we have the following situation with $R_1 \neq \Rule{fg-context}$ and
  $R_2 \neq \Rule{fg-context}$:
  \begin{mathpar}
    \inferrule*[left=fg-context]{
      \inferrule*[left=$R_1$]{
      }{
        \reduce{g_1}{g_1'}
      }
    }{
      \reduce{\BraceBelow{\EvCtx_1[g_1]}{= e}}{\BraceBelow{\EvCtx_1[g_1']}{= e'}}
    }

    \inferrule*[right=fg-context]{
      \inferrule*[right=$R_2$]{
      }{
        \reduce{g_2}{g_2'}
      }
    }{
      \reduce{\BraceBelow{\EvCtx_2[g_2]}{= e}}{\BraceBelow{\EvCtx_2[g_2']}{= e''}}
    }
  \end{mathpar}
  As neither $R_1$ nor $R_2$ are \Rule{fg-context}, we know that $g_1$ and $g_2$ are directly
  reducible. Thus, with $\EvCtx_1[g_1] = \EvCtx_2[g_2]$ and (c) we get $\EvCtx_1 = \EvCtx_2$ and
  $g_1 = g_2$. With $R_1$ and $R_2$ not being \Rule{fg-context}, we have $g_1' = g_2'$, so
  $e' = e''$ as required. \MyQED
\end{proof}

\begin{lemma}[Deterministic evaluation in TL]\label{lem:determ-eval-tl}
  If $\reduceExpTL{\tlMethTable}{E}{E'}$ and $\reduceExpTL{\tlMethTable}{E}{E''}$ then $E' = E''$.
  Further, if $\reduce{E}{E'}$ and $\reduce{E}{E''}$ then $E' = E''$.
\end{lemma}
\begin{proof}
  \renewcommand\LabelQualifier{lem:determ-eval-tl}
  We first prove the first implication of the lemma
  \begin{igather}
    \forall E, E', E'', \tlMethTable ~.~
    \reduceExpTL{\tlMethTable}{E}{E'} \wedge \reduceExpTL{\tlMethTable}{E}{E''} \implies E' = E''
    \QLabel{main}
  \end{igather}
  There are three sublemmas, analogously to the proof of \Cref{lem:determ-eval-fgg}.
  \begin{EnumAlph}
  \item If $E = \EvCtxT_1[\EvCtxT_2[E']]$ then there exists $\EvCtxT_3$ with $E = \EvCtxT_3[E']$.
  \item If $\reduceExpTL{\tlMethTable}{E}{E'}$ then there exists a derivation of
    $\reduceExpTL{\tlMethTable}{E}{E'}$ that ends with
    at most one consecutive application of rule \Rule{tl-context}.
  \item We call a target-language expression \emph{directly reducible} if it reduces but not by rule \Rule{tl-context}.
    If $E_1$ and $E_2$ are now directly reducible and $\EvCtxT_1[E_1] = \EvCtxT_2[E_2]$ then $\EvCtxT_1 = \EvCtxT_2$
    and $E_1 = E_2$.
  \end{EnumAlph}
  The proofs of these lemmas are similar to the proofs of the sublemmas in \Cref{lem:determ-eval-fgg}.
  Then \QRef{main}
  follows with reasoning similar to the proof of \Cref{lem:determ-eval-fgg}.
  If the derivations of
  $\reduceExpTL{\tlMethTable}{E}{E'}$ and $\reduceExpTL{\tlMethTable}{E}{E''}$ both end
  with rule \Rule{tl-case}, then our assumption that the constructors of a case-expression are
  distinct ensures determinacy.

  The second claim of the lemma
  ($\reduce{E}{E'}$ and $\reduce{E}{E''}$ imply $E' = E''$) then follows directly from \QRef{main}.
  Our assumption that the variables of a top-level let-binding are distinct
  ensures that the substitution $\tlMethTable$ built from the
  top-level let-bindings is well-defined.
  \MyQED
\end{proof}

\subsection{Preservation of Static Semantics}
\label{sec:pres-stat-semant}

\begin{proof}[Proof of \Cref{thm:fgg-typing-equiv}]
  We prove (a) and (b) by case distinctions on the last rule of the given derivations;
  (c) and (d) follow by induction on the derivations, using (a) and (b). Claim (e) then
  follows by examining the typing rules, using (c) and (d). \MyQED
\end{proof}

\subsection{Preservation of Dynamic Semantics}
\label{sec:pres-dynam-semant}

\begin{convention}
  We omit $\tlMethTable$ from reductions in the target language, writing
  $\reduce{E}{E'}$ instead of $\reduceExpTL{\tlMethTable}{E}{E'}$.
\end{convention}

\newcommand\fgMakeSubst[3]{\subst{#1}{#2} : #3}
\newcommand\fgTyFormals{\Phi}
\newcommand\fgTyFormalsAux{\Psi}
\newcommand\fgTyActuals{\phi}
\newcommand\fgTyActualsAux{\psi}

\makeatletter
\define@key{fgSyn}{recv}{\def\fgSyn@recv{#1}}
\define@key{fgSyn}{spec}{\def\fgSyn@spec{#1}}
\define@key{fgSyn}{body}{\def\fgSyn@body{#1}}
\newcommand{\fgFunc}[1][]{%
  \begingroup%
  \setkeys{fgSyn}{recv={x\ t_S[\fgTyFormals]}, spec={mM}, body=e}%
  \setkeys{fgSyn}{#1}%
  \FUNC\,(\fgSyn@recv)\,\fgSyn@spec\,\{ \RETURN\,\fgSyn@body \}%
  \endgroup%
}
\makeatother

\begin{definition}
  We make use of some extra metavariables and notations.
  \begin{itemize}[noitemsep,topsep=0pt,parsep=0pt,partopsep=0pt]
  \item $\fgTyFormals, \fgTyFormalsAux$ denote formal type parameters $\ForeachN{\fgTyVar\,\fgType_I}$.
  \item $\fgGetTyVars{\fgTyFormals}$ denotes the type variables of $\fgTyFormals$;
    that is, if $\fgTyFormals = \Multi{\fgTyVar\,\fgType_I}$ then $\fgGetTyVars{\fgTyFormals} = \Multi{\fgTyVar}$.
  \item $\fgTyActuals, \fgTyActualsAux$ denote actual type arguments $\ForeachN{\fgType}$.
  \item $M ::= [\fgTyFormals](\overline{x \ \fgType}) \ \fgType$ denotes the type-part of a method signature $R$.
  \item $L$ denotes a type literal
    $\STRUCT\ \{ \overline{f \ \fgType} \}$ or
    $\INTERFACE\ \{ \overline{R} \}$.
  \item $\fgMakeSubst{\fgTyFormals}{\fgTyActuals}{\fgTySubst}$ create a type substitution
  $\fgTySubst$ form type parameters $\fgTyFormals$ and arguments $\fgTyActuals$. It
  is defined like this:
  $\fgMakeSubst{\Foreach{\fgTyVar\, \fgType}{n}}{\Foreach{\fgTypeAux}{n}}{\Angle{\Foreach{\subst{\fgTyVar_i}{\fgTypeAux_i}}{n}}}$
\end{itemize}
\end{definition}

\subsubsection{The Logical Relation}

\begin{lemma}[Monotonicity for expressions]\label{lem:mono-exp}
  Assume $k' \leq k$.
  If $\LREquiv{e}{E}{\fgType}{k}$ then $\LREquiv{e}{E}{\fgType}{k'}$.
  If $\LREquivVal{v}{V}{\fgType}{k}$ then $\LREquivVal{v}{V}{\fgType}{k'}$.
\end{lemma}

\begin{proof}
  \renewcommand\LabelQualifier{lem:mono-exp}
  We proceed by induction on $(k, s)$ where $s$ is the combined size of $v, V$.
  \begin{CaseDistinction}{on the last rule used in the two derivations}
    \Case{rule \Rule{equiv-exp}}
    We label the two implications in the premise of the rule as (a) and (b).
    \begin{EnumAlph}
    \item Assume $k'' < k'$ and $\reducek{k''}{e}{u}$ for some value $u$.
      From $\LREquiv{e}{E}{\fgType}{k}$
      \begin{igather}
        \exists U \DOT \reduceStar{E}{U} \\
        \LREquivVal{u}{U}{\fgType}{k - k''} \QLabel{red-u}
      \end{igather}
      If $k = k'$ then $\LREquivVal{u}{U}{\fgType}{k'-k''}$. Otherwise, $k' - k'' < k - k''$, so the IH
      (induction hypothesis)
      applied to \QRef{red-u} also yields $\LREquivVal{u}{U}{\fgType}{k'-k''}$.
      This proves implication (a).
    \item Assume $k'' < k'$ and $\reducek{k''}{e}{e'}$ and $\Diverge{e'}$.
      Then we get with $\LREquiv{e}{E}{\fgType}{k}$ and $k' \leq k$ that
      $\Diverge{E}$.
    \end{EnumAlph}
    \Case{rule \Rule{equiv-struct}} Follows from IH.
    \Case{rule \Rule{equiv-iface}} Obvious.
  \end{CaseDistinction} \MyQED
\end{proof}

\begin{lemma}[Monotonicity for method dictionaries]\label{lem:mono-dict}
  If $\LREquiv{\Angle{x, \fgType_S, R, e}}{V}{R'}{k}$ and $k' \leq k$ then
  $\LREquiv{\Angle{x, \fgType_S, R, e}}{V}{R'}{k'}$.
\end{lemma}

\begin{proof}
Obvious. \MyQED
\end{proof}

\begin{lemma}[Monotonicity for type parameters]\label{lem:mono-typarams}
  If $\LREquiv{\fgTyActuals}{V}{\fgTyFormals}{k}$ and $k' \leq k$ then
  $\LREquiv{\fgTyActuals}{V}{\fgTyFormals}{k'}$.
\end{lemma}

\begin{proof}
Obvious. \MyQED
\end{proof}

\begin{lemma}[Monotonicity for method declarations]\label{lem:mono-mdecl}
  Assume declaration $D = \fgFunc[spec=R]$ and $k' \leq k$.  If $\LREquivNoTy{D}{k}{X}$ then
  $\LREquivNoTy{D}{k'}{X}$.
\end{lemma}

\begin{proof}
Obvious. \MyQED
\end{proof}

\begin{lemma}[Monotonicity for programs]\label{lem:mono-prog}
  If $\LREquivNoTy{\Multi{D}}{k}{\tlMethTable}$ and $k' \leq k$ then
  $\LREquivNoTy{\Multi{D}}{k'}{\tlMethTable}$.
\end{lemma}

\begin{proof}
  Follows from \Cref{lem:mono-mdecl}. \MyQED
\end{proof}

\subsubsection{Equivalence Between Source and Translation}
\label{sec:dummy}

\begin{proof}[Proof of \Cref{lem:target-reduce}]
  Straightforward.
\end{proof}

\begin{proof}[Proof of \Cref{lem:source-reduce}]
  We label the two implications in the premise of rule \Rule{equiv-exp} with (a) and (b).
  \begin{EnumAlph}
  \item Assume $k' < k+1$ and $\reducek{k'}{e_2}{v}$.
    Then by \Cref{lem:determ-eval-fgg} $e_2 \reduceSym \reducek{k'-1}{e}{v}$.
    Noting that $k' -1 < k$ we get with the assumption $\LREquiv{e}{E}{\fgType}{k}$
    \begin{igather}
      \exists V \DOT \reduceStar{E}{V} \wedge \LREquivVal{v}{V}{\fgType}{k+1-k'}
    \end{igather}
    But this is exactly what is needed to prove implication (a) for
    $\LREquiv{e_2}{E}{\fgType}{k+1}$.
  \item Assume $k' < k+1$ and $\reducek{k'}{e_2}{e'}$ and $\Diverge{e'}$.
    Then by \Cref{lem:determ-eval-fgg} $e_2 \reduceSym \reducek{k'-1}{e}{e'}$.
    Noting that $k' -1 < k$ we get with the assumption $\LREquiv{e}{E}{\fgType}{k}$
    that $\Diverge{E}$. This proves implication (b).
  \MyQED
  \end{EnumAlph}
\end{proof}

\begin{lemma}[Expression equivalence implies value equivalence]\label{lem:exp-equiv-implies-val-equiv}
  If $k \geq 1$ and $\LREquiv{v}{V}{\fgType}{k}$ then
  $\LREquivVal{v}{V}{\fgType}{k}$.
\end{lemma}

\begin{proof}
  From the first implication of rule \Rule{equiv-exp} we get for $k' = 0 < k$ and with
  $\reducek{0}{v}{v}$ that
  $\exists V' \DOT \reduceStar{V}{V'} \wedge \LREquivVal{v}{V'}{\fgType}{k}$.
  But V is already a value, so $V' = V$.\MyQED
\end{proof}

\begin{lemma}[Value equivalence implies expression equivalence]\label{lem:val-equiv-implies-exp-equiv}
  If $\LREquivVal{v}{V}{\fgType}{k}$ then
  $\LREquiv{v}{V}{\fgType}{k}$ for any $k$.
\end{lemma}

\begin{proof}
  We have $\reducek{0}{v}{v}$, so we get the first implication of rule \Rule{equiv-exp} by
  setting $E = V$ and by assumption $\LREquivVal{v}{V}{\fgType}{k}$.
  The second implication holds vacuously because values do not diverge. \MyQED
\end{proof}

\begin{lemma}
  \label{lem:method-decl-equiv-implies-dict-equiv}
  Assume $\LREquivNoTy{\fgFunc}{k}{X}$.
  Then the following holds:
  \begin{mathparNarrow}
    \forall k' < k, \fgTyActuals, W \DOT \LREquiv{\fgTyActuals}{W}{\fgTyFormals}{k'}
    \implies
    \fgMakeSubst{\fgTyFormals}{\fgTyActuals}{\fgTySubst} \wedge\\
    \LREquiv{
      \Angle{x, t_S[\fgTyActuals], \fgTySubst mM, \fgTySubst e}
    }{
      \tllambda{\Triple{Y_1}{Y_2}{Y_3}} X\,\Quadr{W}{Y_1}{Y_2}{Y_3}
    }{
      \fgTySubst mM
    }{k'}
  \end{mathparNarrow}
\end{lemma}

\begin{proof}
  \renewcommand\LabelQualifier{lem:method-decl-equiv-implies-dict-equiv}
  Let $M = [\fgTyFormals'](\MultiN{x_i\ \fgType_i})\,\fgType$ and assume
  for any $k', \fgTyActuals, W$
  \begin{igather}
    k' < k \QLabel{kprime}\\
    \LREquiv{\fgTyActuals}{W}{\fgTyFormals}{k'} \QLabel{phi-equiv-W}
  \end{igather}
  Obviously
  \begin{igather}
    \fgMakeSubst{\fgTyFormals}{\fgTyActuals}{\fgTySubst} \QLabel{def-eta}
  \end{igather}
  To show
  that
  \begin{igather}
    \LREquiv{
      \Angle{x, t_S[\fgTyActuals], \fgTySubst m[\fgTyFormals'](\MultiN{x_i\ \fgType_i})\,\fgType, \fgTySubst e}
    }{
      \tllambda{\Triple{Y_1}{Y_2}{Y_3}} X\,\Quadr{W}{Y_1}{Y_2}{Y_3}
    }{
      \fgTySubst mM
    }{k'}
  \end{igather}
  holds, we assume the left-hand side of the implication in the premise of rule \Rule{equiv-method-dict-entry}
  for some $k'', \fgTyActuals', W', u, U, \Multi[n]{v}, \Multi[n]{V}$:
  \begin{igather}
    k'' \leq k' \QLabel{def-kpprime}\\
    \fgMakeSubst{\fgTySubst \fgTyFormals'}{\fgTyActuals'}{\fgTySubst'} \QLabel{def-eta-prime}\\
    \LREquiv{\fgTyActuals'}{W'}{\fgTySubst \fgTyFormals'}{k''} \QLabel{phi-equiv-W-prime}\\
    \LREquiv{u}{U}{t_S[\fgTyActuals]}{k''} \QLabel{u-equiv-U}\\
    (\forall i \in [n]) \DOT \LREquiv{v_i}{V_i}{\fgTySubst' \fgTySubst \fgType_i}{k''} \QLabel{vi-equiv-Vi}
  \end{igather}
  We then need to prove \QRef{goal} to show the overall goal.
  \begin{igather}
    \LREquiv{
      \fgSubst\fgTySubst'\fgTySubst e
    }{
      V\,\Triple{U}{W'}{\Tuple{\MultiN{V}}}
    }{
      \fgTySubst' \fgTySubst \fgType
    }{k''} \QLabel{goal}
    \\
    \fgSubst = \Angle{\subst{x}{u}, \MultiN{\subst{x_i}{v_i}}}
    \\
    V = \tllambda{\TripleY} X\, \Quadr{W}{Y_1}{Y_2}{Y_3} \QLabel{def-V}
  \end{igather}
  Let $\fgTyFormals = \Multi[p]{\fgTyVar\,\fgTypeAux}$,
  $\fgTyFormals' = \Multi[q]{\fgTyVarAux\,\fgTypeAux'}$,
  $\fgTyActuals = \Multi[p]{\fgTypeAux''}$,
  $\fgTyActuals' = \Multi[q]{\fgTypeAux'''}$.
  Then by \QRef{def-eta} and \QRef{def-eta-prime}
  \begin{igather}
    \fgTySubst = \Angle{\Multi[p]{\subst{\fgTyVar_i}{\fgTypeAux''_i}}} \QLabel{def2-eta}\\
    \fgTySubst' = \Angle{\Multi[q]{\subst{\fgTyVarAux_i}{\fgTypeAux'''_i}}} \QLabel{def2-eta-prime}
  \end{igather}
  Define
  \begin{igather}
    \fgTyFormalsAux := \Multi[p]{\fgTyVar_i\,\fgTypeAux_i} \Multi[q]{\fgTyVarAux_i\,\fgTypeAux'_i}\\
    \fgTyActuals'' := \Multi[p]{\fgTypeAux''} \Multi[q]{\fgTypeAux'''}\\
    \fgTySubst'' := \Angle{\Multi[p]{\subst{\fgTyVar_i}{\fgTypeAux''_i}} \Multi[q]{\subst{\fgTyVarAux_i}{\fgTypeAux'''_i}}}
    \QLabel{def2-eta-pprime}
  \end{igather}
  Then
  \begin{igather}
    \fgMakeSubst{\fgTyFormalsAux}{\fgTyActuals''}{\fgTySubst''} \QLabel{subst-Psi}
  \end{igather}
  The $\Multi[q]{\fgTyVarAux}$ are sufficiently fresh, so
  $\ftv{\Multi[p]{\fgTypeAux''}} \cap \Multi[q]{\fgTyVarAux} = \emptyset$. Hence
  by \QRef{def2-eta}, \QRef{def2-eta-prime}, \QRef{def2-eta-pprime}
  \begin{igather}
    \fgTySubst'\fgTySubst \Multi[q]{\fgTypeAux'_i} = \fgTySubst'' \Multi[q]{\fgTypeAux'_i} \quad
    \fgTySubst'\fgTySubst \Multi[n]{\fgType_i} = \fgTySubst'' \Multi[n]{\fgType_i} \quad
    \fgTySubst'\fgTySubst \fgType = \fgTySubst'' \fgType \quad
    \fgTySubst'\fgTySubst e = \fgTySubst'' e \QLabel{subst-eqs}
  \end{igather}
  We have from \QRef{phi-equiv-W} and \QRef{phi-equiv-W-prime}
  \begin{igather}
    W = \Tuple{\Multi[p]{W}} \QLabel{def-W}\\
    W' = \Tuple{\Multi[q]{W'}} \QLabel{def-W-prime}
  \end{igather}
  We now prove
  \begin{igather}
  \LREquiv{\fgTyActuals''}{\Pair{\Multi[p]{W}}{\Multi[q]{W'}}}{\fgTyFormalsAux}{k''} \QLabel{phi-equiv-pprime}
  \end{igather}
  by verifying the implication in the premise of rule \Rule{equiv-bounded-typarams}.
  We consider two cases for every $\ell \leq k''$.
  \begin{CaseDistinction}{whether $i$ in $[p]$ or in $[q]$}
    \Case{$i \in [p]$} We need to prove
    $\forall u, U \DOT \LREquiv{u}{U}{\fgTypeAux''_i}{\ell} \implies
    \LREquiv{u}{W_i\,U}{\fgTySubst'' \fgTypeAux_i}{\ell}$.
    From \QRef{phi-equiv-W} we get with $ \LREquiv{u}{U}{\fgTypeAux''_i}{\ell}$
    and $\ell \leq k'' \ReasonAbove{\leq}{\QRef{def-kpprime}} k'$ that
    $\LREquiv{u}{W_i\,U}{\fgTySubst \fgTypeAux_i}{\ell}$.
    By assumption~\ref{conv:fg-decls} $\ftv{\fgTypeAux_i} \subseteq \Multi[p]{\fgTyVar}$, so
    $\fgTySubst'' \fgTypeAux_i = \fgTySubst \fgTypeAux_i$ by \QRef{def2-eta} and \QRef{def2-eta-pprime}.

    \Case{$i \in [q]$} We need to prove
    $\forall u, U \DOT \LREquiv{u}{U}{\fgTypeAux'''_i}{\ell} \implies
    \LREquiv{u}{W'_i\,U}{\fgTySubst'' \fgTypeAux'_i}{\ell}$.
    From \QRef{phi-equiv-W-prime} we get with $\LREquiv{u}{U}{\fgTypeAux'''_i}{\ell}$ that
    $\LREquiv{u}{W'_i\,U}{\fgTySubst'\fgTySubst \fgTypeAux'_i}{\ell}$.
    Also, $\fgTySubst'\fgTySubst \fgTypeAux'_i = \fgTySubst'' \fgTypeAux'_i$
    by \QRef{subst-eqs}.
  \end{CaseDistinction}
  This finishes the proof of \QRef{phi-equiv-pprime}.\\[\smallskipamount]
  From \QRef{u-equiv-U} and \QRef{def2-eta-pprime} we have
  \begin{igather}
    \LREquiv{u}{U}{\fgTySubst'' t_S[\Multi[p]{\fgTyVar}]}{k''} \QLabel{u-equiv-U2}
  \end{igather}
  From \QRef{vi-equiv-Vi} and \QRef{subst-eqs}
  \begin{igather}
    \LREquiv{v_i}{V_i}{\fgTySubst'' \fgType_i}{k''} \QLabel{vi-equiv-Vi2}
  \end{igather}
  From the assumption $\LREquivNoTy{\fgFunc}{k}{X}$, we can invert rule
  \Rule{equiv-method-decl}. Noting that $k'' \ReasonAbove{\leq}{\QRef{def-kpprime}} k' \ReasonAbove{<}{\QRef{kprime}} k$
  and that \QRef{subst-Psi}, \QRef{phi-equiv-pprime},
  \QRef{u-equiv-U2}, \QRef{vi-equiv-Vi2} give us the left-hand side of the implication
  in the premise of the rule, we get by the right-hand side of the implication
  \begin{igather}
    \LREquiv{
      \fgSubst \fgTySubst'' e
    }{
      X\ \Quadr{\Tuple{\Multi[p]{W}}}{U}{\Tuple{\Multi[q]{W'}}}{\Tuple{\Multi[n]{V}}}
    }{
      \fgTySubst'' \fgType
    }{k''}
  \end{igather}
  With \QRef{subst-eqs}, \QRef{def-W}, \QRef{def-W-prime}
  \begin{igather}
    \LREquiv{
      \fgSubst \fgTySubst'\fgTySubst e
    }{
      X\ \Quadr{W}{U}{W'}{\Tuple{\Multi[n]{V}}}
    }{
      \fgTySubst' \fgTySubst \fgType
    }{k''}
    \QLabel{e-equiv}
  \end{igather}
  We have by \QRef{def-V}
  \begin{igather}
    V~\Triple{U}{W'}{\Tuple{\Multi[n]{V}}} =
    (\tllambda{\TripleY} X\, \Quadr{W}{Y_1}{Y_2}{Y_3})~\Triple{U}{W'}{\Tuple{\Multi[n]{V}}}
  \end{igather}
  so
  \begin{igather}
    V~\Triple{U}{W'}{\Tuple{\Multi[n]{V}}} \reduceStarSym
    X\ \Quadr{W}{U}{W'}{\Tuple{\Multi[n]{V}}}
  \end{igather}
  Thus, with \QRef{e-equiv} and \Cref{lem:target-reduce}
  \begin{igather}
    \LREquiv{
      \fgSubst\fgTySubst'\fgTySubst e
    }{
      V\,\Triple{U}{W'}{\Tuple{\MultiN{V}}}
    }{
      \fgTySubst' \fgTySubst \fgType
    }{k''}
  \end{igather}
  as required to prove \QRef{goal}. \MyQED
\end{proof}

\begin{definition}[Domain]
  We write $\dom{\cdot}$ for the domain of a substitution $\fgTySubst$, $\fgSubst$,
  $\tlSubst$ or $\tlMethTable$, of a type environment $\fgTyEnv$,
  of a value environment $\fgEnv$, or some type parameters $\fgTyFormals$.
\end{definition}

\begin{definition}[Free variables]
  We write $\fv{\cdot}$ for the set of free term variables,
  and $\ftv{\cdot}$ for the set of free type variables.
\end{definition}

\begin{lemma}[Subtyping preserves equivalence]\label{subtyping-equiv}
  Let $\LREquivNoTy{\fgDecls}{k}{\tlMethTable}$. Assume
  $\tdICons{\fgTyEnv}{\fgType}{\fgTypeAux}{V}$ and $\LREquiv{\fgTySubst}{\tlSubst}{\fgTyEnv}{k}$
  and $\LREquiv{e}{E}{\fgTySubst\fgType}{k}$.
  Then $\LREquiv{e}{(\tlSubst V)\,E}{\fgTySubst \fgTypeAux}{k}$.
\end{lemma}
We prove \Cref{subtyping-equiv} together with the following two lemmas.

\begin{lemma}\label{lem:method-lookup}
  Assume $\LREquivNoTy{\fgDecls}{k}{\tlMethTable}$ and
  $\LREquiv{\fgTySubst}{\tlSubst}{\fgTyEnv}{k}$.
  Let $ \tdMethods{R}{V}{\fgTyEnv}{t_S[\fgTyActuals]}$ and
  define $U = \tllambda{\TripleY}{\mT{m}{t_S}}\,\Quadr{\tlSubst V}{Y_1}{Y_2}{Y_3}$.
  Then we have for all $k' < k$ that
  $\LREquiv{\methodLookup{m}{\fgTySubst t_S[\fgTyActuals]}}{U}{\fgTySubst R}{k'}$.
\end{lemma}

\begin{lemma}[Substitution preserves equivalence]\label{lem:subst-equiv}
  Assume $\LREquivNoTy{\fgDecls}{k}{\tlMethTable}$ and
  $\LREquiv{\fgTySubst}{\tlSubst}{\fgTyEnv}{k}$.
  If $\tdCheckSubst{\fgTyEnv}{\fgTyFormals}{\fgTyActuals}{\fgTySubst'}{V}$
  then $\LREquiv{\fgTySubst \fgTyActuals}{\tlSubst V}{\fgTySubst \fgTyFormals}{k}$.
\end{lemma}

\begin{proof}[Proof of Lemmas~\ref{subtyping-equiv}, \ref{lem:method-lookup}, and~\ref{lem:subst-equiv}]
  \renewcommand\LabelQualifier{subtyping-equiv}
  We show the three lemmas by induction on the combined height of the derivations for
  $\tdICons{\fgTyEnv}{\fgType}{\fgTypeAux}{V}$ and
  $\tdMethods{R}{V}{\fgTyEnv}{t_S[\fgTyActuals]}$ and
  $\tdCheckSubst{\fgTyEnv}{\fgTyFormals}{\fgTyActuals}{\fgTySubst'}{V}$.

  We start with the proof for \emph{\Cref{subtyping-equiv}}. We have from the assumptions
  \begin{igather}
    \LREquiv{e}{E}{\fgTySubst\fgType}{k} \QLabel{e-equiv-E}
  \end{igather}
  Assume $k' < k$ and $\reducek{k'}{e}{e'}$.
  The second implication in the premise of rule \Rule{equiv-exp} holds obviously, because
  with $\Diverge{e'}$ we get from \QRef{e-equiv-E} $\Diverge{E}$, so also $\Diverge{(\tlSubst V)\ E}$.

  Thus, we only need to prove the first implication. Assume that $e' = v$ for some value $v$.
  Then via \QRef{e-equiv-E} for some $U$
  \begin{igather}
    \reduceStar{E}{U} \QLabel{eval-E}\\
    \LREquivVal{v}{U}{\fgTySubst \fgType}{k-k'} \QLabel{v-equiv-U}
  \end{igather}
  We then need to verify that $\reduceStar{(\tlSubst V)\ U}{U'}$ for some $U'$ with
  $\LREquivVal{v}{U'}{\fgTySubst \fgTypeAux}{k - k'}$.
  In fact, $k' < k$, so with \Cref{lem:exp-equiv-implies-val-equiv}
  it suffices to show that
  $\LREquiv{v}{U'}{\fgTySubst \fgTypeAux}{k - k'}$.

  \begin{CaseDistinction}[repeat]{on the last rule in the derivation of $\tdICons{\fgTyEnv}{\fgType}{\fgTypeAux}{V}$}
    \Case{\Rule{coerce-tyvar}}
    \begin{mathpar}
      \inferrule{
        X \textrm{~fresh}\\
        (a : \fgType_I) \in \fgTyEnv\\
        \tdICons{\fgTyEnv}{\fgType_I}{\fgTypeAux}{W}
      }{
        \tdICons{\fgTyEnv}{\EqualBelow{\fgTyVar}{\fgType}}{\fgTypeAux}{\BraceBelow{\tllambda{X}{W\ (\xTy{\fgTyVar}\ X)}}{= V}}
      }
    \end{mathpar}
    Our goal to show is
    \begin{igather}
      \LREquiv{e}{(\tlSubst \xTy{\fgTyVar})\ U}{\fgTySubst \fgType_I}{k} \QLabel{goal}
    \end{igather}
    With \QRef{goal} and the IH for \Cref{subtyping-equiv} we then get
    \begin{igather}
      \LREquiv{e}{(\tlSubst W)\ ((\tlSubst \xTy{\fgTyVar})\ U)}{\fgTySubst \fgTypeAux}{k}
    \end{igather}
    Then, with $\reducek{k'}{e}{v}$, we get
    $(\tlSubst V)\ U \reduceSym \reduceStar{(\tlSubst W)\ ((\tlSubst \xTy{\fgTyVar})\ U)}{U'}$ for some $U'$ with
    $\LREquivVal{v}{U'}{\fgTySubst \fgTypeAux}{k - k'}$.
    \\
    We now prove \QRef{goal}. From the assumption $\LREquiv{\fgTySubst}{\tlSubst}{\fgTyEnv}{k}$ we have
    \begin{igather}
      \fgTyEnv = \Multi[n]{\fgTyVar_i : \fgType_i}\\
      \fgTySubst = \Angle{\Multi[n]{\subst{\fgTyVar_i}{\fgTypeAux_i}}}\\
      \tlSubst = \Angle{\Multi[n]{\subst{\xTy{\fgTyVar_i}}{V_i}}}\\
      \LREquiv{\Multi[n]{\fgTypeAux}}{\Multi[n]{V}}{\Multi[n]{\fgTyVar_i\ \fgType_i}}{k} \QLabel{eq-typarams}
    \end{igather}
    such that $\fgTyVar = \fgTyVar_j$ and $\fgType_I = \fgType_j$ for some $j \in [n]$.
    Inverting rule \Rule{equiv-bounded-typarams} on \QRef{eq-typarams} yields
    \begin{igather}
      \forall k'' \leq k, w, W' \DOT
      \LREquiv{w}{W'}{\fgTypeAux_j}{k''} \implies
      \LREquiv{w}{V_j\ W'}{\fgTySubst \fgType_j}{k''} \QLabel{forall-kpprime}
    \end{igather}
    From \QRef{v-equiv-U} by $\fgTySubst \fgType = \fgTypeAux_j$ then
    $\LREquivVal{v}{U}{\fgTypeAux_j}{k - k'}$. Thus with \QRef{forall-kpprime} and \Cref{lem:val-equiv-implies-exp-equiv}
    \begin{igather}
      \LREquiv{v}{V_j\ U}{\fgTySubst \fgType_j}{k - k'}
    \end{igather}
    With \Cref{lem:source-reduce} and $\reducek{k'}{e}{v}$ then
    \begin{igather}
      \LREquiv{e}{V_j\ U}{\fgTySubst \fgType_j}{k}
    \end{igather}
    But $\fgType_j = \fgType_I$ and $V_j = \tlSubst \xTy{\fgTyVar}$, so this proves \QRef{goal}.

    \Case{\Rule{coerce-struct-iface}}
    \begin{mathpar}
      \inferrule{
        \LabeledRule{
          X, Y_1, Y_2, Y_3~\textrm{fresh}\\
          \TYPE\ t_I[\fgTyFormals] \ \INTERFACE\ \{ \Foreach{mM}{n} \} \in \ForeachN{D} & \TextLabel{def-I}\\
          \fgMakeSubst{\fgTyFormals}{\fgTyActuals}{\fgTySubst'} & \TextLabel{def-eta-prime}\\
          \tdMethods{\fgTySubst'(m_i M_i)}{V_i}{\fgTyEnv}{t_S[\fgTyActualsAux]} & \TextLabel{methods}\\
           V_i' = \PartialM{\mT{m_i}{t_S}}{V_i}
          \quad\noteForall{i \in [n]} & \TextLabel{def-V-prim}
        }
      }{
        \tdUpcast{\fgTyEnv}{\subtypeOf{\EqualBelow{t_S[\fgTyActualsAux]}{\fgType}}{\EqualBelow{t_I[\fgTyActuals]}{\fgTypeAux}}}
        {\BraceBelow{\tllambda{\xT} \Tuple{\xT\Comma \Tuple{\Foreach{V_i'}{n}}}}{= V}}
      }
    \end{mathpar}
    Hence $\reduce{(\tlSubst V)\ U}{\Pair{U}{\tlSubst\Tuple{\MultiN{V'}}}}$ and
    $U' := \Pair{U}{\tlSubst\Tuple{\MultiN{V'}}}$ is a value.
    We now want to show $\LREquiv{v}{U'}{\fgTySubst \fgTypeAux}{k - k'}$ via
    rule \Rule{equiv-iface}. Define the $\fgTypeAux_S$ in the premise of \Rule{equiv-iface}
    as $\fgTySubst \fgType = t_S[\fgTySubst \fgTyActualsAux]$.

    The first premise of \Rule{equiv-iface}
    \begin{igather}
      \forall k_1 < k - k' \DOT \LREquivVal{v}{U}{\fgTySubst \fgType}{k_1} \QLabel{premise1}
    \end{igather}
    follows from \QRef{v-equiv-U} and \Cref{lem:mono-exp}. From \QRef{def-I} and \QRef{def-eta-prime} we get with
    $\fgTypeAux = t_I[\fgTyActuals]$
    the second premise as
    \begin{igather}
      \methodSpecifications{\fgTySubst \fgTypeAux} = \fgTySubst \fgTySubst' \MultiN{mM} \QLabel{premise2}
    \end{igather}
    We next prove the third premise of \Rule{equiv-iface}. Pick some $j \in [n]$ and $k_2 < k -k'$.
    With the assumptions $\LREquivNoTy{\fgDecls}{k}{\tlMethTable}$ and $\LREquiv{\fgTySubst}{\tlSubst}{\fgTyEnv}{k}$
    and with \QRef{methods}, \QRef{def-V-prim}, and the IH for \Cref{lem:method-lookup} we get
    \begin{igather}
      \LREquiv{
        \methodLookup{m_j}{\fgTySubst t_S[\fgTyActualsAux]}
      }{
        \tlSubst V_j'
      }{
        \fgTySubst \fgTySubst' m_j M_j
      }{k_2} \QLabel{premise3}
    \end{igather}
    \QRef{premise1}, \QRef{premise2}, \QRef{premise3}, and the definition of $U'$
    are the pieces required to derive
    $\LREquiv{v}{U'}{\fgTySubst \fgTypeAux}{k - k'}$ via rule \Rule{equiv-iface}.

    \Case{\Rule{coerce-iface-iface}}
    \begin{mathpar}
      \inferrule{
        \LabeledRule{
          Y,\Foreach{X}{n}~\textrm{fresh} \qquad
          \mapPerm : [q] \to [n]~\textrm{total}\\
          \TYPE\ t_I[\fgTyFormals_1] \ \INTERFACE\ \{ \Foreach{R}{n} \} \in \ForeachN{D}  & \TextLabel{def-tI}\\
          \TYPE\ u_I[\fgTyFormals_2] \ \INTERFACE\ \{ \Foreach{R'}{q} \} \in \ForeachN{D}  & \TextLabel{def-uI}\\
          \fgMakeSubst{\fgTyFormals_1}{\fgTyActuals_1}{\fgTySubst_1} \qquad
          \fgMakeSubst{\fgTyFormals_2}{\fgTyActuals_2}{\fgTySubst_2}\\
          \fgTySubst_2 R'_i = \fgTySubst_1 R_{\mapPerm(i)} \quad\noteForall{i \in [q]} & \TextLabel{Si-eq-Ri}
        }
        }{
        \tdUpcast{\fgTyEnv}{\subtypeOf{\EqualBelow{t_I[\fgTyActuals_1]}{\fgType}}{\EqualBelow{u_I[\fgTyActuals_2]}{\fgTypeAux}}}
        { \BraceBelow{
            \tllambda{\Tuple{Y\Comma \Tuple{\Foreach{X}{n}}}}
            \Tuple{Y\Comma \Tuple{\xT_{\mapPerm(1)}\Comma \ldots\Comma \xT_{\mapPerm(q)}}}
          }{= V ~~~~\TextLabel{def-V}}
        }
      }
    \end{mathpar}
    As $\fgTySubst \fgType = \fgTySubst t_I[\fgTyActuals_1]$ is an interface type, we get from \QRef{v-equiv-U} by inverting
    rule \Rule{equiv-iface} for some $W, \fgTypeAux_S, \MultiN{W}$ that
    \begin{igather}
      \forall k_1 < k - k' \DOT \LREquiv{v}{W}{\fgTypeAux_S}{k_1} \QLabel{impl1}\\
      \methodSpecifications{\fgTySubst \fgType} = \fgTySubst \fgTySubst_1 \MultiN{R} = \MultiN{mM} \QLabel{methods-tau}\\
      \forall i \in [n], k_2 < k - k' \DOT
      \LREquiv{\methodLookup{m_i}{\fgTypeAux_s}}{W_i}{m_i M_i}{k_2} \QLabel{impl2}\\
      U = \Pair{W}{\Tuple{\MultiN{W}}} \QLabel{def-U}
    \end{igather}
    Our goal is to show $\reduceStar{(\tlSubst V)\ U}{U'}$ for some $U'$ with
    $\LREquiv{v}{U'}{\fgTySubst \fgTypeAux}{k - k'}$. Via \QRef{def-V} and \QRef{def-U}
    \begin{igather}
      (\tlSubst V)\ U = V\ U \reduceStarSym
      \Pair{W}{\Tuple{W_{\mapPerm(1)}\Comma \ldots \Comma W_{\mapPerm(q)}}} =: U'
      \QLabel{def-Uprime}
    \end{igather}
    From \QRef{def-tI}, \QRef{def-uI}, \QRef{Si-eq-Ri}, and \QRef{methods-tau} we have
    \begin{igather}
      \begin{array}[t]{@{}r@{~}l@{}}
        \Multi[q]{m' M'}
        = \fgTySubst \fgTySubst_2 \Multi[q]{R'}
        &= \fgTySubst \fgTySubst_1 R_{\mapPerm(1)}, \ldots, \fgTySubst \fgTySubst_1 R_{\mapPerm(q)}\\
        &= m_{\mapPerm(1)} M_{\mapPerm(1)}, \ldots, m_{\mapPerm(q)} M_{\mapPerm(q)}
      \end{array} \QLabel{eq-M}
      \\
      \methodSpecifications{\fgTySubst \fgTypeAux} = \{\Multi[q]{m' M'} \} \QLabel{methods-sigma}
    \end{igather}
    Pick $j \in [q]$. Then via \QRef{eq-M}
    \begin{igather}
      \methodLookup{m_j'}{\fgTypeAux_S} =
      \methodLookup{m_{\mapPerm(j)}}{\fgTypeAux_s}
    \end{igather}
    Hence with \QRef{impl2} and \QRef{eq-M}
    \begin{igather}
      \forall j \in [q], k_2 < k - k' \DOT
      \LREquiv{
        \methodLookup{m_j'}{\fgTypeAux_S}
      }{
        W_{\mapPerm(j)}
      }{m_j' M_j'}{k_2} \QLabel{impl22}
    \end{igather}
    With \QRef{impl1}, \QRef{impl22}, \QRef{methods-sigma} and the definition of $U'$ in \QRef{def-Uprime} we then
    get by applying rule \Rule{equiv-iface}
    $\LREquiv{v}{U'}{\fgTySubst \fgTypeAux}{k - k'}$ and with \QRef{def-Uprime} also
    $\reduceStar{(\tlSubst V)\ U}{U'}$.
  \end{CaseDistinction}
  \\
  This finishes the proof of \Cref{subtyping-equiv}.

  \renewcommand\LabelQualifier{lem:method-lookup}
  We next prove \Cref{lem:method-lookup}. By inverting rule \Rule{methods-struct} for the assumption
  $\Angle{R,V} \in \methodSpecifications{\fgTyEnv, t_S[\fgTyActuals]}$ we get
  \begin{igather}
    \fgFunc \in \fgDecls \QLabel{func} \\
    \tdCheckSubst{\fgTyEnv}{\fgTyFormals}{\fgTyActuals}{\fgTySubst'}{V} \QLabel{subst}\\
    R = \fgTySubst' m M \QLabel{R}
  \end{igather}
  Inverting \QRef{subst} yields
  \begin{igather}
    \fgTyFormals = \MultiN{\fgTyVar\ \fgType}\\
    \fgTySubst' = \Angle{\Multi[n]{\subst{\fgTyVar_i}{\fgTypeAux_i}}}\\
    \fgTyActuals = \MultiN{\fgTypeAux}\\
    \tdICons{\fgTyEnv}{\fgTypeAux_i}{\fgTySubst' \fgType_i}{V_i} \quad (\forall i \in [n])\\
    V = \Tuple{\MultiN{V}}
  \end{igather}
  Define $\fgTySubst'' = \Angle{\MultiN{\subst{\fgTyVar_i}{\fgTySubst\fgTypeAux_i}}}$.
  Then by rule \Rule{method-lookup} and \QRef{func}
  \begin{igather}
    \methodLookup{m}{\fgTySubst t_S[\fgTyActuals]} =
    \Angle{x, \fgTySubst t_S[\fgTyActuals], \fgTySubst'' mM, \fgTySubst'' e}
    \QLabel{method-lookup}
  \end{igather}
  By assumption~\ref{conv:fg-decls}, the $\MultiN{\fgTyVar}$ can be assumed to be fresh,
  $\ftv{\fgTyFormals} \subseteq \MultiN{\fgTyVar}$, and
  $\fgTySubst \fgTyFormals = \fgTyFormals$.
  Applying the IH for \Cref{lem:subst-equiv} on \QRef{subst} yields
  $\LREquiv{\fgTySubst \fgTyActuals}{\tlSubst V}{\fgTySubst \fgTyFormals}{k}$.
  \begin{igather}
    \LREquiv{\fgTySubst \fgTyActuals}{\tlSubst V}{\fgTyFormals}{k} \QLabel{phi-equiv-V}
  \end{igather}
  From the assumption $\LREquivNoTy{\fgDecls}{k}{\tlMethTable}$ we get with \QRef{func}
  \begin{igather}
    \LREquivNoTy{\fgFunc}{k}{\mT{m}{t_S}}
  \end{igather}
  Then for any $k' < k$ by
  \Cref{lem:method-decl-equiv-implies-dict-equiv}, where \QRef{phi-equiv-V} and
  \Cref{lem:mono-typarams} give the left-hand side of the implication
  \begin{igather}
    \fgMakeSubst{\fgTyFormals}{\fgTySubst \fgTyActuals}{\fgTySubst''}\\
    \LREquiv{
      \Angle{x, \fgTySubst t_S[\fgTyActuals], \fgTySubst'' m M, \fgTySubst'' e}
    }{~\\
      \BraceBelow{\tllambda{\TripleY}{\mT{m}{t_S}\ \Quadr{\tlSubst V}{Y_1}{Y_2}{Y_3}}}{= U}
    }{
      \fgTySubst'' m M
    }{k'}
    \QLabel{equiv}
  \end{igather}
  We have $\fgTySubst R \ReasonAbove{=}{\QRef{R}} \fgTySubst \fgTySubst' mM = \fgTySubst'' mM$,
  where the last equality holds because $\ftv{mM} \subseteq \MultiN{\fgTyVar}$ and $\MultiN{\fgTyVar}$
  fresh by assumption~\ref{conv:fg-decls}.
  Hence \QRef{method-lookup} and \QRef{equiv} give
  the desired claim.

  Finally, we prove \Cref{lem:subst-equiv}. By inverting rule \Rule{type-inst-checked} for the
  assumption $\tdCheckSubst{\fgTyEnv}{\fgTyFormals}{\fgTyActuals}{\fgTySubst'}{V}$ we get
  \begin{igather}
    \fgTyFormals = \MultiN{\fgTyVar\,\fgType}\\
    \fgTyActuals = \MultiN{\fgTypeAux}\\
    \fgTySubst' = \Angle{\MultiN{\subst{\fgTyVar_i}{\fgTypeAux_i}}}\\
    \tdICons{\fgTyEnv}{\fgTypeAux_i}{\fgTySubst' \fgType_i}{V_i} \quad (\forall i \in [n]) \QLabel{icons}\\
    V = \Tuple{\MultiN{V}}
  \end{igather}
  Define $\fgTySubst'' = \Angle{\MultiN{\subst{\fgTyVar_i}{\fgTySubst \fgTypeAux_i}}}$.
  To prove $\LREquiv{\fgTySubst \fgTyActuals}{\tlSubst V}{\fgTySubst \fgTyFormals}{k}$
  we need to show the implication
  $\forall j \in [n], k' \leq k \DOT \LREquiv{u}{U}{\fgTySubst \fgTypeAux_j}{k'} \implies
  \LREquiv{u}{(\tlSubst V)\ U}{\fgTySubst''\fgTySubst\fgType_j}{k'}$ from the premise
  of rule \Rule{equiv-bounded-typarams}.
  Assume $j \in [n], k' \leq k$, and $\LREquiv{u}{U}{\fgTySubst \fgTypeAux_j}{k'}$. Applying
  the IH for \Cref{subtyping-equiv} on \QRef{icons} yields together with
  \Cref{lem:mono-typarams} and \Cref{lem:mono-prog} that
  \begin{igather}
    \LREquiv{u}{(\tlSubst V_j)\ U}{\fgTySubst\fgTySubst'\fgType_j}{k'} \QLabel{u-equiv-VU}
  \end{igather}
  As the $\Multi{\fgTyVar}$ are bound in $\fgTyFormals$, we may assume that
  $\dom{\fgTySubst} \cap \Multi{\fgTyVar} = \emptyset = \ftv{\fgTySubst} \cap \Multi{\fgTyVar}$.
  We now argue that
  \begin{igather}
    \fgTySubst\fgTySubst'\fgType_j = \fgTySubst''\fgTySubst\fgType_j \QLabel{type-eq}
  \end{igather}
  by induction on the structure of $\fgType_j$. The interesting case is were $\fgType_j$ is
  a type variable (otherwise the claim follows by the IH). If $\fgType_j \in \Multi{\fgTyVar}$ then
  \begin{igather}
    \fgTySubst\fgTySubst' \fgType_j \ReasonAbove{=}{\textrm{def. of~}\fgTySubst''}
    \fgTySubst'' \fgType_j \ReasonAbove{=}{\dom{\fgTySubst} \cap \Multi{\fgTyVar} = \emptyset}
    \fgTySubst'' \fgTySubst \fgType_j
  \end{igather}
  If $\fgType_j \in \dom{\fgTySubst}$ then
  \begin{igather}
    \fgTySubst\fgTySubst' \fgType_j \ReasonAbove{=}{\dom{\fgTySubst} \cap \Multi{\fgTyVar} = \emptyset}
    \fgTySubst \fgType_j \ReasonAbove{=}{\ftv{\fgTySubst} \cap \Multi{\fgTyVar} = \emptyset}
    \fgTySubst'' \fgTySubst \fgType_j
  \end{igather}
  If $\fgType_j$ is some other type variable, \QRef{type-eq} holds obviously.
  With \QRef{u-equiv-VU} and \QRef{type-eq} we get
  $\LREquiv{u}{(\tlSubst V_j)\ U}{\fgTySubst''\fgTySubst\fgType_j}{k'}$ as required.
  \MyQED
\end{proof}

\begin{lemma}[Free variables of coercion values]\label{lem:free-icons}
  If $\tdICons{\fgTyEnv}{\fgType}{\fgTypeAux}{V}$
  then
  $\fv{V} \subseteq \{ \xTy{\fgTyVar} \mid \fgTyVar \in \dom{\fgTyEnv} \} \cup \mathcal X$
  where $\mathcal X = \{ \mT{m}{t_S} \mid m \textrm{~method name}, t_S \textrm{~struct name}\}$.
\end{lemma}

\begin{proof}
  By straightforward induction on the derivation of $\tdICons{\fgTyEnv}{\fgType}{\fgTypeAux}{V}$.
  \MyQED
\end{proof}

\renewcommand\LabelQualifier{}
\paragraph{Proof of \Cref{lem:exp-equiv}}
\label{sec:proof-exp-equiv}

By induction on the derivation of
$\tdExpTrans{\pair{\fgTyEnv}{\fgEnv}}{e : \fgType}{E}$.
\begin{CaseDistinctionExplicit}{on the last rule in the derivation}{ on the last rule in the derivation of $\tdExpTrans{\pair{\fgTyEnv}{\fgEnv}}{e : \fgType}{E}$}
  \renewcommand\LabelQualifier{exp-equiv-var}
  \Case{\Rule{var}}
  \begin{mathpar}
    \inferrule{
      (x : \fgType) \in \fgEnv
    }{
      \tdExpTrans{\pair{\fgTyEnv}{\fgEnv}}{x : \fgType}{\xT}
    }
  \end{mathpar}
  with $\fgSubst\fgTySubst e = \fgSubst x$ and
  $\tlSubst E = \tlSubst X$. From the assumption
  $\LREquiv{\fgSubst}{\tlSubst}{\fgTySubst \fgEnv}{k}$ we get
  $\LREquiv{\fgSubst x}{\tlSubst X}{\fgTySubst \fgType}{k}$ as required.
  \renewcommand\LabelQualifier{exp-equiv-struct}
  \Case{\Rule{struct}}
  \begin{mathpar}
    \inferrule{
      \LabeledRule{
        \fgTyOk{\fgTyEnv}{t_S[\fgTyActuals]}\\
        \TYPE\ t_S[\fgTyFormals] \ \STRUCT\ \{ \Foreach{f \ \fgType}{n} \} \in  \ForeachN{D} & \TextLabel{def-struct}\\
        \fgMakeSubst{\fgTyFormals}{\fgTyActuals}{\fgTySubst'} & \TextLabel{def-eta-prime}\\
        \tdExpTrans{\pair{\fgTyEnv}{\fgEnv}}{e_i : \fgTySubst'\fgType_i}{\expT_i}
        \quad\noteForall{i \in [n]} & \TextLabel{trans-ei}
      }
    }{
      \tdExpTrans{\pair{\fgTyEnv}{\fgEnv}}
      {\BraceBelow{
          t_S[\fgTyActuals] \{ \Foreach{e}{n} \}
        }{= e}
        : \BraceBelow{t_S[\fgTyActuals]}{= \fgType}
      }{
        \BraceBelow{\Tuple{\Foreach{\expT}{n}}}{= E}
      } \qquad \TextLabel{conclusion}
    }
  \end{mathpar}
  Applying the IH to \QRef{trans-ei} yields
  \begin{igather}
    \LREquiv{
      \fgSubst\fgTySubst e_i
    }{
      \tlSubst E_i
    }{
      \fgTySubst\fgTySubst' \fgType_i
    }{k} \quad(\forall i \in [n]) \QLabel{equiv-ei}
  \end{igather}
  We now consider the two implications in the premise of rule \Rule{equiv-exp}
  \begin{EnumAlph}
  \item Assume $k' < k$ and $\reducek{k'}{\fgSubst\fgTySubst e}{v}$ for some value $v$.
    The goal is to show that there exists some value $V$ with $\reduceStar{\tlSubst E}{V}$
    and $\LREquivVal{v}{V}{\fgTySubst\fgType}{k - k'}$.\\
    With $\reducek{k'}{\fgSubst\fgTySubst e}{v}$ there must
    exist values $\MultiN{v}$ such that
    \begin{igather}
      \fgSubst\fgTySubst e_i \reduceSym^{k_i} v_i \qquad (\forall i \in [n])\\
      k_i \leq k' \qquad (\forall i \in [n]) \QLabel{ki-leq-k}\\
      v = t_S[\fgTySubst\fgTyActuals]\{\MultiN{v}\} \QLabel{shape-v}
    \end{igather}
    Via \QRef{equiv-ei} and $k_i \leq k' < k$ then for all $i \in [n]$
    \begin{igather}
      \tlSubst E_i \reduceStarSym V_i ~\textrm{for some}~V_i \quad(\forall i \in [n]) \QLabel{reduce-Ei}\\
      \LREquivVal{v_i}{V_i}{\fgTySubst\fgTySubst'\fgType_i}{k - k_i} \quad(\forall i \in [n]) \QLabel{equiv-vi}
    \end{igather}
    We have $k - k' \leq k - k_i$ for all $i \in [n]$ by \QRef{ki-leq-k}. Thus with \QRef{equiv-vi}
    and \Cref{lem:mono-exp}
    \begin{igather}
      \LREquivVal{v_i}{V_i}{\fgTySubst\fgTySubst'\fgType_i}{k - k'} \qquad (\forall i \in [n]) \QLabel{equiv-vi2}
    \end{igather}
    We also have with \QRef{reduce-Ei} and the definition of $E$ in \QRef{conclusion}
    \begin{igather}
      \tlSubst E \reduceStarSym \Tuple{\MultiN{V}} \QLabel{reduce-E}
    \end{igather}
    Assume $\fgGetTyVars{\fgTyFormals} = \MultiP{\fgTyVar}$ and
    $\fgTyActuals = \MultiP{\fgTypeAux}$. Then by \QRef{def-eta-prime}
    $\fgTySubst' = \Angle{\MultiP{\subst{\fgTyVar_i}{\fgTypeAux_i}}}$ and for
    $\fgTySubst'' = \Angle{\MultiP{\subst{\fgTyVar_i}{\fgTySubst\fgTypeAux_i}}}$
    we have
    \begin{igather}
      \fgMakeSubst{\fgTyFormals}{\fgTySubst \fgTyActuals}{\fgTySubst''} \QLabel{subst2}
    \end{igather}
    By assumption~\ref{conv:fg-decls} we have $\ftv{\MultiN{\fgType}} \subseteq \{\Multi{\fgTyVar}\}$, so
    \begin{igather}
      \fgTySubst\fgTySubst'\fgType_i = \fgTySubst'' \fgType_i \qquad (\forall i \in [n]) \QLabel{tau-eq}
    \end{igather}
    With \QRef{def-struct}, \QRef{equiv-vi2}, \QRef{shape-v}, \QRef{subst2}, \QRef{tau-eq}, and rule
    \Rule{equiv-struct} then
    \begin{igather}
      \LREquivVal{v}{\Tuple{\MultiN{V}}}{t_S[\fgTySubst\fgTyActuals]}{k - k'}
    \end{igather}
    Together with \QRef{reduce-E}, this finishes subcase (a) for $V = \Tuple{\MultiN{V}}$.
  \item Assume $k' < k$ and $\reducek{k'}{\fgSubst\fgTySubst e}{e'}$ and $\Diverge{e'}$. Then $\Diverge{e_j}$
    for some $j \in [n]$, so with \QRef{equiv-ei} and \QRef{ki-leq-k} also $\Diverge{\tlSubst E_j}$. Thus, by definition of $E$ in
    \QRef{conclusion}, $\Diverge{\tlSubst E}$ as required.
  \end{EnumAlph}

  \renewcommand\LabelQualifier{exp-equiv-access}
  \Case{\Rule{access}}
  \begin{mathpar}
    \inferrule{
      \LabeledRule{
        \tdExpTrans{\pair{\fgTyEnv}{\fgEnv}}{e' : t_S[\fgTyActuals]}{E'} & \TextLabel{type-eprime}\\
        \TYPE\ t_S[\fgTyFormals] \ \STRUCT\ \{ \Foreach{f \ \fgType}{n} \} \in  \ForeachN{D}\\
        \fgMakeSubst{\fgTyFormals}{\fgTyActuals}{\fgTySubst'}
      }
    }{
      \tdExpTrans{\pair{\fgTyEnv}{\fgEnv}}
      {\BraceBelow{e'.f_j}{= e} : \BraceBelow{\fgTySubst' \fgType_j}{= \fgType}}
      { \BraceBelow{\CASE\ E'\ \OF\ \Tuple{\Foreach{\xT}{n}} \rightarrow \xT_i}{= E}}
      \qquad \TextLabel{conclusion}
    }
  \end{mathpar}
  Applying the IH to \QRef{type-eprime} yields
  \begin{igather}
    \LREquiv{
      \fgSubst\fgTySubst e'
    }{
      \tlSubst E'
    }{
      t_S[\fgTySubst\fgTyActuals]
    }{k} \QLabel{IH}
  \end{igather}
  We now consider the two implications in the premise of rule \Rule{equiv-exp}
  \begin{EnumAlph}
  \item Assume $k' < k$ and $\reducek{k'}{\fgSubst\fgTySubst e}{v}$ for some value $v$.
    The goal is to show that there exists some value $V$ with $\reduceStar{\tlSubst E}{V}$
    and $\LREquivVal{v}{V}{\fgTySubst\fgType}{k - k'}$.\\
    With $\reducek{k'}{\fgSubst\fgTySubst e}{v}$ then
    $\reducek{k''}{\fgSubst\fgTySubst e'}{v'}$ for some $v'$ and $k'' < k'$. With \QRef{IH} then
    for some $V'$
    \begin{igather}
      \tlSubst E' \reduceStarSym V' \QLabel{reduce-Eprime} \\
      \LREquivVal{v'}{V'}{t_S[\fgTySubst \fgTyActuals]}{k - k''} \QLabel{equiv-vprime}
    \end{igather}
    Inverting rule \Rule{equiv-struct} on \QRef{equiv-vprime} yields
    \begin{igather}
      v' = t_S[\fgTySubst \fgTyActuals]\{\MultiN{v}\} \QLabel{shape-vprime}\\
      V' = \Tuple{\MultiN{V}} \quad \textrm{for some}~\MultiN{V}\\
      \LREquivVal{v_i}{V_i}{\fgTySubst''\fgType_i}{k - k''} \quad (\forall i \in [n]) \QLabel{equiv-vi}
    \end{igather}
    where $\fgTySubst'' = \Angle{\MultiP{\subst{\fgTyVar_i}{\fgTySubst\fgTypeAux_i}}}$, assuming
    $\fgGetTyVars{\fgTyFormals} = \MultiP{\fgTyVar}$ and $\fgTyActuals = \MultiP{\fgTypeAux}$.
    By assumption~\ref{conv:fg-decls} we have $\ftv{\fgType_j} \subseteq \{\Multi{\fgTyVar}\}$.
    Thus, $\fgTySubst''\fgType_j = \fgTySubst\fgTySubst'\fgType_j \ReasonAbove{=}{\QRef{conclusion}} \fgTySubst\fgType$.
    Also, $k'' \leq k'$, so $k - k' \leq k - k''$.
    Hence with \QRef{equiv-vi} and \Cref{lem:mono-exp}
    \begin{igather}
      \LREquivVal{v_j}{V_j}{\fgTySubst\fgType}{k - k'} \QLabel{equiv-vj}
    \end{igather}
    With \QRef{reduce-Eprime} and the definition of $E$ in \QRef{conclusion} we get
    \begin{igather}
      \tlSubst E \reduceStarSym V_j \QLabel{reduce-E}
    \end{igather}
    With $\reducek{k'}{\fgSubst\fgTySubst e}{v}$ and
    $\reducek{k''}{\fgSubst\fgTySubst e'}{v'}$ and the form of $v'$ in \QRef{shape-vprime}, we get
    $\reducek{k'}{\fgSubst\fgTySubst e}{v_j}$ and $v = v_j$ by rule \Rule{fg-field}.
    Define $V = V_j$ and we are done with subcase (a) by \QRef{equiv-vj} and \QRef{reduce-E}.
  \item Assume $k' < k$ and $\reducek{k'}{\fgSubst\fgTySubst e}{e''}$ and $\Diverge{e''}$.
    Then we must have that $\reducek{k''}{\fgSubst\fgTySubst e'}{e'''}$ for some $k'' < k'$
    and some $e'''$. Thus, $\Diverge{e'''}$ by the definition
    of $e$ in \QRef{conclusion} and the evaluation rules for \FGGm. With \QRef{IH} then
    $\Diverge{\tlSubst E'}$. By definition of $E$ in \QRef{conclusion} then $\Diverge{\tlSubst E}$
    as required.
  \end{EnumAlph}

  \renewcommand\LabelQualifier{exp-equiv-call-struct}
  \Case{\Rule{call-struct}}
  \begin{mathpar}
    \inferrule{
      \LabeledRule{
        \tdExpTrans{\pair{\fgTyEnv}{\fgEnv}}{g : t_S[\fgTyActuals]}{G} & \TextLabel{deriv-g} \\
        \tdMethods{m[\fgTyFormalsAux](\Foreach{x\,\fgTypeAux}{n})\fgTypeAux}{W}{\fgTyEnv}{t_S[\fgTyActuals]} & \TextLabel{methods}\\
        \tdCheckSubst{\fgTyEnv}{\fgTyFormalsAux}{\fgTyActualsAux}{\fgTySubst_1}{W'} & \TextLabel{subst}\\
        \tdExpTrans{\pair{\fgTyEnv}{\fgEnv}}{e_i: \fgTySubst_1 \fgTypeAux_i}{\expT_i}
        \quad(\forall i \in [n]) & \TextLabel{deriv-ei}
      }
    }{
      \tdExpTrans{\pair{\fgTyEnv}{\fgEnv}}{
        \BraceBelow{g.m[\fgTyActualsAux](\Foreach{e}{n})}{= e} : {\BraceBelow{\fgTySubst_1 \fgTypeAux}{=\fgType}}
      }
      {\BraceBelow{\mT{m}{t_S} \ \Quadr{W}{G}{W'}{\Tuple{\Foreach{\expT}{n}}}}{= E}}
      \qqquad \TextLabel{conclusion}
    }
  \end{mathpar}
  From the IH applied to \QRef{deriv-g} and \QRef{deriv-ei}
  \begin{igather}
    \LREquiv{
      \fgSubst \fgTySubst g
    }{
      \tlSubst G
    }{\fgTySubst t_S[\fgTyActuals]}{k} \QLabel{g-equiv}\\
    \LREquiv{
      \fgSubst \fgTySubst e_i
    }{
      \tlSubst E_i
    }{\fgTySubst \fgTySubst_1 \fgTypeAux_i}{k} \quad (\forall i \in [n]) \QLabel{ei-prime-equiv}
  \end{igather}
  Assume $\reducek{k'}{\fgSubst \fgTySubst e}{e'}$  for some $k' < k$. We first consider the following situation
  for some values $u, \MultiN{v}$:
  \begin{igather}
    \reducek{k''}{\fgSubst \fgTySubst g}{u} \QLabel{reduce-g}\\
    \reducek{k_i}{\fgSubst \fgTySubst e_i}{v_i} \QLabel{reduce-ei}\\
    \fgSubst \fgTySubst e \reduceSym^{k''+\smallsum{k_i}} u.m[\fgTySubst \fgTyActualsAux](\MultiN{v})
    \reduceSym^{k' - k'' - \smallsum{k_i}} e' \QLabel{reduce-e}
  \end{igather}
  with $k'' + \smallsum{k_i} \leq k'$. \QRef{reduce-g}, \QRef{reduce-ei}, and \QRef{reduce-e} are intermediate
  assumptions, which become true when we later prove the two implications of rule \Rule{equiv-exp}.\\
  We have from \QRef{g-equiv}, \QRef{reduce-g}, \QRef{ei-prime-equiv}, and \QRef{reduce-ei}
  \begin{igather}
    \reduceStar{\tlSubst G}{U} \textrm{~for some~} U \textrm{~with~}
    \LREquivVal{u}{U}{\fgTySubst t_S[\fgTyActuals]}{k - k''} \QLabel{reduce-G}\\
    (\forall i \in [n])~\reduceStar{\tlSubst E_i}{V_i} \textrm{~for some~} V_i \textrm{~with~}
    \LREquivVal{v_i}{V_i}{\fgTySubst\fgTySubst_1 \fgTypeAux_i}{k - k_i} \QLabel{reduce-Ei}
  \end{igather}
  From \QRef{methods} we get by inverting rule \Rule{methods-struct}:
  \begin{igather}
    \fgFunc[spec=m{[\fgTyFormalsAux']}(\MultiN{x\ \fgTypeAux'})\,\fgTypeAux', body=g'] \in \fgDecls \QLabel{func-def}\\
    \tdCheckSubst{\fgTyEnv}{\fgTyFormals}{\fgTyActuals}{\fgTySubst_2}{W} \QLabel{subst2}\\
    m[\fgTyFormalsAux](\MultiN{x\ \fgTypeAux})\,\fgTypeAux =
    \fgTySubst_2(m[\fgTyFormalsAux'](\MultiN{x\ \fgTypeAux'})\,\fgTypeAux') \QLabel{eq-m}
  \end{igather}
  From the assumption $\LREquivNoTy{\fgDecls}{k}{\tlMethTable}$ and \QRef{func-def}
  \begin{igather}
    \LREquivNoTy{
      \fgFunc[spec=m{[\fgTyFormalsAux']}(\MultiN{x\ \fgTypeAux'})\,\fgTypeAux', body=g']
    }{k}{
      \mT{m}{t_S}
    } \QLabel{m-equiv}
  \end{igather}
  Define
  \begin{igather}
    k''' := min(k - k'', k - \smallsum{k_i}) - 1 \QLabel{def-k3}
  \end{igather}
  We have $k' - k'' - \smallsum{k_i} < k''' + 1$ by the following reasoning:
  \begin{igather}
    \begin{array}{@{}r@{}c@{}l@{}}
      k''' + 1 & \ReasonAbove{=}{\QRef{def-k3}} & min(k - k'', k - \smallsum{k_i}) \\
               & = & k - max(k'', \smallsum{k_i}) \\
               & \geq & k - k'' - \smallsum{k_i} \\
               & \ReasonAbove{>}{k' < k} & k' - k'' - \smallsum{k_i}
    \end{array} \QLabel{ineq-k3}
  \end{igather}
  With \QRef{m-equiv} and $k''' < k$, we now want to use
  the implication from the premise of rule \Rule{equiv-method-decl}. We instantiate
  the universally quantified variables of the implication as follows:
  $k' = k''', \fgTyActuals = \fgTySubst(\fgTyActuals, \fgTyActualsAux), \MultiP{W} = \tlSubst W, \MultiQ{W'} = \tlSubst W', v=u,
  V=U, \MultiN{v} = \MultiN{v}, \MultiN{V} = \MultiN{V}$.
  Next, we prove the
  left-hand side of the implication.
  But first assume (see \QRef{subst2}, \QRef{subst}, \QRef{eq-m})
  \begin{igather}
    \fgTyFormals = \MultiP{\fgTyVar\,\fgType} \qquad
    \fgTyActuals = \MultiP{\fgType'} \qquad W = \MultiP{W} \QLabel{def-typarams1}\\
    \fgTyFormalsAux' = \MultiQ{\fgTyVarAux\,\fgType''} \qquad
    \fgTyActualsAux = \MultiQ{\fgType'''} \qquad W' = \MultiQ{W'} \QLabel{def-typarams2}\\
  \end{igather}
  and define
  \begin{igather}
    \fgTySubst_3 = \Angle{\MultiP{\subst{\fgTyVar_i}{\fgTySubst \fgType'_i}}, \MultiQ{\subst{\fgTyVarAux_i}{\fgTySubst\fgType'''_i}}} \QLabel{def-eta3}
  \end{igather}
  \begin{itemize}
  \item We start by showing the first two conjuncts of the implication's left-hand side.
    \begin{igather}
      \fgMakeSubst{\fgTyFormals, \fgTyFormalsAux'}{\fgTySubst(\fgTyActuals, \fgTyActualsAux)}{\fgTySubst_3} \wedge
      \LREquiv{
        \fgTySubst(\fgTyActuals,\fgTyActualsAux)
      }{
        \tlSubst \Pair{\Multi{W}}{\Multi{W'}}
      }{
        \fgTyFormals,\fgTyFormalsAux'
      }{
        k'''
      } \QLabel{lhs1}
    \end{igather}
    The left part of the conjunction follows from \QRef{def-eta3}.
    We then show
    $\LREquiv{
      \fgTySubst(\fgTyActuals,\fgTyActualsAux)
    }{
      \tlSubst \Pair{\Multi{W}}{\Multi{W'}}
    }{
      \fgTyFormals,\fgTyFormalsAux'
    }{
      k
    }$ by proving the two implications required to fulfill the premise of rule
    \Rule{equiv-bounded-typarams}. The right part of the conjunction in \QRef{lhs1} then follows via \Cref{lem:mono-typarams}.

    \begin{itemize}
    \item First implication:
      $\LREquiv{u_j}{U_j}{\fgTySubst \fgType_j'}{k} \implies \LREquiv{u_j}{(\tlSubst W_j)\,U_j}{\fgTySubst_3 \fgType_j}{k}$
      for all $j \in [p]$ and all $u_j, U_j$.\\
      From \QRef{subst2} and \Cref{lem:subst-equiv} we have
      $\LREquiv{\fgTySubst\fgTyActuals}{\tlSubst W}{\fgTySubst\fgTyFormals}{k}$
      Hence, with $\LREquiv{u_j}{U_j}{\fgTySubst \fgType_j'}{k}$ and the implication in the premise of rule
      \Rule{equiv-bounded-typarams}, we have
      $\LREquiv{u_j}{(\tlSubst W_j)\,U_j}{\Angle{\Multi{\subst{\fgTyVar_i}{\fgTySubst \fgType'_i}}}\fgTySubst\fgType_j}{k}$.
      From assumption~\ref{conv:fg-decls}, \QRef{func-def}, and \QRef{def-typarams1},
      we know that $\ftv{\fgType_j} \subseteq \{\Multi{\fgTyVar}\}$ and $\Multi{\fgTyVar}$ fresh, so
      $\Angle{\Multi{\subst{\fgTyVar_i}{\fgTySubst\fgType'_i}}}\fgTySubst\fgType_j = \fgTySubst_3 \fgType_j$.
      Thus
      $\LREquiv{u_j}{(\tlSubst W_j)\,U_j}{\fgTySubst_3 \fgType_j}{k}$ as required.
    \item Second implication:
      $\LREquiv{u_j}{U_j}{\fgTySubst \fgType_j'''}{k} \implies \LREquiv{u_j}{(\tlSubst W'_j)\,U_j}{\fgTySubst_3 \fgType''_j}{k}$
      for all $i \in [q]$ and all $u_j, U_j$.\\
      From \QRef{subst} and \Cref{lem:subst-equiv} we have
      $\LREquiv{\fgTySubst\fgTyActualsAux}{\tlSubst W'}{\fgTySubst\fgTyFormalsAux}{k}$. Hence, with
      $\LREquiv{u_j}{U_j}{\fgTySubst \fgType_j'''}{k}$, the implication
      in the premise of rule \Rule{equiv-bounded-typarams}, and \QRef{eq-m} then
      $\LREquiv{u_j}{(\tlSubst W_j')\,U_j}{\Angle{\Multi{\subst{\fgTyVarAux_i}{\fgTySubst\fgType_i'''}}}\fgTySubst\fgTySubst_2\fgType_j''}{k}$.
      We have with \QRef{subst2} and \QRef{def-typarams1} that $\fgTySubst_2 = \Angle{\Multi{\subst{\fgTyVar_i}{\fgType_i'}}}$.
      Because of assumption~\ref{conv:fg-decls}, \QRef{func-def}, and \QRef{def-typarams2}, we know that
      $\ftv{\fgType_j''} \subseteq \{\Multi{\fgTyVar}, \Multi{\fgTyVarAux}\}$ and $\Multi{\fgTyVar}, \Multi{\fgTyVarAux}$ fresh.
      Hence, $\Angle{\Multi{\subst{\fgTyVarAux_i}{\fgTySubst\fgType_i'''}}}\fgTySubst\fgTySubst_2\fgType_j'' =
      \Angle{\Multi{\subst{\fgTyVarAux_i}{\fgTySubst\fgType_i'''}}}\Angle{\Multi{\subst{\fgTyVar_i}{\fgTySubst\fgType_i'}}}\fgType_j''
      \ReasonAbove{=}{\QRef{def-eta3}} \fgTySubst_3\fgType_j''$.
      Thus
      $\LREquiv{u_j}{(\tlSubst W_j')\,U_j}{\fgTySubst_3\fgType_j''}{k}$ as required.
    \end{itemize}
    This finishes the proof of \QRef{lhs1}.

  \item We next show the third conjunct of the implication's left-hand side.
    \begin{igather}
      \LREquiv{u}{U}{t_S[\fgTySubst_3 \Multi{\fgTyVar}]}{k'''} \QLabel{lhs2}
    \end{igather}
    We have $t_S[\fgTySubst_3 \Multi{\fgTyVar}] = t_S[\fgTySubst\fgTyActuals]$ by \QRef{def-eta3} and \QRef{def-typarams1}.
    Hence, with \QRef{reduce-G}, \Cref{lem:val-equiv-implies-exp-equiv}, and \Cref{lem:mono-exp}, it suffices
    to show that $k''' \leq k - k''$. But this follows from construction of $k'''$ in \QRef{def-k3}.

  \item Finally, we show the fourth conjunct:
    \begin{igather}
      \LREquiv{v_i}{V_i}{\fgTySubst_3\fgTypeAux_i'}{k'''} (\forall i \in [n]) \QLabel{lhs3}
    \end{igather}
    By \QRef{subst}, \QRef{eq-m}, \QRef{def-typarams2} we have $\fgTySubst_1 = \Angle{\MultiQ{\subst{\fgTyVarAux_i}{\fgType'''_i}}}$.
    By \QRef{subst2} and \QRef{def-typarams1} we have $\fgTySubst_2 = \Angle{\MultiP{\subst{\fgTyVar_i}{\fgType'_i}}}$.
    Thus,
    \begin{igather}
      \fgTySubst\fgTySubst_1\fgTypeAux_i \ReasonAbove{=}{\QRef{eq-m}}
      \fgTySubst\fgTySubst_1\fgTySubst_2\fgTypeAux_i' \ReasonAbove{=}{\QRef{def-eta3}}
      \fgTySubst_3\fgTypeAux_i'
    \end{igather}
    For the last equation, note that $\ftv{\fgTypeAux_i'} \subseteq \{\Multi{\fgTyVar}, \Multi{\fgTyVarAux}\}$
    by assumption~\ref{conv:fg-decls} and \QRef{func-def}, \QRef{def-typarams1}, \QRef{def-typarams2}.
    Hence, with \QRef{reduce-Ei}, \Cref{lem:val-equiv-implies-exp-equiv}, and \Cref{lem:mono-exp}, it suffices
    to show that $k''' \leq k - k_i$. But this follows from construction of $k'''$ in \QRef{def-k3}.
  \end{itemize}
  Now \QRef{lhs1}, \QRef{lhs2}, and \QRef{lhs3} are the left-hand side of the implication of rule
  \Rule{equiv-method-decl}, which we get from \QRef{m-equiv}. The right-hand side of the implication then yields
  \begin{igather}
    \LREquiv{
      \BraceBelow{\Angle{\subst{x}{u}, \MultiN{\subst{x_i}{v_i}}}}{=: \fgSubst'}\fgTySubst_3 g'
    }{
      \mT{m}{t_S}\,\Quadr{\tlSubst W}{U}{\tlSubst W'}{\Tuple{\Multi{V}}}
    }{
      \fgTySubst_3\fgTypeAux'
    }{
      k'''
    } \QLabel{equiv-call1}
  \end{igather}
  From \QRef{reduce-G} we have $u = t_S[\fgTySubst\fgTyActuals]$ by inverting rule \Rule{equiv-struct}. Hence by
  \QRef{func-def}, \QRef{def-eta3}, and rule \Rule{fg-call}
  \begin{igather}
    u.m[\fgTySubst \fgTyActualsAux](\Multi{v}) \reduceSym \fgSubst'\fgTySubst_3 g' \QLabel{reduce-call}
  \end{igather}
  Also we have
  \begin{igather}
    \fgTySubst \fgType \ReasonAbove{=}{\QRef{conclusion}}
    \fgTySubst\fgTySubst_1\fgTypeAux \ReasonAbove{=}{\QRef{eq-m}}
    \fgTySubst\fgTySubst_1\fgTySubst_2\fgTypeAux'
    = \fgTySubst_3\fgTypeAux'
  \end{igather}
  where the last equation follows from
  \QRef{subst}, \QRef{eq-m}, \QRef{def-typarams2}, \QRef{subst2}, \QRef{def-typarams1} and
  $\ftv{\fgTypeAux'} \subseteq \{\Multi{\fgTyVar}, \Multi{\fgTyVarAux}\}$ with
  assumption~\ref{conv:fg-decls}.
  Thus, with \QRef{equiv-call1}, \QRef{reduce-call}, and \Cref{lem:source-reduce}
  \begin{igather}
    \LREquiv{
      u.m[\fgTySubst \fgTyActualsAux](\Multi{v})
    }{
      \mT{m}{t_S}\,\Quadr{\tlSubst W}{U}{\tlSubst W'}{\Tuple{\Multi{V}}}
    }{
      \fgTySubst\fgType
    }{
      k''' + 1
    } \QLabel{equiv-call2}
  \end{igather}
  By definition of $E$ in \QRef{conclusion} and with \QRef{reduce-G} and \QRef{reduce-Ei} we have
  \begin{igather}
    \tlSubst E \reduceStarSym
    \tlMethTable(\mT{m}{t_S})\,\Quadr{\tlSubst W}{U}{\tlSubst W'}{\Tuple{\Multi{V}}} \QLabel{eval-E}
  \end{igather}
  Also, we have by rules \Rule{tl-context} and \Rule{tl-method}
  \begin{igather}
    \mT{m}{t_S}\,\Quadr{\tlSubst W}{U}{\tlSubst W'}{\Tuple{\Multi{V}}} \reduceSym
    \tlMethTable(\mT{m}{t_S})\,\Quadr{\tlSubst W}{U}{\tlSubst W'}{\Tuple{\Multi{V}}} \QLabel{eval-Xmts}
  \end{igather}
  So far, we proved everything under the assumptions
  \QRef{reduce-g}, \QRef{reduce-ei}, \QRef{reduce-e}. We next consider
  the two implications of rule \Rule{equiv-exp}.
  \begin{EnumAlph}
  \item Assume $e' = v$ for some value $v$. Our goal is to prove that there exists some value $V$
    such that $\reduceStar{\tlSubst E}{V}$ and $\LREquivVal{v}{V}{\fgTySubst\fgType}{k-k'}$.
    Noting that \QRef{reduce-g}, \QRef{reduce-ei}, \QRef{reduce-e} hold, we have together with \QRef{reduce-call}
    \begin{igather}
      \fgSubst\fgTySubst e \reduceSym^{k'' + \smallsum{k_i}}
      u.m[\fgTySubst \fgTyActualsAux](\Multi{v}) \reduceSym^{k' - k'' - \smallsum{k_i}} v \QLabel{reduce-steps}
    \end{igather}
    with $k'' + \smallsum{k_i} < k'$. We have $k' - k'' - \smallsum{k_i} < k''' + 1$ by \QRef{ineq-k3}.
    Hence with \QRef{equiv-call2} and \QRef{reduce-steps}
    we know that there exists some value $V$ with
    \begin{igather}
      \mT{m}{t_S}\,\Quadr{\tlSubst W}{U}{\tlSubst W'}{\Tuple{\Multi{V}}} \reduceStarSym V  \QLabel{eval-Xmts2}\\
      \LREquivVal{v}{V}{\fgTySubst\fgType}{k''' + 1 - k' + k'' + \smallsum{k_i}} \QLabel{equiv-v-V}
    \end{igather}
    We have $k - k' \leq k''' + 1 - k' + k'' + \smallsum{k_i}$ by the following reasoning:
    \begin{igather}
      \begin{array}{@{}r@{}c@{}l@{}}
        k''' + 1 - k' + k'' + \smallsum{k_i}
        & \ReasonAbove{=}{\QRef{def-k3}} & min(k - k'', k - \smallsum{k_i}) - k' + k'' + \smallsum{k_i} \\
        & = & k - max(k'', \smallsum{k_i}) - k' + k'' + \smallsum{k_i} \\
        & \geq & k - k'' - \smallsum{k_i} - k' + k'' + \smallsum{k_i} \\
        & = & k - k'
      \end{array}
    \end{igather}
    With \QRef{equiv-v-V} and \Cref{lem:mono-exp} then
    $\LREquivVal{v}{V}{\fgTySubst\fgType}{k - k'}$.
    And from \QRef{eval-E}, \QRef{eval-Xmts}, \QRef{eval-Xmts2}, and \Cref{lem:determ-eval-tl} we have that
    $\reduceStar{\tlSubst E}{V}$.
  \item Assume $\Diverge{e'}$. We then have to show $\Diverge{\tlSubst E}$.
    \begin{CaseDistinction}{whether receiver, argument or method call diverges}
      \Case{receiver diverges} Then
      $\reducek{k'}{\fgSubst\fgTySubst g}{g''}$ and $\Diverge{g''}$.
      With \QRef{g-equiv} and $k' < k$ then $\Diverge{\tlSubst G}$,
      so by the definition of $E$ in \QRef{conclusion} we get
      $\Diverge{\tlSubst E}$.

      \Case{$j$-th argument diverges} Then
      $\reducek{k''}{\fgSubst\fgTySubst g}{u}$ and
      $\reducek{k_i}{\fgSubst\fgTySubst e_i}{v_i}$ for all $i < j$ and
      $\reducek{k_j}{\fgSubst\fgTySubst e_j}{e''}$ and
      $\Diverge{e''}$. With \QRef{ei-prime-equiv} and $k_j \leq k' < k$ we get $\Diverge{\tlSubst E_j}$.
      By definition of $E$ in \QRef{conclusion} then $\Diverge{\tlSubst E}$.

      \Case{method call diverges} Then we are in the situation that \QRef{reduce-g}, \QRef{reduce-ei}, and
      \QRef{reduce-e} hold. We then have
      \begin{igather}
        u.m[\fgTySubst \fgTyActualsAux](\MultiN{v}) \reduceSym^{k' - k'' - \smallsum{k_i}} e'
      \end{igather}
      Hence, with \QRef{ineq-k3}, \QRef{equiv-call2}, and the second implication in the premise of rule
      \Rule{equiv-exp}, we have that $\Diverge{\mT{m}{t_S}\,\Quadr{\tlSubst W}{U}{\tlSubst W'}{\Tuple{\Multi{V}}}}$.
      With \QRef{eval-E} and \QRef{eval-Xmts} and \Cref{lem:determ-eval-tl} then also $\Diverge{\tlSubst E}$ as required.
    \end{CaseDistinction}
  \end{EnumAlph}
  This finishes the proof for rule \Rule{call-struct}.
  \renewcommand\LabelQualifier{exp-equiv-call-iface}
  \Case{\Rule{call-iface}}
  \begin{mathpar}
    \inferrule{
      \LabeledRule{
        \tdExpTrans{\pair{\fgTyEnv}{\fgEnv}}{g : \fgType_I}{G} &\TextLabel{deriv-eprime}\\
        \methodSpecifications{\fgType_I} = \Foreach{R}{q} &\TextLabel{methods}\\
        R_j = m[\fgTyFormalsAux](\Foreach{x\,\fgTypeAux}{n})\fgTypeAux \quad(\textrm{for some}~j \in [q]) &\TextLabel{def-Sj}\\
        \tdCheckSubst{\fgTyEnv}{\fgTyFormalsAux}{\fgTyActualsAux}{\fgTySubst_1}{V} & \TextLabel{subst}\\
        \tdExpTrans{\pair{\fgTyEnv}{\fgEnv}}{e_i : \fgTySubst_1 \fgTypeAux_i}{\expT_i}
        \quad\noteForall{i \in [n]} & \TextLabel{deriv-ei}\\
        Y, \Foreach{X}{q}\textrm{~fresh}
      }
    }{
      \LabeledRule{
        \tdExpTrans{\pair{\fgTyEnv}{\fgEnv}}
        {\BraceBelow{g.m[\fgTyActualsAux](\Foreach{e}{n})}{= e} : \BraceBelow{\fgTySubst_1 \fgTypeAux}{= \fgType}}
        {E}
        &\TextLabel{conclusion}
      }
    }
  \end{mathpar}
  with
  \begin{igather}
    E = \CASE\ G \ \OF\ \Tuple{Y\Comma \Tuple{\Foreach{X}{q}}} \to X_j\ \Triple{Y}{V}{\Tuple{\Foreach{E}{n}}} \QLabel{def-E}
  \end{igather}
  From the IH applied to \QRef{deriv-eprime}, \QRef{deriv-ei}
  \begin{igather}
    \LREquiv{
      \fgSubst \fgTySubst g
    }{
      \tlSubst G
    }{\fgTySubst \fgType_I}{k} \QLabel{g-equiv}\\
    \LREquiv{
      \fgSubst \fgTySubst e_i
    }{
      \tlSubst E_i
    }{\fgTySubst \fgTySubst_1 \fgTypeAux_i}{k} \quad (\forall i \in [n]) \QLabel{ei-prime-equiv}
  \end{igather}
  Assume $\reducek{k'}{\fgSubst \fgTySubst e}{e'}$  for some $k' < k$. We first consider the following situation
  for some values $u, \MultiN{v}$:
  \begin{igather}
    \reducek{k''}{\fgSubst \fgTySubst g}{u} \QLabel{reduce-g}\\
    \reducek{k_i}{\fgSubst \fgTySubst e_i}{v_i} \QLabel{reduce-ei}\\
    \fgSubst \fgTySubst e \reduceSym^{k''+\smallsum{k_i}} u.m[\fgTySubst \fgTyActualsAux](\MultiN{v})
    \reduceSym^{k' - k'' - \smallsum{k_i}} e' \QLabel{reduce-e}
  \end{igather}
  with $k'' + \smallsum{k_i} \leq k'$. \QRef{reduce-g}, \QRef{reduce-ei}, and \QRef{reduce-e} are intermediate
  assumptions, which become true when we later prove the two implications of rule \Rule{equiv-exp}.\\
  We have from \QRef{g-equiv}, \QRef{ei-prime-equiv}, \QRef{reduce-g}, and \QRef{reduce-ei}
  \begin{igather}
    \reduceStar{\tlSubst G}{U} \textrm{~for some~} U \textrm{~with~}
    \LREquivVal{u}{U}{\fgTySubst \fgType_I}{k - k''} \QLabel{reduce-G}\\
    (\forall i \in [n])~\reduceStar{\tlSubst E_i}{V_i} \textrm{~for some~} V_i \textrm{~with~}
    \LREquivVal{v_i}{V_i}{\fgTySubst\fgTySubst_1 \fgTypeAux_i}{k - k_i} \QLabel{reduce-Ei}
  \end{igather}
  From \QRef{reduce-G} and \QRef{methods} we get by inverting rule \Rule{equiv-iface}
  \begin{igather}
    \exists \fgTypeAux_S = t_S[\fgTyActuals] \QLabel{def-tauS}\\
    U = \Pair{U'}{\Tuple{\MultiQ{U}}} \QLabel{U-eq}\\
    \forall \ell_1 < k - k'' \DOT \LREquivVal{u}{U'}{\fgTypeAux_S}{\ell_1} \QLabel{u-equiv}\\
    \forall \ell_2 < k - k'' \DOT \LREquiv{\methodLookup{m_j}{\fgTypeAux_S}}{U_j}{\fgTySubst R_j}{\ell_2} \QLabel{methodLookup-equiv}
  \end{igather}
  Hence we have by \QRef{def-tauS}, \QRef{methodLookup-equiv}, \QRef{def-Sj}, and rule \Rule{method-lookup}
  \begin{igather}
    \fgFunc[spec={\BraceBelow{m[\fgTyFormalsAux'](\MultiN{x\ \fgTypeAux'})\,\fgTypeAux'}{=: R'}}, body=e''] \in \fgDecls \QLabel{func}\\
    \fgMakeSubst{\fgTyFormals}{\fgTyActuals}{\fgTySubst_2} \QLabel{def-eta2}\\
    \methodLookup{m_j}{\fgTypeAux_S} = \Angle{x, t_S[\fgTyActuals], \fgTySubst_2 R', \fgTySubst_2 e''} \QLabel{methodLookup-eq}\\
    \fgTySubst_2 R' = \fgTySubst R_j = \fgTySubst (m[\fgTyFormalsAux](\MultiN{x_i\ \fgTypeAux_i})\,\fgTypeAux) \QLabel{Sprime-eq-Sj}
  \end{igather}
  Then by \QRef{methodLookup-equiv} and \QRef{methodLookup-eq}
  \begin{igather}
    \LREquiv{\Angle{x, t_S[\fgTyActuals], \fgTySubst_2 R', \fgTySubst_2 e''}}{U_j}{\fgTySubst R_j}{k - k'' - 1}
    \QLabel{star}
  \end{igather}
  Define $k''' := \min(k - k'' - 1, k - \smallsum{k_i} - 1)$. Then
  \begin{igather}
    k''' \leq k - k'' - 1 \QLabel{k3-ineq0}\\
    k''' < k \QLabel{k3-ineq1}\\
    k''' < k - k'' \QLabel{k3-ineq2}\\
    k''' < k - k_i \quad(\forall i \in [n]) \QLabel{k3-ineq3}\\
    k' - k'' - \smallsum{k_i} < k''' + 1 \QLabel{k3-ineq4}
  \end{igather}
  The first four of these claims are straightforward to verify. The last can be shown with
  the following reasoning:
  \begin{igather}
    \begin{array}{@{}r@{}c@{}l@{}}
      k''' + 1
      & = & k - \max(k'' + 1, \smallsum{k_i} + 1) + 1 \\
      & > & k - (k'' + 1 + \smallsum{k_i} + 1) + 1\\
      & = & k - 1 - k'' - \smallsum{k_i} \\
      & \ReasonAbove{\geq}{k' < k} & k' - k'' - \smallsum{k_i}
    \end{array}
  \end{igather}
  From \QRef{star} we get the implication in the premise of rule \Rule{equiv-method-dict-entry}.
  We now show that the left-hand side of the implication holds.
  The universally quantified variables of the rule's premise are instantiated as follows:
  $k' = k''', \fgTyActuals = \fgTySubst \fgTyActualsAux, W = \tlSubst V, v = u, V = U', \MultiN{v} = \MultiN{v}, \MultiN{V} = \MultiN{V}$.
  The variables in the conclusion are instantiated as follows:
  $x = x, \fgType_S = t_S[\fgTyActuals], m[\fgTyFormals](\MultiN{x_i\ \fgType_i})\,\fgType = \fgTySubst_2 R', e = \fgTySubst_2 e''$.
  The requirement $k''' \leq k - k'' - 1$ follows from \QRef{k3-ineq0}.\\
  We have from \QRef{subst} and \QRef{Sprime-eq-Sj} the first conjunct:
  \begin{igather}
    \fgMakeSubst{\fgTySubst \fgTyFormalsAux}{\fgTySubst \fgTyActualsAux}{
      \BraceBelow{\Angle{\Multi{\subst{\fgTyVar}{\fgTySubst\fgType}}}}{= \fgTySubst_4}
    } ~(\textrm{assuming } \fgTySubst_1 = \Angle{\Multi{\subst{\fgTyVar}{\fgType}}},
    \fgGetTyVars{\fgTyFormalsAux} = \Multi{\fgTyVar}, \fgTyActualsAux = \Multi{\fgType})\QLabel{eta4}
  \end{igather}
  From \QRef{subst} we get the second conjunct by \Cref{lem:subst-equiv}, \QRef{k3-ineq1}, by the assumptions
  $\LREquivNoTy{\fgDecls}{k}{\tlMethTable}$ and
  $\LREquiv{\fgTySubst}{\tlSubst}{\fgTyEnv}{k}$, and by \Cref{lem:mono-typarams}:
  \begin{igather}
    \LREquiv{
      \fgTySubst \fgTyActualsAux
    }{
      \tlSubst V
    }{
      \fgTySubst \fgTyFormalsAux
    }{k'''}
  \end{igather}
  With \QRef{k3-ineq2}, \QRef{u-equiv}, \QRef{def-tauS},
  and \Cref{lem:val-equiv-implies-exp-equiv} we get the third conjunct:
  \begin{igather}
    \LREquiv{u}{U'}{t_S[\fgTyActuals]}{k'''}
  \end{igather}
  With \QRef{k3-ineq3}, \QRef{reduce-Ei}, \Cref{lem:val-equiv-implies-exp-equiv}, and \Cref{lem:mono-exp} we have
  \begin{igather}
    \LREquiv{v_i}{V_i}{\fgTySubst\fgTySubst_1\fgTypeAux_i}{k'''} \quad (\forall i \in [n]) \QLabel{vi-equiv-Vi-1}
  \end{igather}
  We next prove
  \begin{igather}
    \fgTySubst \fgTySubst_1 \fgTypeAux_i = \fgTySubst_4 \fgTySubst \fgTypeAux_i \quad (\forall i \in [n]) \QLabel{subst-eq1}\\
    \fgTySubst \fgTySubst_1 \fgTypeAux = \fgTySubst_4 \fgTySubst \fgTypeAux \QLabel{subst-eq2}
  \end{igather}
  by induction on $\fgTypeAux_i$ or $\fgTypeAux$.
  The interesting case is where $\fgTypeAux_i$ or $\fgTypeAux$ is a type variable $\fgTyVar \in \dom{\fgTySubst_1} \cup \dom{\fgTySubst}$.
  As the $\Multi{\fgTyVar} = \dom{\fgTySubst_1} = \dom{\fgTySubst_4}$ are bound in $\fgTyFormalsAux$
  (see \QRef{eta4}), we may assume that
  $\Multi{\fgTyVar} \cap \dom{\fgTySubst} = \emptyset = \Multi{\fgTyVar} \cap \ftv{\fgTySubst}$.
  If $\fgTyVar \in \dom{\fgTySubst_1}$ then
  \begin{igather}
    \fgTySubst \fgTySubst_1 \fgTyVar \ReasonAbove{=}{\QRef{eta4}} \fgTySubst_4 \fgTyVar
    \ReasonAbove{=}{\dom{\fgTySubst_1} \cap \dom{\fgTySubst} = \emptyset} \fgTySubst_4 \fgTySubst \fgTyVar
  \end{igather}
  If $\fgTyVar \in \dom{\fgTySubst}$ then
  \begin{igather}
    \fgTySubst\fgTySubst_1 \fgTyVar \ReasonAbove{=}{\dom{\fgTySubst} \cap \dom{\fgTySubst_1} = \emptyset}
    \fgTySubst \fgTyVar \ReasonAbove{=}{\dom{\fgTySubst_1} \cap \ftv{\fgTySubst} = \emptyset}
    \fgTySubst_4\fgTySubst\fgTyVar
  \end{igather}

  We now get with
  \QRef{Sprime-eq-Sj} and \QRef{subst-eq1}
  that $\fgTySubst\fgTySubst_1 \fgTypeAux_i = \fgTySubst_4\fgTySubst_2\fgTypeAux_i'$. Hence
  with \QRef{vi-equiv-Vi-1} the fourth conjunct:
  \begin{igather}
    (\forall i \in [n])\quad \LREquiv{v_i}{V_i}{\fgTySubst_4\fgTySubst_2\fgTypeAux'_i}{k'''}
  \end{igather}
  Now the right-hand side of the implication of rule \Rule{equiv-method-dict-entry} yields with \QRef{star}
  \begin{igather}
    \LREquiv{
      \Angle{\subst{x}{u}, \MultiN{\subst{x_i}{v_i}}}\fgTySubst_4\fgTySubst_2 e''
    }{
      U_j\ \Triple{U'}{\tlSubst V}{\Tuple{\MultiN{V}}}
    }{
      \fgTySubst_4 \fgTySubst_2 \fgTypeAux'
    }{k'''} \QLabel{eta4-eta2-epprime}
  \end{igather}
  Define $\fgTySubst_3$ such that
  \begin{igather}
    \fgMakeSubst{\fgTyFormals,\fgTyFormalsAux'}{\fgTyActuals, \fgTySubst\fgTyActualsAux}{\fgTySubst_3} \QLabel{def-eta3}
  \end{igather}
  Then with \QRef{def-eta2} and \QRef{eta4}
  \begin{igather}
    \fgTySubst_4\fgTySubst_2 e'' = \fgTySubst_3 e'' \QLabel{eq-epprime}
  \end{igather}
  by induction on $e''$. The interesting case is the one for a type variable $\fgTyVar$. By assumption~\ref{conv:fg-decls}
  and \QRef{func}, we know that $\fgTyVar \in \fgGetTyVars{\fgTyFormals} \cup \fgGetTyVars{\fgTyFormalsAux'}$.
  Further we may assume that the type variables $\fgGetTyVars{\fgTyFormalsAux'}$ are fresh, and we have
  $\dom{\fgTySubst_2} = \fgGetTyVars{\fgTyFormals}$ by \QRef{def-eta2} and
  $\dom{\fgTySubst_4} = \fgGetTyVars{\fgTyFormalsAux'}$ by \QRef{eta4}.
  Thus, if $\fgTyVar \in \fgGetTyVars{\fgTyFormals}$ then $\fgTySubst_4\fgTySubst_2\fgTyVar = \fgTySubst_2\fgTyVar$
  because $\fgGetTyVars{\fgTyFormalsAux'}$ fresh, and $\fgTySubst_3\fgTyVar = \fgTySubst_2\fgTyVar$ by
  \QRef{def-eta2} and \QRef{def-eta3}. If $\fgTyVar \in \fgGetTyVars{\fgTyFormalsAux'}$ then
  $\fgTySubst_4\fgTySubst_2\fgTyVar = \fgTySubst_4\fgTyVar$ because $\fgGetTyVars{\fgTyFormalsAux'}$ fresh, and
  $\fgTySubst_3\fgTyVar = \fgTySubst_4\fgTyVar$ by \QRef{def-eta3} and \QRef{eta4}.
  \\
  With \QRef{subst-eq2} and \QRef{Sprime-eq-Sj} $\fgTySubst_4\fgTySubst_2 \fgTypeAux' = \fgTySubst\fgTySubst_1\fgTypeAux$.
  Hence we have with \QRef{eq-epprime}, \QRef{eta4-eta2-epprime}
  \begin{igather}
    \LREquiv{
      \Angle{\subst{x}{u}, \MultiN{\subst{x_i}{v_i}}}\fgTySubst_3 e''
    }{
      U_j\ \Triple{U'}{\tlSubst V}{\Tuple{\MultiN{V}}}
    }{
      \fgTySubst\fgTySubst_1 \fgTypeAux
    }{k'''} \QLabel{equiv-body}
  \end{igather}
  From \QRef{def-tauS} and \QRef{u-equiv} we get by inverting rule \Rule{equiv-struct} that
  $u = t_S[\fgTyActuals]\{\ldots\}$.
  Hence by rule \Rule{fg-call} with \QRef{func} and \QRef{def-eta3}
  \begin{igather}
    u.m[\fgTySubst\fgTyActualsAux](\MultiN{v}) \reduceSym
    \Angle{\subst{x}{u}, \MultiN{\subst{x_i}{v_i}}}\fgTySubst_3 e''
  \end{igather}
  Then with \QRef{equiv-body} and \Cref{lem:source-reduce}
  \begin{igather}
    \LREquiv{
      u.m[\fgTySubst\fgTyActualsAux](\MultiN{v})
    }{
      U_j\ \Triple{U'}{\tlSubst V}{\Tuple{\MultiN{V}}}
    }{
      \fgTySubst\fgTySubst_1 \fgTypeAux
    }{k'''+1} \QLabel{equiv-call}
  \end{igather}
  We also have
  \begin{igather}
    \tlSubst E
    \ReasonAbove{\reduceStarSym}{\QRef{reduce-G}, \QRef{def-E}}
    \CASE\ U \ \OF\ \Tuple{Y\Comma \Tuple{\Foreach{X}{q}}} \to X_j\ \Triple{Y}{V}{\Tuple{\Foreach{E}{n}}}\\
    \ReasonAbove{\reduceSym}{\QRef{U-eq}}
    U_j\ \Triple{U'}{\tlSubst V}{\tlSubst \Tuple{\MultiN{E}}} \\
    \ReasonAbove{\reduceStarSym}{\QRef{reduce-Ei}}
    U_j\ \Triple{U'}{\tlSubst V}{\Tuple{\MultiN{V}}}
    \QLabel{eval-E}
  \end{igather}
  So far, we proved everything under the assumptions
  \QRef{reduce-g}, \QRef{reduce-ei}, \QRef{reduce-e}. We next consider
  the two implications of rule \Rule{equiv-exp}.
  \\
  \begin{EnumAlph}
  \item Assume $e' = v$ for some value $v$.
    Then \QRef{reduce-g}, \QRef{reduce-ei}, and \QRef{reduce-e} hold.
    We now need to show that there exists some $W$ with
    $\tlSubst E \reduceStarSym W$ and $\LREquivVal{v}{W}{\fgTySubst \fgType}{k - k'}$.
    We have $k' - k'' - \smallsum{k_i} < k''' + 1$ by \QRef{k3-ineq4}
    Also, we have with \QRef{reduce-e} that
    \begin{igather}
      u.m[\fgTySubst \fgTyActualsAux](\MultiN{v}) \reduceSym^{k' - k'' - \smallsum{k_i}} v
    \end{igather}
    Hence,  \QRef{equiv-call} gives us the existence of some $W$ such that
    \begin{igather}
      U_j\ \Triple{U'}{\tlSubst V}{\Tuple{\MultiN{V}}} \reduceStarSym W \QLabel{eval-to-W}\\
      \LREquiv{v}{W}{\fgTySubst \fgTySubst_1 \fgTypeAux}{k''' + 1 - (k' - k'' - \smallsum{k_i})}
    \end{igather}
    We get $k - k' \leq k''' + 1 - (k' - k'' - \smallsum{k_i})$ by
    \begin{igather}
      \begin{array}{@{}r@{~}l@{}}
        & k''' + 1 - (k' - k'' - \smallsum k_i) \\
        = & k - max(k'' + 1, \smallsum{k_i} + 1) + 1 - k' + k'' + \smallsum{k_i} \\
        = & k - max(k'', \smallsum{k_i}) - k' + k'' + \smallsum{k_i} \\
        \geq & k - (k'' + \smallsum{k_i}) - k' + (k'' + \smallsum{k_i}) \\
        = & k - k'
      \end{array}
    \end{igather}
    Hence by \Cref{lem:mono-exp}
    \begin{igather}
      \LREquiv{v}{W}{\fgTySubst \fgTySubst_1 \fgTypeAux}{k - k'}
    \end{igather}
    By \QRef{conclusion} $\fgTySubst_1 \fgTypeAux = \fgType$ so
    $\LREquiv{v}{W}{\fgTySubst \fgType}{k - k'}$ and with
    \QRef{eval-E} and \QRef{eval-to-W} $\tlSubst E \reduceStarSym W$.
  \item Assume $\Diverge{e'}$. We then have to show $\Diverge{\tlSubst E}$.
    \begin{CaseDistinction}{whether receiver, argument or method call diverges}
      \Case{receiver diverges} Then
      $\reducek{k'}{\fgSubst\fgTySubst g}{g'}$ and $\Diverge{g'}$.
      With \QRef{g-equiv} and $k' < k$ then $\Diverge{\tlSubst G}$,
      so by the definition of $E$ in \QRef{def-E} we get
      $\Diverge{\tlSubst E}$.
      \Case{$j$-th argument diverges} Then
      $\reducek{k''}{\fgSubst\fgTySubst g}{u}$.
      By \QRef{g-equiv} and rule \Rule{equiv-iface} we know that
      $U = \Pair{U'}{\MultiQ{U}}$ for some $U', \MultiQ{U}$.
      Hence
      \begin{igather}
        \tlSubst E \reduceStarSym U_j\ \Triple{U'}{\tlSubst V}{\tlSubst \Tuple{\MultiN{E}}}
        \QLabel{eval-subst-E}
      \end{igather}
      Because the $j$-th argument diverges, we also have
      $\reducek{k_i}{\fgSubst\fgTySubst e_i}{v_i}$ for all $i < j$ and
      $\reducek{k_j}{\fgSubst\fgTySubst e_j}{e''}$ and
      $\Diverge{e''}$.
      With \QRef{ei-prime-equiv} we get $\Diverge{\tlSubst E_j}$, so with \QRef{eval-subst-E} also
      $\Diverge{\tlSubst E}$.
      \Case{method call diverges} Then we are in the situation that \QRef{reduce-g}, \QRef{reduce-ei}, and
      \QRef{reduce-e} hold. Thus, we get with \QRef{reduce-e}, \QRef{equiv-call}, \QRef{k3-ineq4}
      \begin{igather}
        u.m[\fgTySubst \fgTyActualsAux](\MultiN{v}) \reduceSym^{k' - k'' - \smallsum{k_i}} e'\\
        \LREquiv{
          u.m[\fgTySubst\fgTyActualsAux](\MultiN{v})
        }{
          U_j\ \Triple{U'}{\tlSubst V}{\Tuple{\MultiN{V}}}
        }{
          \fgTySubst\fgTySubst_1 \fgTypeAux
        }{k'''+1}
        \\
        k' - k'' - \smallsum{k_i} < k''' + 1
      \end{igather}
      Hence $\Diverge{U_j\ \Triple{U'}{\tlSubst V}{\Tuple{\MultiN{V}}}}$ by the implication in the
      premise of rule \Rule{equiv-exp}. So by \QRef{eval-E} also $\Diverge{\tlSubst E}$ as required.
    \end{CaseDistinction}
  \end{EnumAlph}
  \renewcommand\LabelQualifier{exp-equiv-sub}
  \Case{\Rule{sub}}
  \begin{mathpar}
    \inferrule[sub]{
      \tdExpTrans{\pair{\fgTyEnv}{\fgEnv}}{e : \fgTypeAux}{E'}\\
      \tdUpcast{\fgTyEnv}{\subtypeOf{\fgTypeAux}{\fgType}}{V}
    }{
      \tdExpTrans{\pair{\fgTyEnv}{\fgEnv}}{e : \fgType}{\BraceBelow{V \ E'}{= E}}
    }
  \end{mathpar}
  From the IH then $\LREquiv{\fgSubst\fgTySubst e}{\tlSubst E'}{\fgTySubst\fgTypeAux}{k}$.
  From \Cref{subtyping-equiv} we get
  $\LREquiv{\fgSubst\fgTySubst e}{(\tlSubst V)\,\tlSubst E'}{\fgTySubst\fgType}{k}$
  with $\tlSubst E = (\tlSubst V)\,(\tlSubst E')$ as required.
\end{CaseDistinctionExplicit} \MyQED

\paragraph{Proof of \Cref{lem:method-equiv}}
\label{sec:lem:method-equiv}

\renewcommand\LabelQualifier{lem:method-equiv}
We proceed by induction on $k$. For $k = 0$,
we first note that $\LREquiv{e}{E}{\fgType}{0}$ holds for any $e, E, \fgType$ because
the two implications in the premise of rule \Rule{equiv-exp} hold trivially.
Thus, we get
$\LREquivNoTy{D}{0}{\mT{m}{t_S}}$ for all $D = \fgFunc[spec=mM] \in \fgDecls$
by rule \Rule{equiv-method-decl}.
Hence $\LREquivNoTy{\fgDecls}{0}{\tlMethTable}$ by rule \Rule{equiv-decls}.

Now assume
$\LREquivNoTy{\fgDecls}{k}{\vbMethTL}$ (IH)
for some $k$ and prove $\LREquivNoTy{\fgDecls}{k+1}{\vbMethTL}$.
By rule \Rule{equiv-decls}, we need to show $\LREquivNoTy{D}{k+1}{\mT{m}{t_S}}$ for all
\begin{igather}
  D = \fgFunc[spec=m{[\fgTyFormals']}(\MultiN{x\ \fgType})\,\fgType] \in \fgDecls  \QLabel{D}
\end{igather}
Thus, we assume the left-hand side of the implication in the premise of rule \Rule{equiv-method-decl}
and then show the right-hand side of the implication. More specifically, let
\begin{igather}
  \fgTyFormals = \MultiP{\fgTyVar_i\,\fgTypeAux_i} \qquad
  \fgTyFormals' = \MultiQ{\fgTyVarAux_i\,\fgTypeAux'_i} \qquad
  \fgTyFormalsAux = \fgTyFormals, \fgTyFormals' = \MultiP{\fgTyVar_i\,\fgTypeAux_i}\,\MultiQ{\fgTyVarAux_i\,\fgTypeAux'_i}
\end{igather}
and assume for arbitrary
$k' < k + 1, \fgTyActuals = \MultiP{\fgTypeAux''}, \fgTyActuals' = \MultiQ{\fgTypeAux'''}, \MultiP{W}, \MultiQ{W'}, u, U, \MultiN{v}, \MultiN{V}$
the left-hand side of the implication:
\begin{igather}
  \fgMakeSubst{\fgTyFormalsAux}{\fgTyActuals, \fgTyActuals'}{\fgTySubst}
  \textrm{~with~}
  \fgTySubst = \Angle{\MultiP{\subst{\fgTyVar_i}{\fgTypeAux_i''}}\,\MultiQ{\subst{\fgTyVarAux_i}{\fgTypeAux_i'''}}} \QLabel{def-eta}\\
  \LREquiv{\fgTyActuals,\fgTyActuals'}{\Pair{\MultiP{W}}{\MultiQ{W'}}}{\fgTyFormalsAux}{k'} \QLabel{typarams-equiv}\\
  \LREquiv{u}{U}{t_S[\fgTySubst\MultiP{\fgTyVar}]}{k'} \QLabel{u-equiv}\\
  \LREquiv{v_i}{V_i}{\fgTySubst \fgType_i}{k'} \quad (\forall i \in [n]) \QLabel{vi-equiv}
\end{igather}
From this we need to prove the following goal:
\begin{igather}
  \LREquiv{
    \BraceBelow{\Angle{\subst{x}{u}, \MultiN{\subst{x_i}{v_i}}}}{=: \fgSubst} \fgTySubst e
  }{
    \mT{m}{t_S}\ \Quadr{\Tuple{\MultiP{W}}}{U}{\Tuple{\MultiQ{W'}}}{\Tuple{\MultiN{V}}}
  }{
    \fgTySubst\fgType
  }{
    k'
  } \QLabel{goal}
\end{igather}
Define
\begin{igather}
  \tlSubst = \Angle{\MultiP{\subst{\xTy{\fgTyVar_i}}{W_i}}, \MultiQ{\subst{\xTy{\fgTyVarAux_i}}{W_i'}},
    \subst{X}{U}, \MultiN{\subst{X_i}{V_i}}} \QLabel{def-rho} \\
  \fgTyEnv = \{ \MultiP{\fgTyVar_i : \fgTypeAux_i},\,\MultiQ{\fgTyVarAux_i : \fgTypeAux_i'} \} \\
  \fgEnv = \{ x: t_S[\MultiP{\fgTyVar}], \MultiN{x_i : \fgType_i} \}
\end{igather}
Then with \QRef{def-eta}, \QRef{typarams-equiv}, and rule \Rule{equiv-ty-subst}
\begin{igather}
  \LREquiv{\fgTySubst}{\tlSubst}{\fgTyEnv}{k'}  \QLabel{tyenv-equiv}
\end{igather}
And with \QRef{u-equiv}, \QRef{vi-equiv}, the definition of $\fgSubst$ in \QRef{goal}, and rule \Rule{equiv-val-subst}
\begin{igather}
  \LREquiv{\fgSubst}{\tlSubst}{\fgTySubst \fgEnv}{k'} \QLabel{valenv-equiv}
\end{igather}
From the assumption $\tdMethTrans{D}{\mT{m}{t_S} = V}$ we get by inverting rule \Rule{method}
\begin{igather}
  \tdExpTrans{\Angle{\fgTyEnv,\fgEnv}}{e : \fgType}{E} \QLabel{type-e}\\
  V = \tllambda{\Quadr{\Tuple{\MultiP{\xTy{\fgTyVar_i}}}}{X}{\Tuple{\MultiQ{\xTy{\fgTyVarAux_i}}}}{\Tuple{\MultiN{X}}}}{E} \QLabel{def-V}
\end{igather}
With $k' < k+1$ we have $k' \leq k$. With the IH and~\Cref{lem:mono-prog} then
\begin{igather}
  \LREquivNoTy{\fgDecls}{k'}{\tlMethTable} \QLabel{prog-equiv}
\end{igather}
\QRef{prog-equiv}, \QRef{tyenv-equiv}, \QRef{valenv-equiv} and \QRef{type-e} are the requirements
of \Cref{lem:exp-equiv}. The lemma then yields
\begin{igather}
  \LREquiv{\fgSubst\fgTySubst e}{\tlSubst E}{\fgTySubst\fgType}{k'} \QLabel{equiv-E}
\end{igather}
We also have
\begin{igather}
  \mT{m}{t_S}\ \Quadr{\Tuple{\MultiP{W}}}{U}{\Tuple{\MultiQ{W'}}}{\Tuple{\MultiN{V}}} \reduceSym
  V\ \Quadr{\Tuple{\MultiP{W}}}{U}{\Tuple{\MultiQ{W'}}}{\Tuple{\MultiN{V}}} \reduceStarSym
  \tlSubst E
\end{igather}
where the first reduction follows from assumption $\tlMethTable(\mT{m}{t_S}) = V$ and rule \Rule{tl-method},
the remaining steps by \QRef{def-V} and \QRef{def-rho}.
With \QRef{equiv-E} and \Cref{lem:target-reduce} we then get \QRef{goal} as required. \MyQED

\paragraph{Proof of \Cref{thm:prog-equiv}}
\label{sec:proof-crefthm:pr-equ}

We first prove that the assumptions of the theorem imply $\LREquiv{e}{E}{\fgType}{k}$ for any $k$.
$\fgDecls$ and $\tlMethTable$ are the declarations and the substitution whose existence we assumed
globally. Obviously, they meet the requirements of Assumption~\ref{conv:fg-decls}.

Assume $k \in \Nat$.
From \Cref{lem:method-equiv} we get $\LREquivNoTy{\fgDecls}{k}{\vbMethTL}$. By the assumption
$\tdProgTrans{\ForeachN{D} \ \FUNC\ \MAIN () \{ \_ = e \}} {\LET\ \Multi{X_i = V_i} \ \IN\ E}$,
by inverting rule \Rule{prog}, and by the assumption that $e$ has type $\fgType$, we find
$\tdExpTrans{\pair{\fgEmptyEnv}{\fgEmptyEnv}}{e : \fgType}{E}$.
\Cref{lem:exp-equiv} then yields $\LREquiv{e}{E}{\fgType}{k}$ as required.

From $\LREquiv{e}{E}{\fgType}{k}$ for any $k$ and the two implications in the premise
of rule \Rule{equiv-exp}, we then get the two claims needed to show. \MyQED

\subsubsection{Equivalence Between Different Translations}
\label{sec:pres-dynam-semant-2-app}

\begin{lemma}
  \label{lem:no-lambda}
  If $\LREquivVal{v}{\Tuple{\Multi{V}}}{\fgType}{k}$ then none of the $V_i$ is a lambda.
\end{lemma}

\begin{proof}
  \begin{CaseDistinction}{on the last rule in the derivation of $\LREquivVal{v}{\Tuple{\Multi{V}}}{\fgType}{k}$}
    \Case{rule \Rule{equiv-struct}}
      Then we know that $v = \fgType_S\{\Multi{v}\}$ and $\fgType = \fgType_S$ and for all $i$ exists some
      $\fgTypeAux_i$ with $\LREquivVal{v_i}{V_i}{\fgTypeAux_i}{k}$. But then obviously $V_i$ cannot
      be a lambda.
      \Case{rule \Rule{equiv-iface}} Obvious.
  \end{CaseDistinction} \MyQED
\end{proof}

\begin{lemma}
  \label{lem:all-lambda}
  If $\LREquivVal{v}{\Pair{U}{\Tuple{\Multi{W}}}}{\fgType_I}{k}$ with $k > 0$, then
  all $W_i$ are lambdas.
\end{lemma}
\begin{proof}
  \renewcommand\LabelQualifier{lem:all-lambda}
  The derivation of $\LREquivVal{v}{\Pair{U}{\Tuple{\Multi{W}}}}{\fgType_I}{k}$ ends with rule
  \Rule{equiv-iface}. The premise of the rule gives us for each $W_i$
  \begin{igather}
      \LREquiv{\methodLookup{\methodName{R_i}}{\fgTypeAux_S}}{W_i}{R_i}{k_2} \QLabel{eq:1}
  \end{igather}
  for some method signature $R_i$, struct type $\fgTypeAux_S$ and all $k_2 < k$.
  As $k > 0$ we know that \QRef{eq:1} holds for at least one $k_2$. Further, the derivation of \QRef{eq:1}
  ends with rule \Rule{equiv-method-dict-entry} and this rule requires that $W_i$ is a lambda. \MyQED
\end{proof}

\begin{lemma}
  \label{lem:form-equiv-iface}
  If $\LREquivVal{v}{V}{t_I[\Multi{\fgType}]}{k}$ for all $k \in \Nat$, then
  $V = \Pair{U}{\Tuple{\Multi[n]{W}}}$ where $n$ is the number of methods defined by $t_I$,
  $v = \fgType_S\{\Multi{v}\}$, and
  $\LREquivVal{v}{U}{\fgType_S}{k}$ for all $k \in \Nat$.
\end{lemma}
\begin{proof}
  \renewcommand\LabelQualifier{lem:form-equiv-iface}
  The derivation of $\LREquivVal{v}{V}{t_I[\Multi{\fgType}]}{k}$ ends with
  rule \Rule{equiv-iface} for any $k \in \Nat$. Also for all $k \in \Nat$,
  the conclusion of this rule requires
  $V = \Pair{U}{\Tuple{\Multi[n]{W}}}$,
  the premise of this rule states that interface $t_I$ has $n$ methods and further gives us
  \begin{igather}
    \exists \fgTypeAux_s . \forall k_1 < k . \LREquivVal{v}{U}{\fgTypeAux_S}{k_1} \QLabel{eq:1}
  \end{igather}
  Obviously, $\LREquivVal{v}{U}{\fgTypeAux_S}{k_1}$ ends with rule \Rule{equiv-struct}.
  Because value $v$ must have the form $v = \fgType_S\{\Multi{v}\}$, we then know that
  the existentially quantified $\fgTypeAux_S$ is the same as $\fgType_S$. Because
  \QRef{eq:1} holds for any $k \in \Nat$, we then get
  $\LREquivVal{v}{U}{\fgType_S}{k}$ for all $k \in \Nat$ as required. \MyQED
\end{proof}

\begin{lemma}
  \label{lem:fgg-equiv-implies-erase-eq}
  If $\LREquivVal{v}{V}{\fgType}{k}$ and $\LREquivVal{v}{V'}{\fgType}{k}$ for any $k \in \Nat$, then
  $\Erase{V} = \Erase{V'}$.
\end{lemma}

\begin{proof}
  \renewcommand\LabelQualifier{lem:fgg-equiv-implies-val-equiv}
  Define a measure function
  \[
    \Measure(v, \fgType) =
  \begin{cases}
    (|v|, 0) & \textrm{if }\fgType\textrm{ is a struct type}   \\
    (|v|, 1) & \textrm{if }\fgType\textrm{ is an interface type} \\
    (|v|, 2) & \textrm{if }\fgType\textrm{ is a type variable}
  \end{cases}
  \]
  and proceed by induction on $\Measure(v, \fgType)$. We first note that
  the derivations of $\LREquivVal{v}{V}{\fgType}{k}$ all end with the same rule, independent from $k \in \Nat$.
  \begin{CaseDistinction}{on the last rule in the derivations of $\LREquivVal{v}{V}{\fgType}{k}$}
    \Case{rule \Rule{equiv-struct}}
    Then $\fgType$ is a struct type, so
    the derivations of $\LREquivVal{v}{V'}{\fgType}{k}$ also all end with \Rule{equiv-struct}. Thus we have
    \begin{igather}
      v = t_S[\Multi{\fgType}]\{ \Multi[n]{v} \} \\
      V = \Tuple{\Multi[n]{V}}\\
      V' = \Tuple{\Multi[n]{V'}} \\
      \TYPE\ t_S[\Multi{\fgTyVar\,\fgType_I}] \ \STRUCT\ \{ \Foreach{f \ \fgTypeAux}{n} \} \in  \ForeachN{D}\\
      \fgTySubst = \MultiSubst{\fgTyVar}{\fgType}\\
      (\forall i \in [n])~\LREquivVal{v_i}{V_i}{\fgTySubst \fgTypeAux_i}{k} \QLabel{eq:1}\\
      (\forall i \in [n])~\LREquivVal{v_i}{V_i'}{\fgTySubst \fgTypeAux_i}{k} \QLabel{eq:2}
    \end{igather}
    As this holds for any $k \in \Nat$ and we have $\Measure(v_i, \fgTySubst\fgTypeAux_i) < \Measure(v, \fgType)$,
    we may apply the IH to \QRef{eq:1} and \QRef{eq:2} and get
    $\Erase{V_i} = \Erase{V_i'}$ for all $i \in [n]$.
    Then $\Erase{V} = \Erase{V'}$ follows by definition of $\EraseSym$.

    \Case{rule \Rule{equiv-iface}}
    Then $\fgType$ is interface type. With \Cref{lem:form-equiv-iface} then for some $\fgTypeAux_S$
    and $n$
    \begin{igather}
      V = \Pair{U}{\Tuple{\Multi[n]{W}}}  \\
      V' = \Pair{U'}{\Tuple{\Multi[n]{W'}}} \\
      (\forall k \in \Nat) . \LREquivVal{v}{U}{\fgTypeAux_S}{k} \QLabel{eq:3}\\
      (\forall k \in \Nat) . \LREquivVal{v}{U'}{\fgTypeAux_S}{k} \QLabel{eq:4}
    \end{igather}
    Noting that $\Measure(v, \fgTypeAux_S) < \Measure(v, \fgType)$, we apply the IH to
    \QRef{eq:3} and \QRef{eq:4} and get
    $\Erase{U} = \Erase{U'}$.
    With \Cref{lem:all-lambda} applied to assumptions
    $(\forall k \in \Nat) . \LREquivVal{v}{V}{\fgType}{k}$ and
    $(\forall k \in \Nat) . \LREquivVal{v}{V'}{\fgType}{k}$,
    we know that all
    $W_i, W_i'$ are lambdas. Thus by definition of $\EraseSym$
    \begin{igather}
      \Erase{V} = \Pair{\Erase{U}}{\Tuple{\Multi[n]{\EraseLam}}} = \Erase{V'}
    \end{igather}
  \end{CaseDistinction} \MyQED
\end{proof}

\begin{lemma}
  \label{lem:fgg-equiv-implies-erase-eq2}
  If $\LREquivVal{v}{V}{\fgType}{k}$ and $\LREquivVal{v}{V'}{\fgType'}{k}$ for any $k \in \Nat$, then
  $\EraseTy{\fgType}{V} = \EraseTy{\fgType'}{V'}$.
\end{lemma}

\begin{proof}
  \renewcommand\LabelQualifier{lem:fgg-equiv-implies-erase-eq2}
  We label the assumptions:
  \begin{igather}
    (\forall k \in \Nat) . \LREquivVal{v}{V}{\fgType}{k}   \QLabel{eq:ass1} \\
    (\forall k \in \Nat) . \LREquivVal{v}{V'}{\fgType'}{k} \QLabel{eq:ass2}
  \end{igather}
  We then perform a case distinction on the form of $\fgType$ and $\fgType'$. Note that
  neither of them can be a type variable, otherwise \QRef{eq:ass1} and \QRef{eq:ass2} would
  not be derivable.
  \begin{CaseDistinction}{on the forms of $\fgType$ and $\fgType'$}
    \Case{$\fgType$ and $\fgType'$ are both struct types}
    Then all derivations of \QRef{eq:ass1} and \QRef{eq:ass2} end with rule \Rule{equiv-struct}.
    Hence $\fgType = \fgType'$. Thus $\Erase{V} = \Erase{V'}$ by \Cref{lem:fgg-equiv-implies-erase-eq}.
    But by definition of $\EraseSym$, we also have $\EraseTy{\fgType}{V} = \Erase{V}$ and
    $\EraseTy{\fgType'}{V'} = \Erase{V}$.

    \Case{$\fgType$ is a struct type and $\fgType'$ is an interface type}
    Then all derivations of \QRef{eq:ass1} end with rule \Rule{equiv-struct},
    so we know that $v = \fgType\{\Multi{v}\}$.
    Then we get with \Cref{lem:form-equiv-iface} and \QRef{eq:ass2}
    \begin{igather}
      V' = \Pair{U}{W}\\
      (\forall k \in \Nat) . \LREquivVal{v}{U}{\fgType}{k}
    \end{igather}
    With \QRef{eq:ass1} and \Cref{lem:fgg-equiv-implies-erase-eq} and the definition of $\EraseSym$ then
    \begin{igather}
      \EraseTy{\fgType}{\Erase{V}} = \Erase{V} = \Erase{U} = \EraseTy{\fgType'}{V'}
    \end{igather}

    \Case{$\fgType$ is an interface type and $\fgType'$ is a struct type}
    Analogously to the preceding case.

    \Case{$\fgType$ and $\fgType'$ are both interface types}
    Then with \Cref{lem:form-equiv-iface} and \QRef{eq:ass1} and \QRef{eq:ass2}
    \begin{igather}
      v = \fgTypeAux_S\{\Multi{v}\}\\
      V = \Pair{U}{W}\\
      V' = \Pair{U'}{W'}\\
      (\forall k \in \Nat) . \LREquivVal{v}{U}{\fgTypeAux_S}{k}   \\
      (\forall k \in \Nat) . \LREquivVal{v}{U'}{\fgTypeAux_S}{k}
    \end{igather}
    Now \Cref{lem:fgg-equiv-implies-erase-eq} and the definition of $\EraseSym$
    \begin{igather}
      \EraseTy{\fgType}{V} = \Erase{U} = \Erase{U'} = \EraseTy{\fgType'}{V'}
    \end{igather}
    as required.
  \end{CaseDistinction} \MyQED
\end{proof}

\paragraph{Proof of \Cref{thm:determ-result}}
\label{sec:proof-thm-determ-result}

From $\tdProgTrans{P}{\LET\ \Multi{X_i = V_i} \ \IN\ E}$ and
$\tdProgTrans{P}{\LET\ \Multi{X_i' = V_i'} \ \IN\ E'}$ and
$e$ having type $\fgType$ and $\fgType'$, respectively, we get
\begin{igather}
  \tdExpTrans{\pair{\EmptyFgEnv}{\EmptyFgEnv}}{e : \fgType}{E} \\
  \tdExpTrans{\pair{\EmptyFgEnv}{\EmptyFgEnv}}{e : \fgType'}{E'}
\end{igather}
With Corollary~\ref{cor:fgg-soundness}, we get that either $e$ reduces to some value $v$ or diverges.

We now start with the first claim. Assume
$\reduceTLN{\vbMethTL}{E}{V}$ for some $V$. Then $e$ must reduce to some value $v$ because of
\Cref{thm:prog-equiv}. Again with \Cref{thm:prog-equiv} and with \Cref{lem:determ-eval-tl}:
\begin{igather}
  \LREquivVal{v}{V}{\fgType}{k} \quad (\forall k \in \Nat) \QLabel{eq:redV}\\
  \reduceTLN{\vbMethTL'}{E}{V'} \quad\textrm{for some }V'\\
  \LREquivVal{v}{V'}{\fgType'}{k} \quad (\forall k \in \Nat) \QLabel{eq:redVprime}\\
\end{igather}
Applying \Cref{lem:fgg-equiv-implies-erase-eq2} yields
$\EraseTy{\fgType}{V} = \EraseTy{\fgType'}{V'}$ as required.

For the second claim, we assume that $E$ diverges. With \Cref{thm:prog-equiv}, we know that $e$ must
diverge as well. Again with \Cref{thm:prog-equiv} we get that $E'$ also diverges. \MyQED

\label{lastpage01}
\end{document}